\newif\ifarxiv                                                                
\arxivtrue  

\ifarxiv
\documentclass[aps,prl,twocolumn,superscriptaddress,nobibnotes,10pt,floatfix]{revtex4-2} 
\else
\documentclass[pdflatex,sn-nature]{sn-jnl}
\fi

\ifarxiv
\RequirePackage{times}  
\RequirePackage{newtx}  
\def\bibcommenthead{}%
\newcommand{\abst}[1]{\begin{abstract}%
		#1
\end{abstract}}
\newcommand{\cwidth}{\columnwidth}

\else
\usepackage{amsmath,amssymb,amsfonts}%
\usepackage{amsthm}%
\usepackage{mathrsfs}%
\usepackage[title]{appendix}%
\newcommand{\onlinecite}[1]{\cite{#1}}
\newenvironment{widetext}{\vspace{2ex}}{\vspace{2ex}}
\newcommand{\cwidth}{3in}

\newcommand{\abst}[1]{\abstract{#1}}
\fi

\usepackage{graphicx}
\usepackage{braket}
\usepackage{xcolor}

\usepackage{pagecolor}
\pagecolor{white}

\newcommand{\figsubref}[2]{Fig.~\ref{#1}{#2}}
\newcommand{\efigref}[2]{Extended Data Fig.~\ref{#1}}
\usepackage[hidelinks]{hyperref}
\newcommand{\SIlink}[2]{\textit{see SI: ``\hyperref[#2]{#1}''}}

\newcommand{\firstSIlink}[2]{\textit{see Supplementary Information (SI): ``\hyperref[#2]{#1}"}}

\newcommand{\e}[1]{\ensuremath{\times 10^{#1}}}

\newcommand{\secref}[1]{Sec.~\ref{#1}}
\newcommand{\figref}[1]{Fig.~\ref{#1}}

\newif\ifusepng
\usepngtrue 

\newcommand{\smartincludegraphics}[2][]{%
	\ifusepng
	\includegraphics[#1]{#2.png}%
	\else
	\includegraphics[#1]{#2.pdf}%
	\fi
}

\usepackage{newunicodechar}
\newunicodechar{≈}{\ensuremath{\approx}}
\newunicodechar{≤}{\ensuremath{\le}}
\usepackage{upgreek}
\newunicodechar{μ}{\ensuremath{{\upmu}}}
\newunicodechar{Ω}{\ensuremath{\Omega}}
\newunicodechar{ϕ}{\ensuremath{\phi}}
\newunicodechar{π}{\ensuremath{\pi}}

\newcommand{\ts}[2]{{#1}_{\textnormal{#2}}} 
\newcommand{\tpp}{\ts{t}{pp}}
\newcommand{\Nosc}{\ensuremath{\ts{N}{osc}}}

\begin{document}
	
	\title{A digitally controlled silicon quantum processing unit}
	\author{Members of the HRL Quantum Team and Collaborators}
	\newcommand{\emailstatement}{Jacob Z. Blumoff (jzblumoff@hrl.com), Thaddeus D. Ladd (tdladd@hrl.com), Matthew D. Reed (mdreed@hrl.com)}
	\ifarxiv
	\thanks{Corresponding authors: \emailstatement}
	\else
	\affil{\orgname{HRL Laboratories, LLC}, \orgaddress{\street{3011 Malibu Canyon Rd.}, \city{Malibu}, \postcode{90265}, \state{California}, \country{USA}}}
	\fi
	
	\abst{
		Commercially-relevant quantum computers will require large numbers of high-performing qubits that can be manufactured, integrated, and controlled at scale. 
		Silicon exchange-only (EO) qubits~\cite{%
			bacon_universal_2000,%
			divincenzo_universal_2000,%
			fong_universal_2011,%
			langrock_reset-if-leaked_2020,%
			andrews_quantifying_2019,%
			ha_flexible_2022,%
			weinstein_universal_2023,%
			ha_two-dimensional_2025,%
			sun_full-permutation_2024,%
			madzik_operating_2025}
		are a strong candidate modality due to their control-signal simplicity and  compatibility with advanced semiconductor manufacturing~\cite{%
			zwerver_qubits_2022,%
			steinacker_industry-compatible_2025,%
			xue_cmos-based_2021}, 
		but questions remain around the achievability of sufficiently low noise and a scalable control and wiring solution~\cite{%
			xue_cmos-based_2021,%
			patra_cryo-cmos_2018,%
			mehrpoo_benefits_2019,%
			geck_control_2019,%
			pauka_cryogenic_2021,%
			bartee_spin-qubit_2025,%
			kiene_134_2021}. 
		Here we introduce a quantum processing unit composed of a custom-designed cryogenic CMOS controller, a novel high-density superconducting ribbon cable, and a low-noise EO qubit device. 
		The quantum chip features a three-rail array of 54 exchange-coupled quantum dots, configurable to host up to 18 EO qubits. 
		We integrate and use these components to demonstrate qubit performance for both single-qubit and entangling operations that advances the EO state of the art~\cite{weinstein_universal_2023,ha_two-dimensional_2025,madzik_operating_2025} by an order of magnitude. 
		We further validate this system by implementing a distance-5 repetition code~\cite{google_exponential_2021} and a \mbox{distance-2} quantum error detecting code~\cite{%
			linke_fault-tolerant_2017,%
			dam_end--end_2024,%
			takita_IBMrep_2017,%
			zhang_422_2026,%
			reichardt_422_2025,%
			undseth_weight-four_2026}
		then make detailed comparisons with simulations. 
		Our approach facilitates a utility-scale quantum computer with manageable operational and capital requirements.
	}
	
	\maketitle
	
	
	From transistors to integrated circuits to Si CMOS, the dominant computing technologies of the modern era have been determined by advantage in cost-effective manufacturability. 
	Building a commercially-viable \emph{quantum} computer will demand the same---efficient fabrication not just of the qubit chip, but also of the systems responsible for control signal generation and delivery. 
	The rapid advancement of the scale and fidelity of Si spin qubits~\cite{%
		mills_two-qubit_2022,%
		stano_review_2022,%
		burkard_semiconductor_2023,%
		madzik_operating_2025,%
		steinacker_industry-compatible_2025,%
		wu_simultaneous_2025,%
		takeda_quantum_2022,%
		undseth_weight-four_2026}, 
	driven in part by their compatibility with advanced semiconductor processes, has outpaced that of other system components~\cite{%
		xue_cmos-based_2021,%
		patra_cryo-cmos_2018,%
		mehrpoo_benefits_2019,%
		geck_control_2019,%
		pauka_cryogenic_2021,%
		bartee_spin-qubit_2025,%
		kiene_134_2021}.  
	In resolving this disconnect, a system architect must choose between placing control signal hardware at room temperature (burdened by the complexity of routing numerous high-fidelity analog links)~\cite{%
		mills_two-qubit_2022,%
		zwerver_qubits_2022,%
		steinacker_industry-compatible_2025,%
		madzik_operating_2025,%
		steinacker_industry-compatible_2025,%
		wu_simultaneous_2025,%
		takeda_quantum_2022,%
		undseth_weight-four_2026}, 
	co-locating it with the mK qubits (with tyrannical cooling requirements)~\cite{bartee_spin-qubit_2025,jordan_quantum_2026}, 
	or using an intermediate temperature~\cite{%
		xue_cmos-based_2021,%
		patra_cryo-cmos_2018,%
		mehrpoo_benefits_2019,%
		geck_control_2019,%
		franke_rents_2019,%
		pauka_cryogenic_2021,%
		tien_222_2026}.
	This third option enjoys greater Carnot efficiency and avoids the room temperature ``wiring bottleneck,''~\cite{franke_rents_2019} but must transmit numerous signals to mK while maintaining thermal isolation.
	
	Our system follows the third option, using a fully-custom low-power CMOS chip operating at 4~K to generate all time-varying qubit control signals. 
	We resolve the thermal isolation problem by routing those signals through a high-density, low-thermal-conductance superconducting ribbon cable to the quantum chip at mK (\figsubref{qpufig}{a}).
	As integrated, this ``quantum processing unit'' (QPU, photograph in \efigref{iqpu_ed} \ ) has modest requirements for its connections to room temperature: only digital communication, qubit readout, and static biases (\figsubref{qpufig}{b}). 
	The superconducting interconnect maintains excellent signal integrity and bandwidth while also serving as a thermal standoff, preventing the 4~K controller from unduly heating the mK stage. 
	EO qubits also pair well with power-constrained CMOS signal generation since their required control waveforms (shape-insensitive baseband voltage pulses) are similar to standard digital signals. 
	The qubit chip, ribbon, and controller are each fabricated with semiconductor wafer processes, a strong indicator for manufacturability. 
	The QPU is designed for the function that will most occupy any utility-scale quantum computer: error correction. 
	Below, we verify this capability with repeated syndrome measurements in {[}3,1,3{]} and {[}5,1,5{]} repetition codes as well as the {[}{[}4,2,2{]}{]} quantum error detecting code.
	
	\section{Quantum chip}\label{quantum-chip}
	
\begin{figure*}[t]
		\centering
		\smartincludegraphics[width=\textwidth]{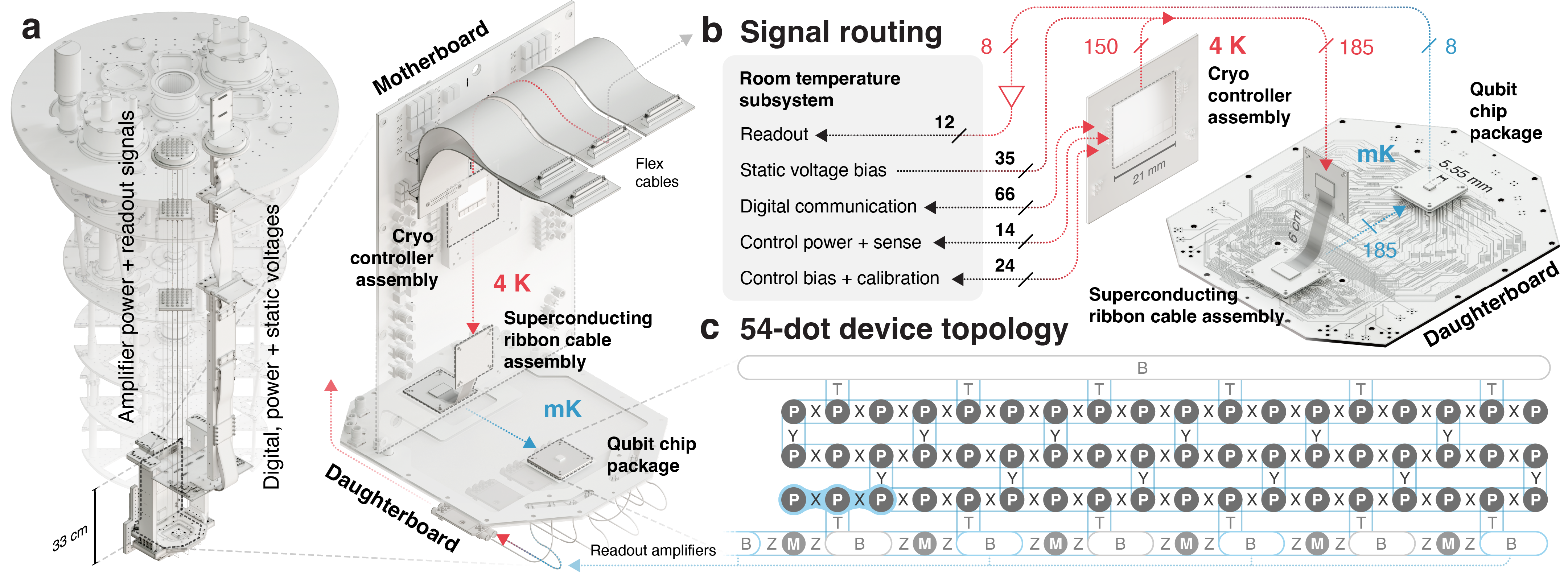}
		\caption{
			\label{qpufig}
			\textbf{An integrated quantum processor.}
			\textbf{a,} 
			The QPU is installed at the bottom of a commercial dilution refrigerator and connected to room temperature with flex cables and coaxial lines (\SIlink{Flex cable}{flex-cable}).
			It is a tightly integrated system consisting of a controller motherboard at 4~K, qubit daughterboard at mK, and superconducting ribbon cable bridging the two.
			Custom-designed transimpedance amplifiers (TIAs, \SIlink{4~K transimpedance amplifier}{transimpedance-amplifier}) are thermalized to 4~K but located below the daughterboard to minimize capacitance. 
			\textbf{b,} 
			The room temperature subsystem provides digital signaling and static current and voltage biases. 
			Qubit readout signals are also digitized at room temperature, though we anticipate future systems will incorporate that function inside the cryostat~\cite{kiene_134_2021}.
			The ribbon cable transmits all 150 time-varying qubit control signals and 35 static device biases from 4~K to mK with minimal thermal load. 
			\textbf{c,} 
			Topology schematic of the quantum chip. 
			The 54 dots are arrayed in three sparsely-coupled rails and support up to 18 encoded qubits arrayed in a 3$\times$6 square lattice with nearest-neighbor coupling. 
			We flexibly partition connected trios of dots into qubits, motivated by yield, performance, or desired connectivity~\cite{ha_two-dimensional_2025}. 
			``P'' gates control dot occupancy and adjacent ``X'' and ``Y'' gates control the tunnel coupling between dots. 
			Dot loading, measurement, and qubit initialization utilize reservoirs accumulated by eight ``B'' gates coupled to the array by twelve ``T'' gates. 
			``M'' and ``Z'' gates form six dot charge sensors used for device tune-up and qubit state readout~\cite{blumoff_fast_2022}. 
			Various field gates deplete charge under inactive regions. 
			The biases applied to these electrodes are determined with a combination of manual and automated methods (\SIlink{Tune-up and calibration}{tune-up-and-calibration}).
		}
	\end{figure*}

	Our qubit devices each feature 54 low-noise exchange-coupled quantum dots distributed over three rails (\figsubref{qpufig}{c}). 
	Qubits are encoded in the joint spin state of three individual electrons in three dots (see \hyperref[methods-section]{Methods}). 
	Electrons are vertically confined in a Si/SiGe quantum well which is engineered to increase valley splitting.  
	Si and Ge in the heterostructure are isotopically enhanced to reduce magnetic noise.
	Lateral confinement is achieved using electrostatic gates patterned on top of the heterostructure through a proprietary 200-mm wafer foundry process.
	The intrinsic charge noise of these devices is more than an order of magnitude lower than our previous Single-Layer Etch-Defined Gate Electrode (SLEDGE) technology~\cite{ha_flexible_2022,weinstein_universal_2023,ha_two-dimensional_2025}. 
	Additional device details are found in Methods. 
	Following fabrication, wafers are probed to select high-yielding die then diced and bump bonded to a fine-pitch laminate LGA package (\efigref{packagedchip} \ ). 
	Devices are installed on a daughterboard thermalized to the mixing chamber of a dilution refrigerator. 
	When fully integrated, we observe an average electron temperature of 150~mK (\firstSIlink{Environmental control}{environmental-control}).

	\section{Superconducting interconnect}\label{superconducting-interconnect}
	
	A high-density superconducting ribbon cable routes signals from the controller to the qubit chip. 
	The cable is fabricated from Nb on polyimide~\cite{tuckerman_flexible_2016,smith_improved_2024} using standard semiconductor wafer processes. 
	Though 296 coaxial signal lines are arranged in a single layer approximately 1 cm wide, crosstalk within the cable up to 10 GHz is limited to $<-80$~dB. 
	The ribbon maintains an effective thermal standoff due to its use of superconducting metal and small feature sizes, conducting less than 10~$\mu$W from the 4~K stage to the mixing chamber. 
	Laminate adapters at both ends allow the cable to be reversibly attached to both the daughterboard and motherboard (\efigref{cablephoto} \ ). 
	The round-trip electrical length from controller to qubit is roughly 3~ns.
	
	\section{Cryogenic CMOS controller}\label{cryogenic-cmos-controller}
	
	\begin{figure}[t]
		\centering
		\smartincludegraphics[width=\cwidth]{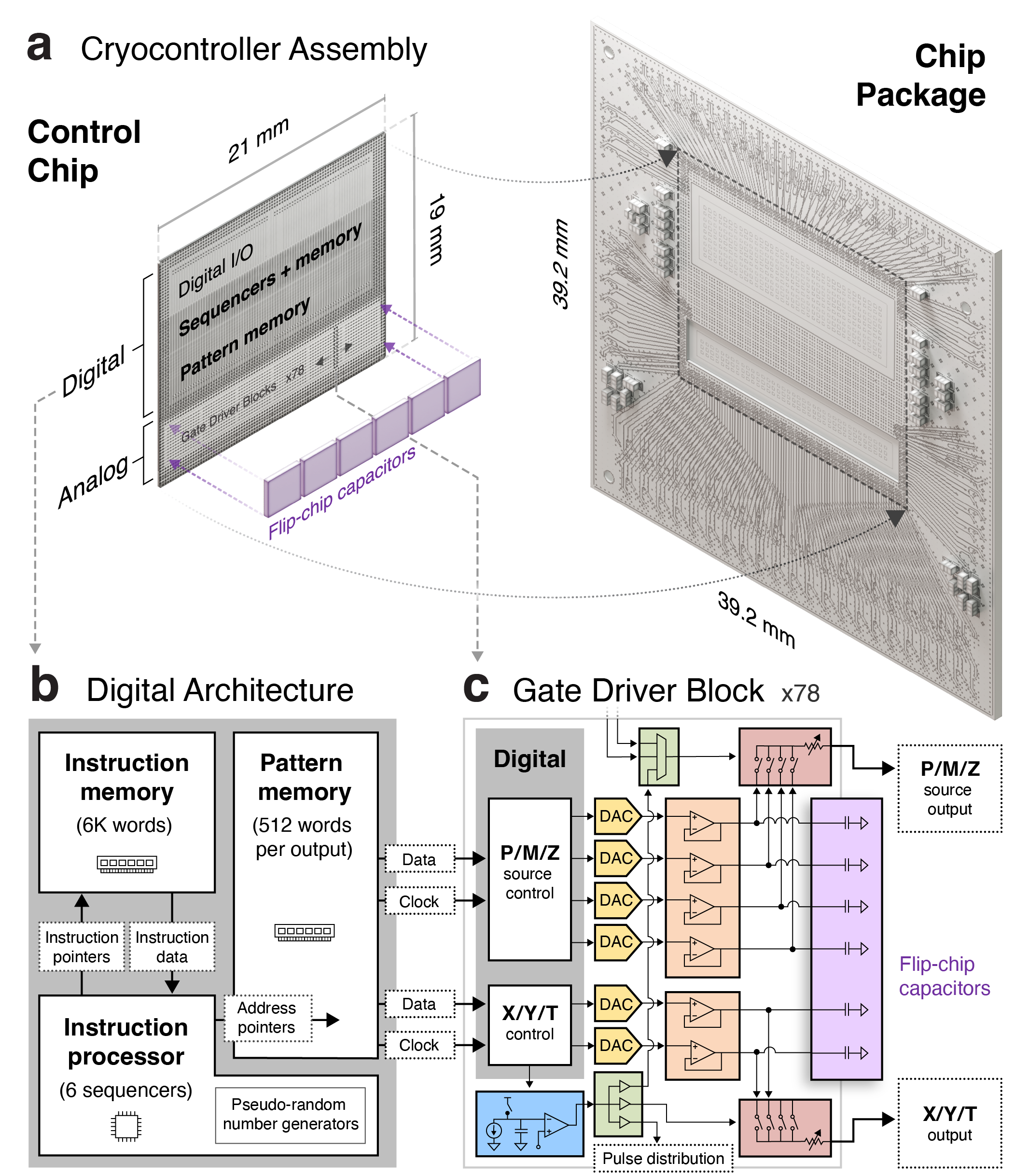}
		\caption{
			\label{swfig}
			\textbf{The cryogenic CMOS controller.}
			\textbf{a,} 
			The controller is a multi-component assembly: custom high-density capacitor arrays are bump-bonded to the CMOS die which is then integrated into a fine-pitch laminate LGA package (\efigref{controllerphoto} \ ). 
			The CMOS die comprises roughly 70 million transistors and has four distinct functional areas. 
			\textbf{b,}~The digital architecture is responsible for interpreting instructions and supplying every analog block with a pattern memory pointer every clock cycle.
			\textbf{c,}~The gate driver layer is composed of an array of 78 nearly-identical analog blocks (one pictured) collectively comprising 156 output channels, 366 DACs, and 78 pulse generators (\SIlink{Cryo-CMOS qubit controller}{cryo-cmos-qubit-controller}). 
			The DACs (yellow, 0 to 1~V, RMS step size \textless{}~10~$\mu$V) are buffered by amplifiers (orange) with output capacitance provided by the capacitor arrays (purple). 
			Each exchange gate has a uniquely-associated pulse generator (blue) which emits timing signals (typical duration 400~ps to 6~ns, with approximately picosecond resolution). 
			To actuate symmetric exchange operation \cite{reed_reduced_2016}, these signals are distributed (green) to switch drivers (red) for the X gate and its neighboring P gates, generating synchronized voltage waveforms ($T_{10\%-90\%}\approx 150$~ps). 
			Each gate driver block has a local digital module to store configurable parameters such as output impedance and filtering.
		}
	\end{figure}
	
	The cryo-controller (\figsubref{swfig}{a}) is a mixed-signal system-on-chip fabricated in a commercial 130-nm RF CMOS process. 
	Circuit designs are optimized at the transistor level for performance at 4~K. 
	The controller has three primary functions: device tune-up, qubit state preparation and measurement (SPAM), and qubit state manipulation. 
	The first two functions are achieved by dynamically changing DAC output voltages and the third with fast, time-modulated, gate-synchronized switching between fixed DAC voltages.
	
	The controller's digital architecture (\figsubref{swfig}{b}) features an autonomous, multi-sequencer engine that operates at up to 250 MHz and uses a custom instruction set architecture for qubit control. 
	On-chip pseudo-random number generators enable memory-efficient characterization of qubit performance. 
	The controller consumes $\le$ 3.5~W in typical operation (\SIlink{Power consumption}{power-consumption}). 
	Pattern memory associated with each output stores parameters for voltages and pulse widths. 
	User programs are converted into machine code using a co-designed compiler and stored in a shared pool of 6144 words of instruction memory. 
	Every clock cycle, each of the sequencers interprets an instruction and updates a pattern memory pointer specifying analog behavior to user-assigned output blocks (\figsubref{swfig}{c}). 
	Once initialized, the program runs without further communication with room temperature. 
	The digital architecture uses a 52-line parallel bus and a Serial Peripheral Interface (SPI) for configuration, diagnostics, and memory readback.
	
	\section{Qubit performance}\label{qubit-performance}
	
	\begin{figure}[h]
		\centering
		\smartincludegraphics[width=\cwidth]{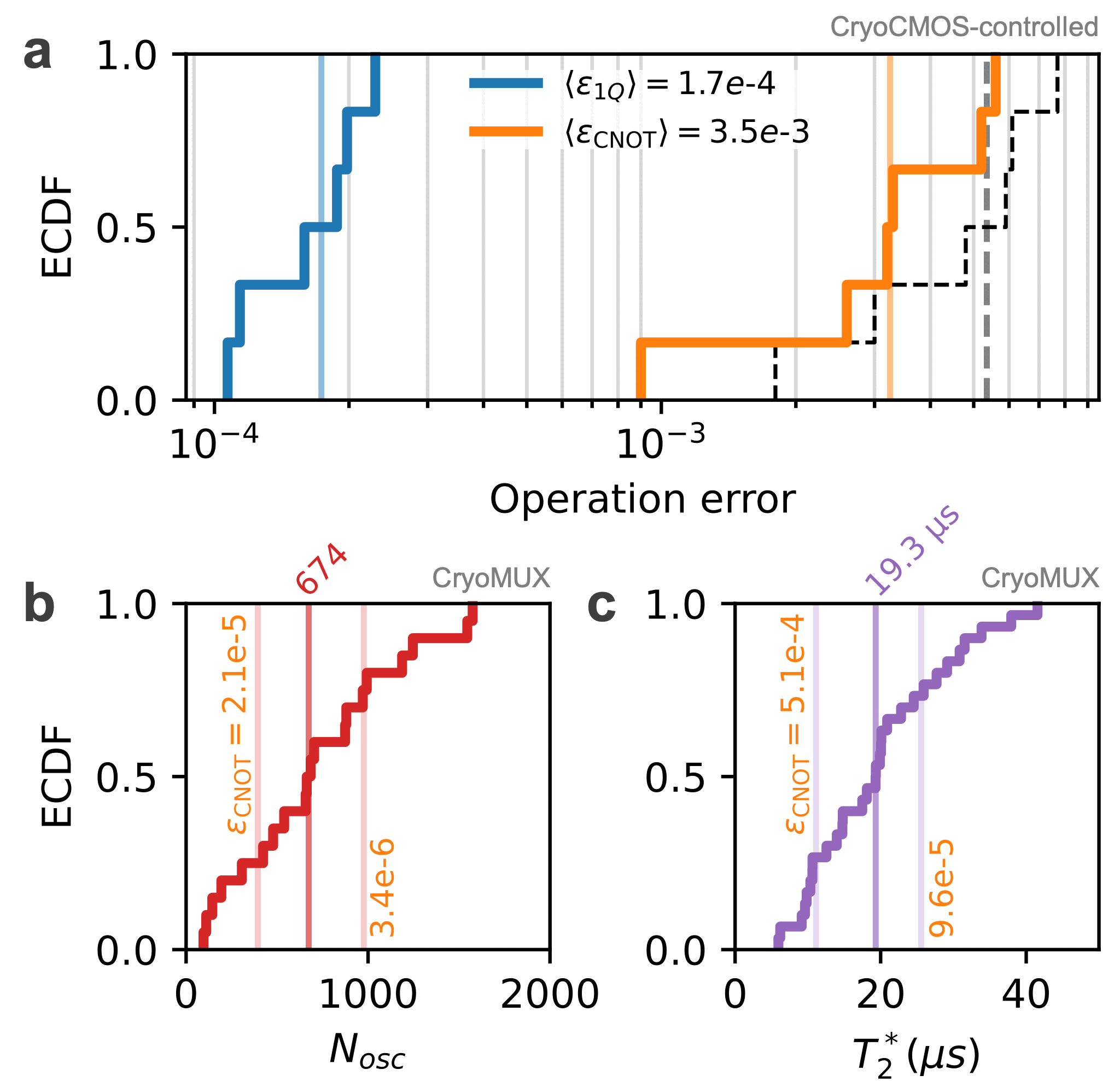}
		\caption{
			\label{performancefig}
			\textbf{Qubit performance.} 
			\textbf{a,} 
			Empirical cumulative distribution functions (ECDF) for gate error of cryo-controlled EO qubits. 
			Mean errors of 2\e{-4} for single-qubit gates (blue) and 3\e{-3} for CNOT (orange) are indicated by vertical bars. 
			Single-qubit gates, each composed of 4 exchange pulses, are characterized with blind randomized benchmarking (BRB)~\cite{andrews_quantifying_2019}. CNOT sequences~\cite{divincenzo_universal_2000,fong_universal_2011,weinstein_universal_2023}, composed of 37 pulses in 45 timesteps, are characterized with an interleaved BRB variant (\SIlink{Blind randomized benchmarking (BRB)}{blind-randomized-benchmarking-brb}). 
			The lowest reproducible CNOT error was 9\e{-4}. 
			The dashed line is the fidelity of CNOT gates with simultaneous dynamical decoupling applied to all other qubits; this degradation is consistent with their slightly slower pulse cadence. 
			\textbf{b,}~Representative 
			charge noise data acquired using a cryoMUX system from a single device similar to that used in panel a.
			Vertical lines indicate the median (674) and quartile (394 and 977) number of exchange oscillations before 1/$e$ amplitude decay in an experiment sweeping evolution duration at constant exchange energy, $N_{\text{osc}}$~\cite{reed_reduced_2016}. 
			Orange vertical text along the quartiles represent the simulated error contribution to CNOT error due to those values of charge noise. 
			\textbf{c,} Magnetic noise data taken from unique dot pairings on the same device in the same cryoMUX system. 
			Vertical lines indicate median (19.3~$\mu$s) and quartile (11.1~$\mu$s and 25.6~$\mu$s) decay envelope of a singlet decay experiment, \(T_{2}^{*}\)~\cite{kerckhoff_magnetic_2021}. 
			Simulated contributions to CNOT error are again annotated at quartile values.
		}
	\end{figure}
	
	Verifying gate performance is a principal consideration as it strongly influences the overhead required for quantum error correction (QEC). 
	The QPU used for the [[4,2,2]] experiment exhibits single-qubit and CNOT performance metrics (\figsubref{performancefig}{a}) that improve on the prior state of the art by an order of magnitude~\cite{ha_two-dimensional_2025,madzik_operating_2025,weinstein_universal_2023}. 
	For budgeting purposes, we characterize noise intrinsic to the qubit chips using a purpose-designed ``cryoMUX'' system. 
	Those systems utilize a custom cryogenic demultiplexer (MUX), similar in functionality but different in operation to those in Refs.~\onlinecite{puddy_multiplexed_2015,paquelet_wuetz_multiplexed_2020,pauka_characterizing_2020,eastoe_method_2025,schmidt_135_2025,thomas_rapid_2025}, that is paired with room-temperature electronics and are capable of controlling a configurable subset of nine dots. 
	In \figsubref{performancefig}{b} and c, we show charge and magnetic noise data from a qubit chip of the same type as those used for the multiqubit experiments. 
	The measured intrinsic charge noise also improves on the prior state of the art by approximately a factor of ten~\cite{ha_two-dimensional_2025,%
		madzik_operating_2025}.
	Mean values of intrinsic device charge and magnetic noise would collectively contribute only 0.02\% to absolute CNOT error. 
	Instead, modeling and additional system characterization indicate that ``extrinsic'' terms such as static magnetic field gradients and contextual pulse miscalibration dominate observed error.

	\section{Repetition codes}\label{repetition-codes}
	
\begin{figure*}
		\smartincludegraphics[width=\textwidth]{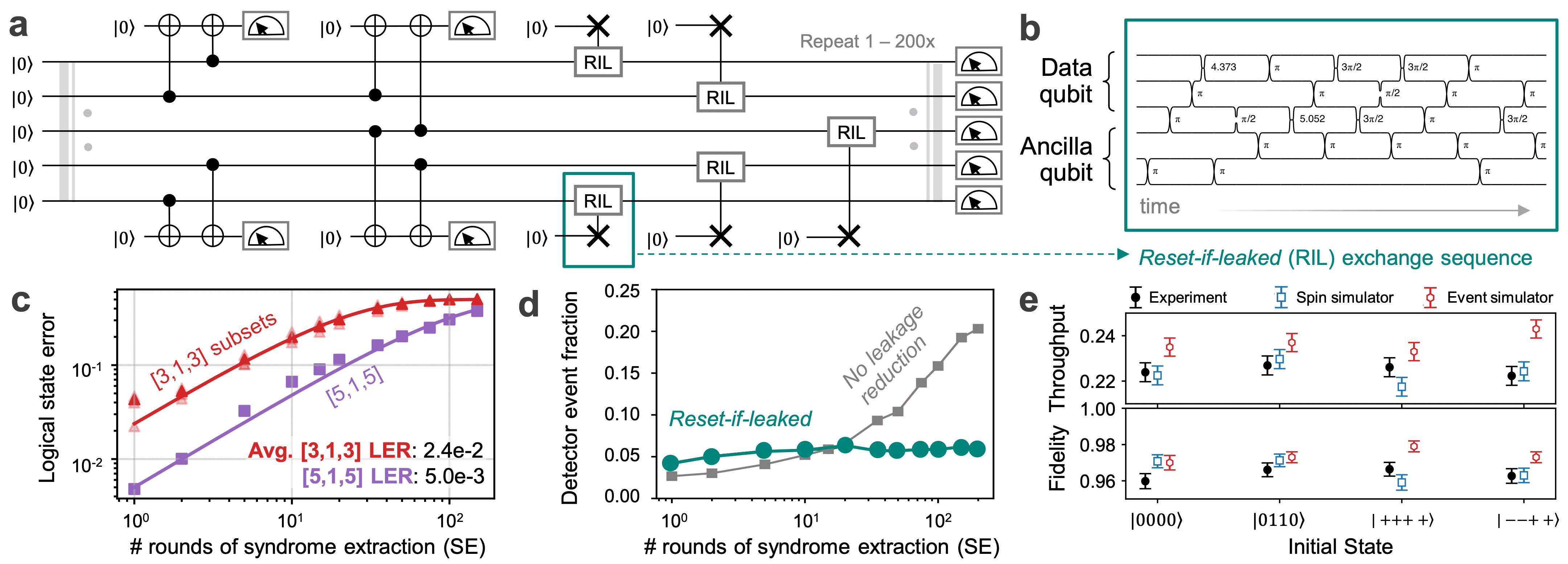}
		\caption{
			\label{mqfig}
			\textbf{Multiqubit validation results.}
			\textbf{a,} 
			Simplified circuit for the distance-5 repetition code.
			Classical information is encoded in five data qubits which are repeatedly interrogated for bit flips using two ancilla qubits. 
			Each syndrome extraction (SE) round is broken into two waves and ancillae are re-used. 
			A single instance of the 200-round experiment consists of 1,805 initializations, 805 measurements, and 335,422 exchange pulses.
			\textbf{b,}~The 
			leakage reduction units (LRUs) used here are composed of an ancilla initialization and a Reset-If-Leaked (RIL) control sequence~\cite{langrock_reset-if-leaked_2020} consisting of 24 or 21 exchange pulses in 30 timesteps (24-pulse variant depicted here). 
			Each angle in the figure represents a partial swap between neighboring spins that is actuated by corresponding controller-generated voltage pulses. 
			\textbf{c,}~We 
			sweep the number of syndrome extraction rounds per shot in the {[}5,1,5{]} sequence quasi-logarithmically from 1 to 200 and repeat each 50,000 times. 
			The syndrome data, interpreted with a naïve parity check decoder, indicate a logical error rate (LER) of 5.0$\e{-3}$ for the distance 5 code (purple) and 2.3\e{-2} for the average of distance-3 subsets (red), yielding a distance improvement factor \(\Lambda_{5/3} = 4.7\). 
			\textbf{d,}~The 
			detector event fraction of a fidelity-optimized {[}3,1,3{]} experiment with (green) and without (gray) LRUs included in each syndrome extraction wave show the effectiveness of LRUs at preventing the build-up of leakage. 
			The LER of this instantiation of the distance-3 code was 3.2\e{-3} when LRUs were included.
			\textbf{e,}~[[4,2,2]]-code 
			logical fidelity (defined as the projection of the post-selected state onto the expected state) and throughput after three rounds of SE are comparable for all four input states. 
			Spin- and event-level simulations show relatively good model-hardware agreement (\SIlink{Multiqubit experiments}{multiqubit-experiments}).
		}
	\end{figure*}

	High-fidelity one- and two-qubit operations are necessary but not sufficient for effective computation---we must also show that performance does not degrade as circuit complexity grows. 
	We first validate the multiqubit system performance by executing ``classical'' repetition codes. 
	Repetition codes are of interest because they enable qubit-efficient study of low-probability correlated errors~\cite{google_exponential_2021} which may unacceptably degrade QEC performance. 
	We realize a distance-5 repetition code (simplified circuit illustrated in \figsubref{mqfig}{a}, \SIlink{[5,1,5] demonstration}{demonstration-515}) using seven EO qubits. 
	Critically, this circuit applies first-order ``NZ1'' full-permutation dynamical decoupling~\cite{sun_full-permutation_2024} to all idling qubits during all SPAM, two-qubit, and leakage-reduction operations. 
	The circuit also employs parallel state preparation, measurement, and exchange operations~\cite{madzik_operating_2025}. Comparing the distance-5 logical performance (LER $\approx 5\e{-3}$) to the average performance of distance-3 subsets assessed from the same underlying data, we observe a scaling factor $\Lambda_{5/3} = 4.7$ (\figsubref{mqfig}{c}). 
	Data from this experiment show as-yet unexplained fluctuations in error rate which were not observed in either the distance-3 or [[4,2,2]] data discussed below (\SIlink{Spikes in detection events}{spikes-in-detection-events}).
	
	The three-spin encoding of this qubit minimizes any distinction between bit- or phase-flips in repetition codes but introduces the important factor of spin-state leakage out of the computational subspace~\cite{%
		divincenzo_universal_2000,%
		fong_universal_2011,%
		andrews_quantifying_2019,%
		langrock_reset-if-leaked_2020,%
		weinstein_universal_2023}.
	We mitigate that concern in all multiqubit circuits shown here with the addition of LRUs (\figsubref{mqfig}{b}). 
	These gadgets conditionally reset data qubits if and only if they have leaked out of the computational space. 
	We confirm the effectiveness of this mitigation with a distance-3 repetition code experiment using a selection of qubits optimized for {[}3,1,3{]} performance (\SIlink{[3,1,3] demonstration}{demonstration-313}). 
	As expected, we observe that the inclusion of LRUs in the syndrome extraction (SE) circuit leads to a stable detector event fraction (DEF), but their omission leads to leakage accumulation and a DEF that grows with round (\figsubref{mqfig}{d}). 
	This stability is likely a prerequisite for effective QEC.
	
	We improve confidence in our error model by exhaustively comparing the {[}3,1,3{]} results to theoretical predictions. 
	In that case, a qubit-level event simulator (calibrated with measured gate error rates from Interleaved BRB experiments performed in-situ and assignment fidelity measurements for SPAM) agreed with the experimental LER to 15\% relative inaccuracy. 
	We found similarly good agreement using a spin-level simulator calibrated on measured dot-level noise terms ($N_{\text{osc}}, T_{2}^{*}$ static field gradients, etc.) and assignment fidelity. 
	An ersatz miscalibration term (chosen to match simulated and measured CNOT BIRB error) was added to reflect errors arising from deterministic imperfections in signal generation and transmission. 
	In this simulation, that term is responsible for approximately 80\% of detection events. 
	The agreement between these two simulators and experiment is an encouraging indication that our system exhibits emergent Markovianity and is well-described by a simple and scalable performance model. 
	More discussion of that measurement, the associated simulations, and error attribution can be found in the SI sections ``\hyperref[simulation]{Simulation}" and  ``\hyperref[multiqubit-experiments]{Multiqubit experiments}."
	
	\section{Quantum error detection}\label{quantum-error-detection}
	
	We perform a final system validation by implementing a quantum error detecting (QED) code that is potentially sensitive to error types not detected by classical codes. 
	The [[4,2,2]] code, previously demonstrated in trapped ion qubits~\cite{linke_fault-tolerant_2017,dam_end--end_2024}, superconducting qubits~\cite{takita_IBMrep_2017}, and neutral atom qubits~\cite{reichardt_422_2025} embeds two logical qubits into four physical qubits and enables the detection of any single-qubit (weight-1) error. 
	Similar syndrome measurements~\cite{undseth_weight-four_2026} and logical operation~\cite{zhang_422_2026} were recently demonstrated in silicon qubits.
	Each round of syndrome extraction entails two weight-4 measurements (\(XXXX\) and \(ZZZZ\)), interleaved flag qubit measurements, and leakage reduction.
	
	We implement this code with six physical qubits on a system similar to, but slightly improved from, that of the {[}5,1,5{]} experiment (\SIlink{{[}{[}4,2,2{]}{]} demonstration}{demonstration-422}). 
	During syndrome extraction, we compile all single-qubit gates into composite two-qubit operations.
	We prepare the data qubits into each of four different initial states $\ket{0000}, \ket{0110}, \ket{{+}{+}{+}{+}},$ and $\ket{{-}{-}{+}{+}}$ and perform three rounds of syndrome extraction before measuring their final states. 
	The first round of syndrome measurements predictably projects these into different logical states ($\ket{\bar{0}\bar{0}}, \ket{\bar{1}\bar{1}}, \ket{\bar{+}\bar{+}},$ and $\ket{\bar{-}\bar{+}}$, respectively). 
	When preparing in the $X$-basis we switch the ordering of two stabilizer measurements to project into a logical state more quickly.
	
	We assess code performance by post-selecting on all error-detecting measurements: syndrome results, flag measurements, and a last syndrome computed from the final data qubit measurements. 
	This process ultimately rejects roughly 77\% of the 10,000 experimental shots acquired per preparation basis. 
	We then analyze the final measurements and assess a two-logical-qubit state fidelity after three rounds, finding $F_L=0.95$ largely independent of the initial state (\figsubref{mqfig}{e}). 
	If we instead ignore the error-detecting measurements, we find an average $F_L=0.59$, indicating the effectiveness of QED.
	
	To examine correlations between error events, we turn to detector-error-model (DEM) analysis~\cite{chen_decoders_2022,arms_estimating_2026,blume-kohout_estimating_2025}. 
	As shown in \hyperref[dem-analysis-422]{SI}, the DEM analysis of the empirical data and of spin simulations show no statistically-relevant unexpected events in the [[4,2,2]] experiment after appropriate grouping of events with the same weight. 
	This indication of ``well-behaved'' errors is encouraging for larger-scale QEC.
	
	\section{Towards commercial relevance}\label{towards-commercial-relevance}
	
	Our demonstration marks a milestone in the technological maturity of semiconductor spin qubits. 
	Relative to prior reporting on these qubits \cite{weinstein_universal_2023,madzik_operating_2025}, we have shown transformative improvements in operational fidelity and power- and space-efficient cryogenic control. 
	Conventional concerns such as charge noise, valley splitting, and leakage out of the computational space are no longer limiting. 
	Rather, the quality of our demonstration is set by how well each of the system sub-components are made to work together as a whole. 
	Development now shifts to integration improvements such as magnetic hygiene, power delivery, internal signal integrity, and device calibration, all having highly feasible engineering solutions. 
	Our ability to predict system-level performance based only on local measures like gate error and noise parametrics also indicates that the fundamental design decisions of the technology are sound.
	
	We believe that a system design integrating 4~K cryogenic control, low-noise mK EO qubits, and superconducting interconnection provides a practical foundation for a utility-scale fault-tolerant quantum computer in a single, commercial cryostat. 
	To get there, work still remains in maturing both the components and their integration. 
	Necessary advances include a manufacturable interconnect and qubit back-end routing solution, controllers with even lower power consumption per qubit, and improved device uniformity to reduce tune-up overhead. 
	A system-level quantum architecture tailored to the connectivity, noise environment, and other properties of EO qubits must also be elaborated. 
	Taken together, the scale and cost of a system anticipated by the prototype reported here position semiconductor spin qubits as a leading technology to realize commercially-relevant quantum computing.
	
	\ifarxiv
	\begin{widetext}
\renewcommand{\author}[1]{#1}
\newcommand{\HRL}{\textsuperscript{\scriptsize 1}}
\newcommand{\Boeing}{\textsuperscript{\scriptsize 2}}
\noindent
\textbf{Members of HRL Quantum Team and Collaborators}
\\
\noindent
\author{Michael Abraham}{\HRL},  
\author{Edwin Acuna}{\HRL},
\author{Tower S. Adams}{\HRL},
\author{Moonmoon Akmal}{\HRL},
\author{Matthew R. Alfaro}{\HRL},   
\author{I. Alvarado}{\HRL},
\author{Jacob Amontree}{\HRL},
\author{Carter Andrews}{\HRL},
\author{Reed W. Andrews}{\HRL},
\author{Michael Antcliffe}{\HRL},   
% ...
\vspace{1ex}\\
\noindent {\HRL}HRL Laboratories, LLC
\\
\noindent {\Boeing}The Boeing Company
	\end{widetext}
	\fi
	
	\section{Methods}\label{methods-section}
	
	\subsection{Qubit Chip}\label{qubit-chip}
	
	SiGe heterostructures for the qubit chip were grown on 200~mm silicon wafers by chemical vapor deposition (CVD). 
	A strain-relaxed SiGe buffer layer was grown on top of a Si wafer~\cite{deelman_metamorphic_2016}, terminating with a SiGe layer that matches the Ge composition of the SiGe barriers to the quantum well. 
	The SiGe surface of the buffer received chemical mechanical polishing to planarize the surface before CVD growth of the quantum well and barriers. 
	The barrier layers are Si\textsubscript{$1-x$}Ge\textsubscript{$x$} grown using SiH\textsubscript{4} and GeH\textsubscript{4} with alloy compositions in the range $x=0.25$ to $0.35$. 
	The SiGe heterostructure, enriched with $^{28}$Si and depleted of $^{73}$Ge, was engineered to increase valley splitting energy~\cite{%
		zhang_genetic_2013,%
		marcks_valley_2025,%
		mcjunkin_sige_2022,%
		stehouwer_engineering_2025}.
	
	Qubit chips were fabricated on the SiGe heterostructure wafers described above, leveraging CMOS-compatible process integration. 
	This combines the proprietary, qubit-specific front-end-of-line integration with industry-standard back-end-of-line interconnect integration. 
	Ohmic contacts and electrically inactive regions were defined by optical lithography and ion implantation of P and Ar, respectively. 
	The qubit gate design, with a minimum feature pitch of 70 nm, was patterned with electron-beam lithography. 
	Gate contacts and routing were formed using a self-aligned dual damascene process with a total of three minimum-pitch routing levels. 
	At each level, designs for vias and for routing are pattern-transferred into a hard mask using electron beam lithography.
	Dry etching opened vias and routing trenches within an SiO\textsubscript{2} interlayer dielectric. 
	CVD W metallization, with a TiN liner, filled the vias and trenches which were subsequently isolated using chemical mechanical planarization.
	
	Next, three additional larger-pitch metal layers were formed to contact the W layers and fan out the routing lines to the pad pitch. 
	These three layers were formed in a co-planar waveguide stack, with each layer patterned by optical lithography. 
	Each layer was formed with a single damascene via made of W followed by a routing or ground plane layer composed of Nb using a subtractive integration. 
	Al bond pads were formed to contact the Nb routing and allow for wafer probing of the qubit devices. 
	Following wafer probing, In bumps were added to the Al pads and the wafers were diced to singulate each qubit chip for later flip-chip packaging.
	
	\subsection{Exchange-only qubits}\label{exchange-only-qubits}
	
	Exchange-only (EO)~\cite{%
		bacon_universal_2000,%
		divincenzo_universal_2000,%
		fong_universal_2011,%
		langrock_reset-if-leaked_2020,%
		madzik_operating_2025,%
		ha_flexible_2022,%
		weinstein_universal_2023,%
		ha_two-dimensional_2025,%
		andrews_quantifying_2019,%
		sun_full-permutation_2024}
	means that the only physical interactions employed on electrons trapped in quantum dots are 
	(1) initialization into antisymmetric spin-singlet states, i.e. $\ket{\uparrow\downarrow}-\ket{\downarrow\uparrow}$, 
	(2) execution of partial spin swaps $U(\theta) = \cos\theta + i \sin\theta\times\text{SWAP}$, where SWAP acts between a pair of spins, and 
	(3) pairwise measurement of singlet vs. triplet (i.e. $\ket{\uparrow\uparrow}, \ket{\uparrow\downarrow}+\ket{\downarrow\uparrow}, \ket{\downarrow\downarrow}$. 
	All such interactions are available from direct-current (DC) or ``baseband'' pulsing using the Pauli-spin blockade mechanism, which uses electrode voltages that push electrons closer or further from another. 
	The resulting ``exchange interaction'' reduces the energy of the singlet state relative to all triplet states due to a combination of Coulomb repulsion and the Pauli-exclusion principle. 
	No magnetic fields of any kind are required, and the entire spin-system, barring errors, would maintain total spherical symmetry.
	
	Qubits are formed from three spins in three dots using the symmetry-respecting encoding of a decoherence free subsystem~\cite{bacon_universal_2000,%
		divincenzo_universal_2000,%
		fong_universal_2011,%
		langrock_reset-if-leaked_2020}.
	The eight resultant spin states may be described in the basis denoted as \(|S_{12},S;m\rangle\) where \emph{S}\textsubscript{12} encodes the total spin angular momentum of the first two spins (\emph{S}\textsubscript{12}=0, singlet; \emph{S}\textsubscript{12}=1, triplet), \emph{S} encodes the total spin of all three spins, and \emph{m} encodes the spin projection. 
	Ideal exchange-only initialization, logic, and read-out address only on the \emph{S}\textsubscript{12} degree of freedom, and \emph{S} would remain, for all encoded qubits, with constant value \emph{S}=1/2. 
	The projection m=±1/2 is a random and inconsequential gauge degree of freedom. 
	Global magnetic fields \emph{B} impart global phases which are also inconsequential. 
	The angles $\theta$ for each exchange ``gate'' $U(\theta)$, which we call exchange angles or ``exchangles'', are determined by the time integral of the exchange energy, and as such are calibrated against DC voltage pulses with no direct dependence on pulse shape. 
	Accordingly, the control of this qubit can be realized by (precise) switching between pairs of voltage levels---highly reminiscent of digital logic---and so is amenable to the energy-efficient cryocontrol system we have described.
	
	The dominant qubit errors can be categorized into two categories: (1) miscalibration of exchangles $\theta$, or (2) dot-level gradients of the magnetic field. 
	In multiqubit operation, \emph{both} error types cause some fraction of encoded error (i.e. bit-flips and phase flips on the \emph{S}\textsubscript{12} number) and some fraction of leakage error (i.e. flips from \emph{S}=1/2 to \emph{S}=3/2). 
	Quantitative modeling of these contributions, as well as the translation of physical processes into exchangle miscalibrations and magnetic gradients, are discussed in the SI section ``\hyperref[simulation]{Simulation}".
	Exchange miscalibration can be incoherent (i.e. dynamically fluctuating charge noise from the qubit chip, interconnect, or controller) or coherent (i.e. static or contextual miscalibration from imperfect signal generation or transmission); the last of these dominates in the demonstrations presented. 
	Similarly, gradients of the magnetic field can arise from fluctuating hyperfine interactions with \textsuperscript{29}Si and \textsuperscript{73}Ge nuclear spins or from static magnetic screening and flux-trapping arising from local superconductivity. 
	The former source of gradients is mitigated using isotopic enhancement, the limitations of which determine the $T_2^*$ values of approximately 20 $\mu$s (\figsubref{performancefig}{c}).
	
	The latter source is mitigated by limiting the magnetic field of the device to be close to zero, for which we use current coils external to the fridge to vector-cancel the Earth's magnetic field. 
	As such, all experiments described in this paper are at approximately zero magnetic field. 
	Dephasing due to residual magnetic gradients of both types is reduced for idling qubits in our demonstrations using permutational dynamical decoupling. 
	This permutes sets of three spins using SWAP operations $U(\pi)$ in a way which causes local gradients across triple-dots to average to global fields impacting only the irrelevant gauge number \emph{m}~\cite{sun_full-permutation_2024}.


	
	\setcounter{figure}{0}
	\renewcommand{\figurename}{Extended Data Fig.}
	
	\renewcommand{\theHfigure}{ED.\arabic{figure}}
	
	\begin{figure*}[h!]
		\section{Extended data figures}\label{extended-figures} 
		
		\centering
		\includegraphics[width=90mm]{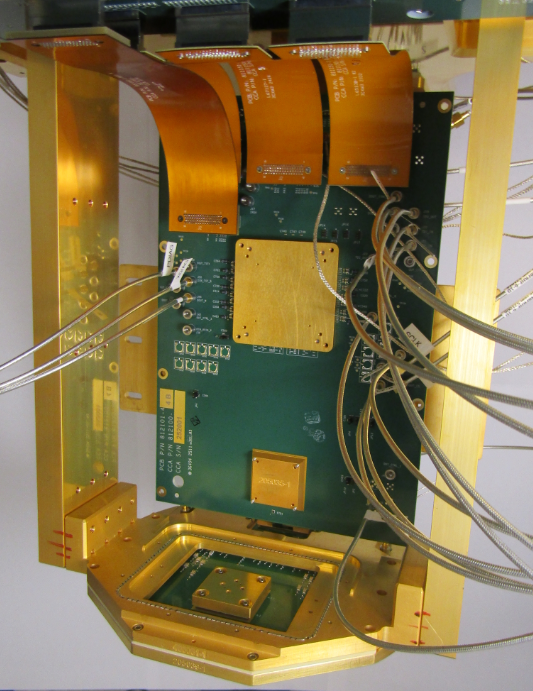}
		\caption{
			\label{iqpu_ed}
			\textbf{Integrated QPU}. 
			The motherboard (vertical) houses the cryo-controller and the daughterboard (horizontal) hosts the qubit chip. 
			A light-tight metal box encloses the daughterboard to shield the qubit chip from ambient radiation. 
			The two are bridged by a superconducting ribbon cable (largely obscured here by the motherboard; the ribbon connects to the side opposite the camera's perspective.) 
			This photograph is of a slightly earlier generation of hardware and lacks the 4~K TIA amplifiers located beneath the daughterboard that are depicted in \figsubref{qpufig}{a}.
		}
	\end{figure*}
	
	\begin{figure*}[h]
		\centering
		\includegraphics[width=90mm]{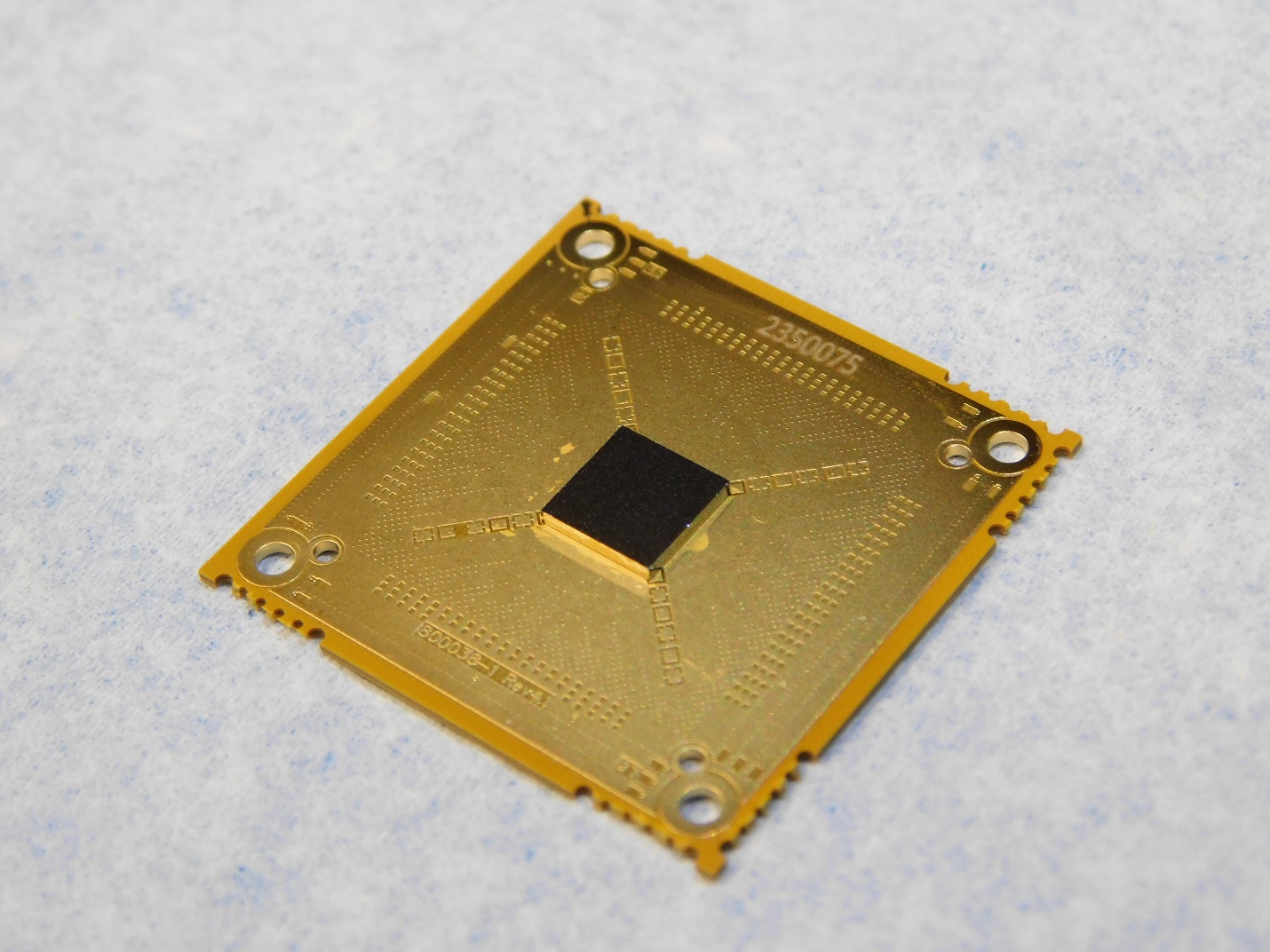}
		\caption{\label{packagedchip}
			\textbf{Photograph of a packaged qubit chip}. 
			The Si qubit chip die (black square) is bump bonded to a fine-pitch laminate LGA package (gold).
		}
	\end{figure*}
	
	\begin{figure*}[h]
		\includegraphics[width=140mm]{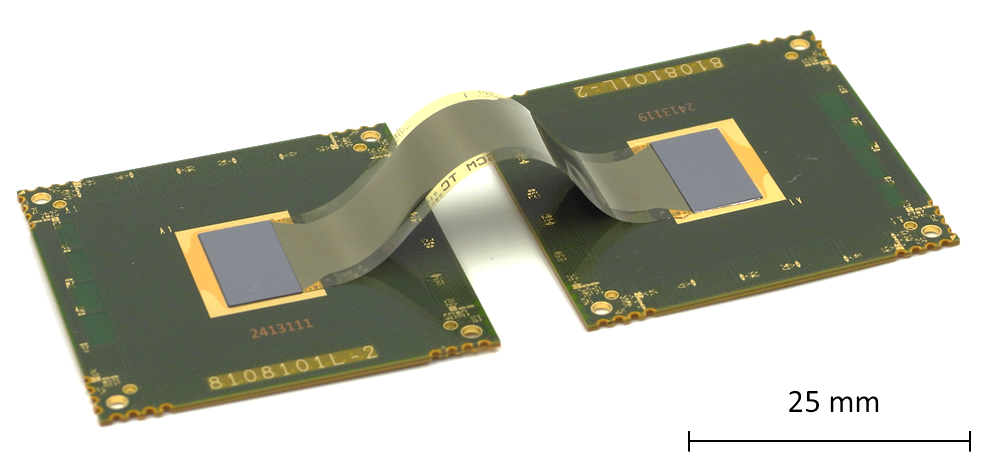}
		\caption{\label{cablephoto}
			\textbf{Photograph of ribbon cable assembly.} 
			The ribbon cable is indium-bump bonded to a 10-layer laminate board on each end that interfaces with printed circuit boards using a land-grid spring-probe array.
		}
	\end{figure*}
	
	\begin{figure*}[h]
		\centering
		\includegraphics[width=90mm]{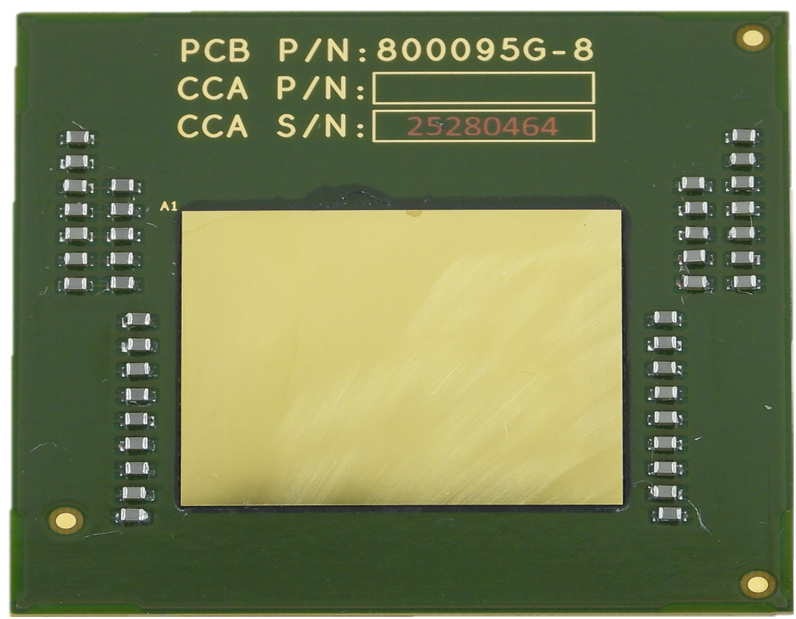}
		\caption{\label{controllerphoto}
			\textbf{Photograph of a cryo-controller assembly.} 
			The bonded combination of CMOS die and high-density capacitor arrays is attached to a fine-pitch laminate LGA package. 
			Additional surface-mount capacitors on the package improve the performance of the power distribution network. 
		}
	\end{figure*}


\clearpage

\renewcommand{\thesection}{S\arabic{section}}
\renewcommand{\figurename}{Supplementary Fig.}
\renewcommand{\thefigure}{S\arabic{figure}}
\renewcommand{\theHfigure}{S\arabic{figure}}
\renewcommand{\theequation}{S\arabic{equation}}
\renewcommand{\thetable}{S\arabic{table}}
\renewcommand{\thepage}{S\arabic{page}}

\begin{widetext}
	


\setcounter{section}{0}
\setcounter{figure}{0}
\setcounter{equation}{0}
\setcounter{table}{0}
\setcounter{page}{1}

\makeatletter
\def\@seccntformat#1{\csname the#1\endcsname.\quad}
\setcounter{secnumdepth}{2}
\setcounter{enumiv}{0} 
\makeatother

\begin{center}
	{\large \bfseries A digitally controlled silicon quantum processing unit: \\ Supplementary Information}
	%
\end{center}

\end{widetext}
\tableofcontents
\clearpage

\section{Cryo-CMOS qubit controller}\label{cryo-cmos-qubit-controller}

\begin{figure*}[ht]
	\centering
	\includegraphics[width=120mm]{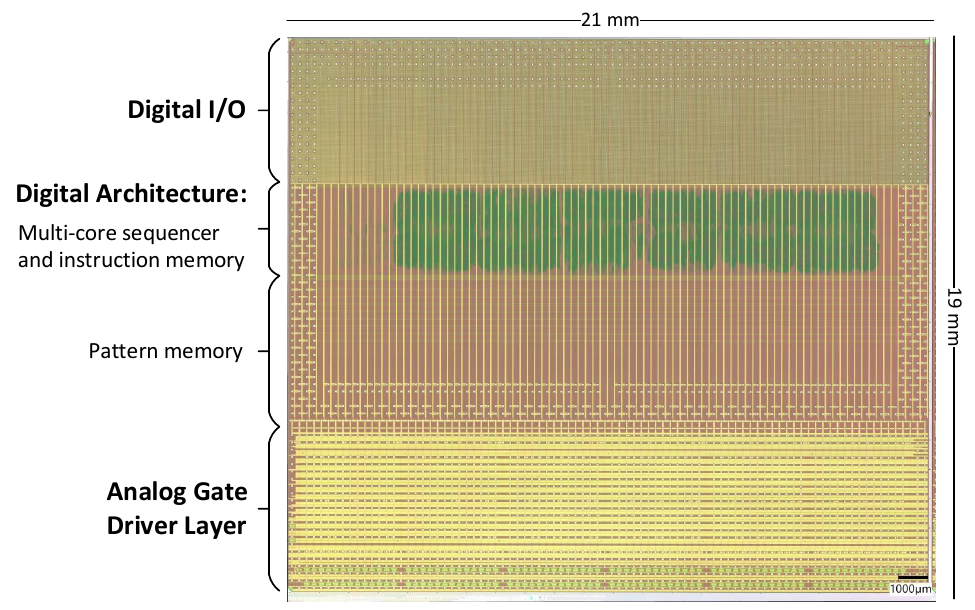} 
	\caption{
		\textbf{False-color die photo of the cryo-CMOS qubit controller.}
	}
	\label{figS1}
\end{figure*}

\begin{figure*}[ht]
	\centering
	\smartincludegraphics[width=120mm]{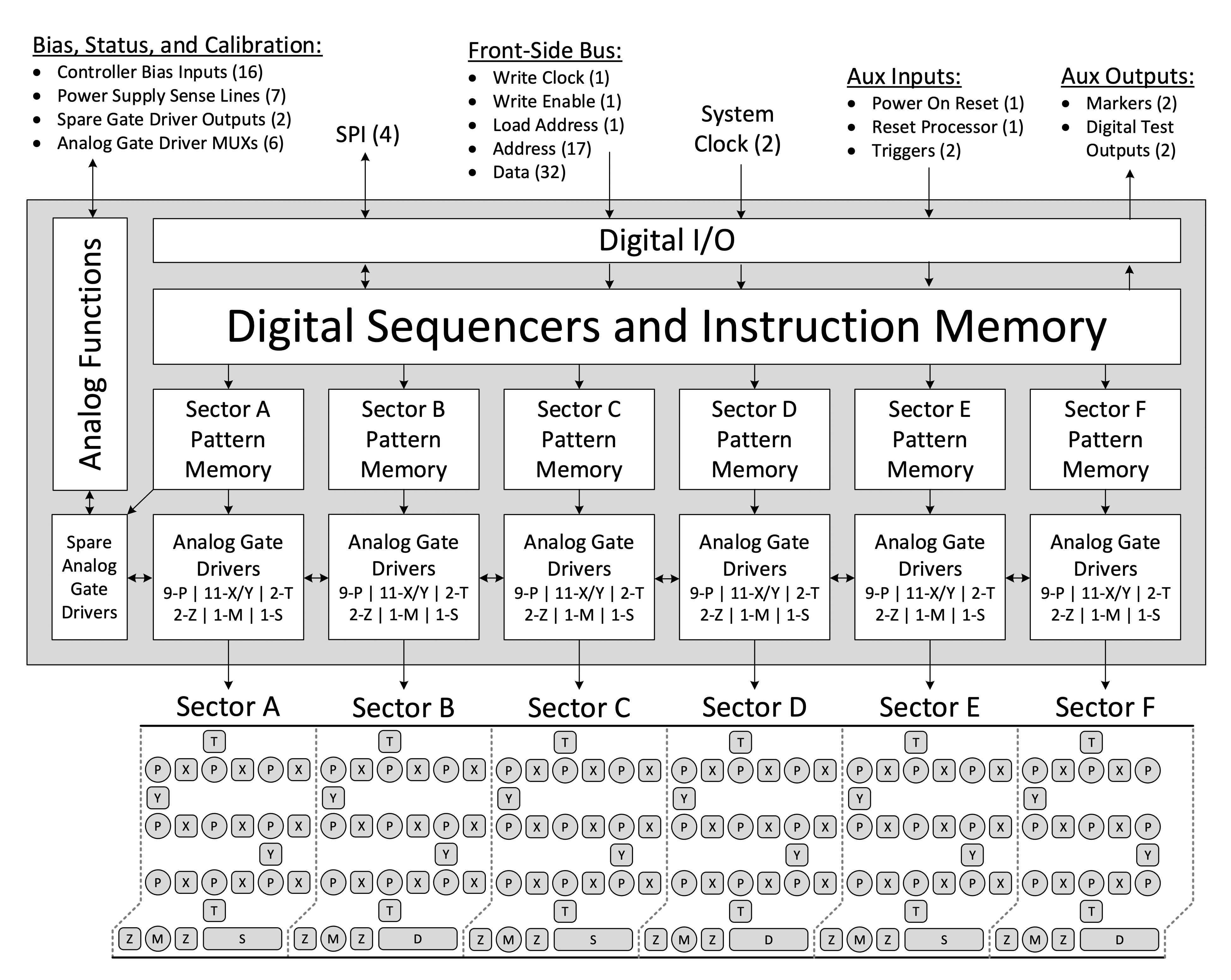} 
	\caption{
		\textbf{Top-level block diagram of the cryo-CMOS qubit controller.}
	}
	\label{figS2}
\end{figure*}

The cryo-CMOS qubit controller is a high-performance, mixed-signal system-on-chip fabricated in a commercial 130-nm RF CMOS technology.
Circuit designs are optimized at the transistor level for cryogenic performance.
The controller is designed to control an array of exchange-only silicon spin qubits from the 4~K stage of a commercial dilution refrigerator.
It generates all time-varying voltage signals required to perform tune-up, state preparation and measurement (SPAM), and qubit state manipulation.
The controller die contains 70 million transistors, measures 21 $\times$ 19~mm\textsuperscript{2}, and has 3054 C4-style pins arranged in a ground-signal/power-ground pattern with a minimum pad pitch of 250~$\mu$m.
Figure~\ref{figS1} 
depicts a photograph of a die with the main sections of the design annotated and 
\figref{figS2} shows a top-level block diagram of the controller.

The cryo-controller can source signals for up to 54 quantum dot (``P'') gates, 66 exchange (``X'' and ``Y'') gates, 12 tunnel (``T'') gates, and 24 gates used for initialization and readout (``Z'', ``M'', and ``S'').
There is a 1:1 mapping of 150 of 156 outputs to the physical qubit gates they control which are naturally arranged into ``sectors'' labeled A-F.
A sector is the set of gates used to form 3 physical qubits (PQB) vertically aligned in 3 rails of the device (9 P-gates, 11 X/Y-gates, 2 T-gates, and 4 SPAM gates per sector). 
To bias quiescent gates such as bath and field gates (not shown in \figref{figS2}), low noise voltages are routed directly to the qubit device from room temperature electronics. 
Three exchange gate outputs in Sector F (XF3, XF6, and XF9) are unused by the qubit and routed to diagnostic connectors on the motherboard, and 3 of 6 available S-gate outputs are not connected. 
The chip also includes two spare gate drivers routed out of the system for diagnostics or characterization. 
The controller only requires power, a handful of DC analog biases, and a manageable number of intermediate-speed lines for digital communication to be provided from room temperature.

\subsection{Digital architecture}\label{digital-architecture}

The cryo-controller's digital architecture consists of six independent instruction sequencers, a shared bank of instruction memory (6144 words, each 32 bits wide), and, for each gate driver output, a custom ``pattern memory'' register that store up to 512 commands that are each 24 bits wide. 
The controller's digital interface consists of 66 signals: a 2-wire differential system clock, a 52-line front-side parallel bus (including write clock, address, and data), an industry-standard 4-wire serial peripheral interface (SPI), and 8 auxiliary inputs and outputs.
The front-side bus transmits program data to instruction and pattern memories using the write clock and executes at the operating frequency of the system clock. 
The SPI is used for reading and writing a variety of configuration registers within the digital architecture and analog gate driver layer. 
The auxiliary I/O includes signals that may be used to interface with external readout electronics, inputs for real-time user-selectable instruction branching (``Triggers''), outputs for general purpose use (``Markers''), and specific digital diagnostic outputs (``Digital Test Outputs''). 
Any of the digital input signals can be multiplexed to these diagnostic outputs to be routed to room temperature. 
This provides the capability to perform in-situ loop-back signal connectivity testing within the dilution refrigerator.

The multi-sequencer core is responsible for providing a memory address to each pattern memory cell every clock cycle. 
The addresses sent to pattern memory are computed according to a customized qubit-control instruction set architecture (ISA).
The ISA includes instructions for stepping, pausing, jumping, looping, branching, memory stack manipulation, toggling of external outputs, and more. 
Each sequencer has integrated pseudorandom number generators that can be accessed with specialized instructions for memory-efficient randomized benchmarking (see \secref{blind-randomized-benchmarking-brb}).
There is also the capability to enforce inter-sequencer coordination using real-time interrupts and interrupt resolution logic.

Addresses computed by the sequencers are used to retrieve pattern memory words which are routed to each analog driver every clock cycle. 
Pattern memory words specify information such as DAC voltage, pulse duration, and other analog functionality that is decoded and executed by local digital modules within the gate driver blocks themselves. 
It is possible to configure the mapping of any pattern memory cell (and therefore any analog driver) to receive its address pointer from any sequencer using the SPI. 
This functionality provides an enormous degree of flexibility in assigning independent or coordinated control to arbitrary parts of the qubit device. 
This may be advantageous in particular for performing experiments on the qubit array in the presence of physical defects.
Collectively, the cryo-controller's digital architecture is capable of autonomously executing effectively arbitrary-length randomized benchmarking sequences and complex multi-qubit experiments.

The cryo-controller contains several on-chip pseudorandom number generators. 
The output of each is a deterministic function of a user-defined seed. 
Each random number generator is a maximum length linear feedback shift register (LFSR) modified to shift out \emph{M} pseudorandom bits in a single clock cycle. 
On-chip logic divides the set of \emph{2\textsuperscript{M}} possible bit-vector values into \emph{K} equal sized subdivisions and selects the subdivision \emph{R} that contains the random bit-vector. 
The resulting random integer \emph{R} is uniformly distributed between \emph{0} and \emph{K-1}.

We implement randomized benchmarking (RB) with on-chip instructions that jump to an address + offset where the offset is the value of an on-chip random number. 
For 1-qubit RB we compile each of the 24 Clifford gates into subroutines with the first instruction of each subroutine located in a single software-defined jump table. 
The first instruction of each subroutine jumps out of the table, and the final instruction of each subroutine is the instruction to jump to address + offset where the address is the start of the jump table and the offset is a random number between 0 and 23. 
Additionally, we load an on-chip register with the desired RB sequence length and decrement this register on completion of each Clifford gate. 
When this register is decremented to zero, the jump to address + offset behavior is modified to jump to the first address past the jump table where we load the first instruction for the inverse Clifford gate subroutine. 
Calculation of the inverse Clifford gate is determined during program compilation by simulating the pseudorandom number generation process using the same seed value. 
We execute more complex RB protocols (2-qubit RB, interleaved RB, etc.) with a similar procedure utilizing multiple different sized random numbers and software-defined jump tables (see \secref{random-clifford-sampling}).

\subsection{Analog gate driver layer}\label{analog-gate-driver-layer}

\begin{figure*}
	\centering
	\smartincludegraphics[width=165mm]{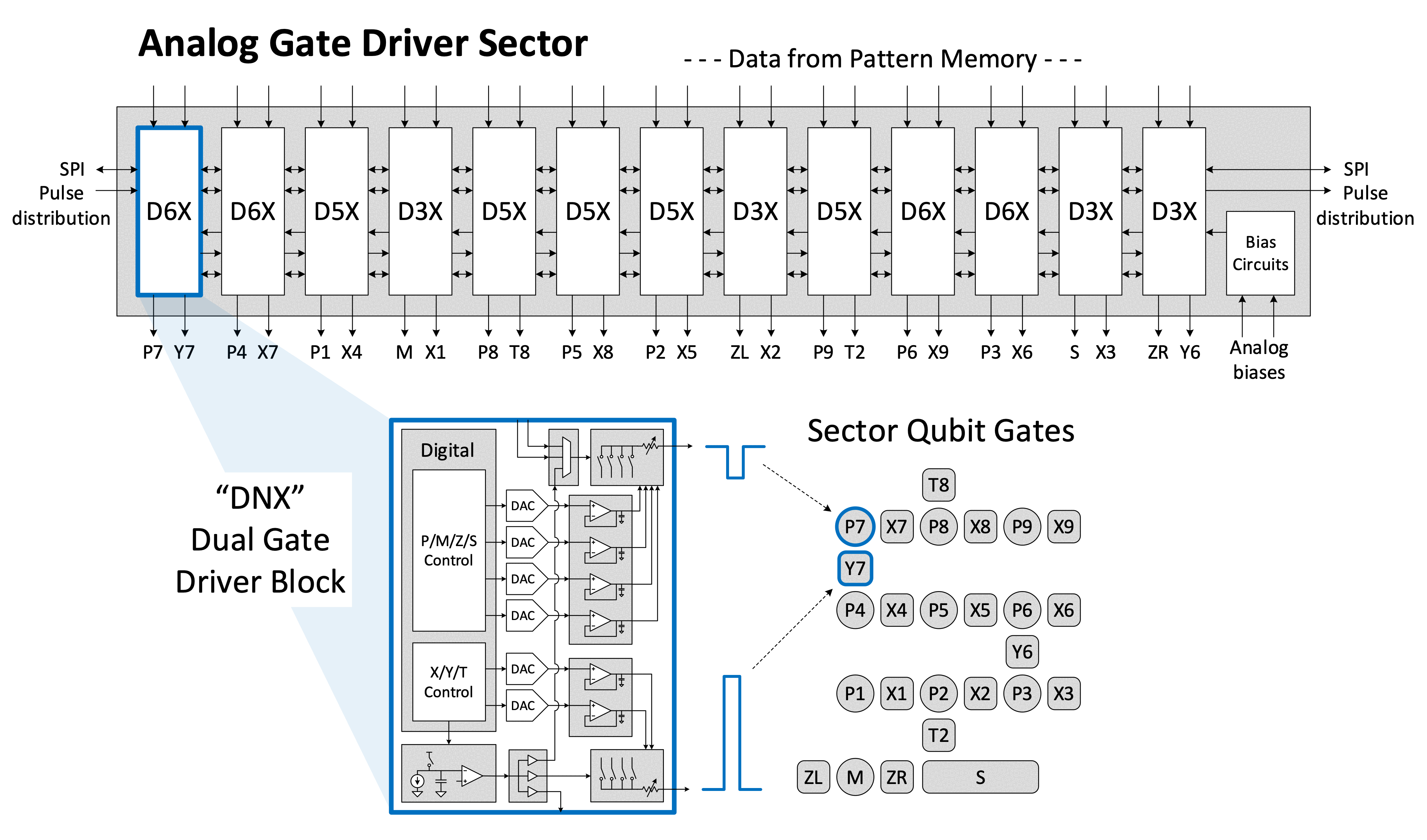}
	\caption{
		\textbf{Block diagram of an analog gate driver sector and the variations of gate driver blocks within it.} 
		The gate driver layer contains six sectors.
	}
	\label{figS3}
\end{figure*}

The ``sector'' is a natural way to organize both the qubit and the cryo-controller's analog gate driver layer. 
A block diagram showing this organization for the controller in detail is shown in \figref{figS3}.
The fundamental analog driver building block is called the ``DNX'' dual gate driver which controls one P, M, Z, or S-gate and one X, Y, or T-gate. 
There are 3 variants of this building block (``D6X'', ``D5X'', and ``D3X'') where ``N" equals the number of DACs in the block, determined by the number of unique high-precision voltage levels required to control the gate pair. 
The ``X'' in the name denotes an exchange pulse generator. 
To control a sector of qubit gates, 13 DNX blocks are integrated together along with driver biasing circuitry. 
The gate driver sector is replicated 6 times to form the cryo-controller's full analog layer consisting of 78 DNX blocks containing a total of 366 DACs and 78 pulse generators.

Device tune-up and SPAM are performed by changing DAC output voltages.
The default DAC bandwidth is about 10~MHz, which balances SPAM switching speed ($T$\textsubscript{10\%-90\%} ≈ 35 ns) and output noise for exchange.
DACs can dynamically change their switching speed in order to generate both accelerated ($T$\textsubscript{10\%-90\%} ≈ 10~ns) transitions as well as slow, pseudo-linear ramps with tightly-controlled slew rates intended for traversing charge transitions (the ``tie bar'') during qubit initialization.

GHz-bandwidth, time-modulated exchange gate pulses are generated for qubit state manipulation by switching between fixed, buffered DAC values. 
The precise timing information comes from the pulse generator, a high-resolution digital-to-time converter (DTC). 
The exchange gate driver's local digital module receives a digital input word from pattern memory and converts a portion of it into a variable-width, logic-level pulse by the DTC. 
The logic-level pulse is translated into an output pulse between the appropriate voltage levels with a high-speed output switch and associated switch driver. 
Each exchange gate driver's pulse generator output is distributed to neighboring dot gate switch drivers with low timing skew to form time-aligned, triplicate pulse waveforms for symmetric exchange operation \cite{s_reed_reduced_2016}.

Qubit gates act like open circuits (i.e. are high impedance), so any pulse sent from the cryo-controller is sent back as a near-perfect reflection that must be sufficiently attenuated before the next pulse is launched. 
Failure to do so results in inter-symbol interference (ISI) error. 
To this end, the gate driver's output impedance is designed to match the nominal 50~Ω characteristic impedance of the interconnect in the system. 
The output switch features configurable resistance control to tune the termination to 50~±~1~Ω using the SPI. 
The parasitic capacitance of the output is minimized to provide at least 20~dB return loss up to the 500~MHz analog bandwidth of instrumentation. 
The output switch also features a configurable switched-capacitor output pulse filter to help mitigate reflections. 
An adjustable capacitor is connected to the output for the duration of a pulse to add to its rise and fall times by up to 400~ps, and it is disconnected before the reflected pulse arrives to avoid degrading the quality of the output termination.

DACs are buffered with unity-gain voltage amplifiers that provide approximately 60~MHz bandwidth for restoring the exchange and compensation pulse voltages to their calibrated values in between pulses in a sequence. 
The buffer amplifier is designed in conjunction with a sizable off-chip load capacitor that provides the initial charge needed to form high-speed pulse rising and falling edges. 
The buffer limits voltage droop on the pulse amplitude for the duration of the pulse and restores the charge on the output capacitor for the next pulse. 
These charge reservoir and stabilization capacitors are fabricated as custom, high-density flip-chip capacitor arrays that are bonded directly to the CMOS die to minimize parasitic inductance. 
Flip-chip capacitors that are not mapped to buffer amplifier outputs are connected to the analog power supply rails to lower the impedance and increase the effectiveness of the power delivery network (PDN).

\begin{figure}
	\centering
	\smartincludegraphics[width=\columnwidth]{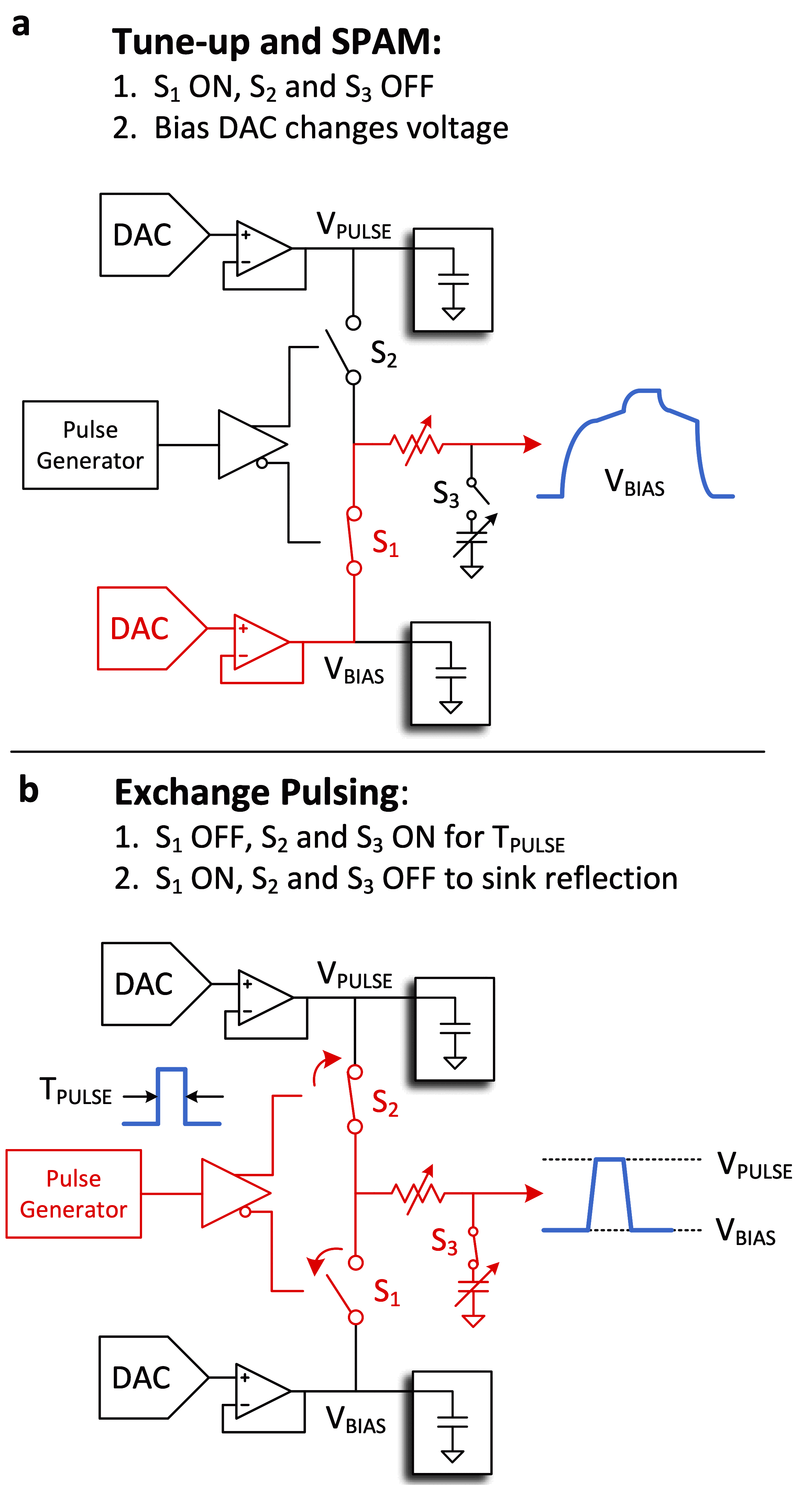} 
	\caption{
		\textbf{Illustrations of how the analog gate driver \textbf{a,} executes tune-up and SPAM, and \textbf{b,} exchange pulsing operations.}
	}
	\label{figS4}
\end{figure}

\begin{figure}
	\centering
	\smartincludegraphics[width=\columnwidth]{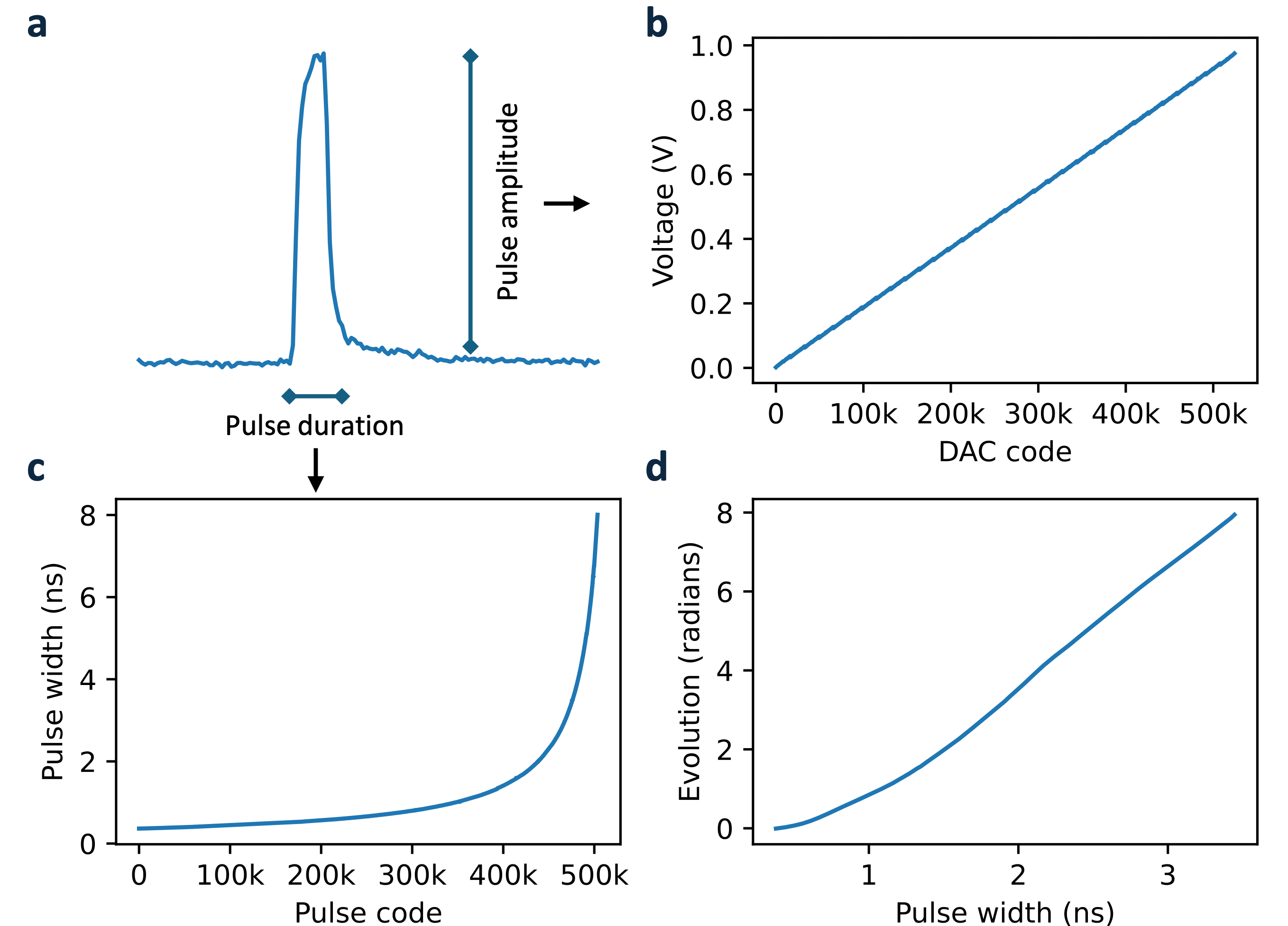} 
	\caption{
		\textbf{Pulse waveform and calibration curves}.  
		\textbf{a,} An example voltage pulse waveform. 
		The duration of this pulse is set by a code-to-time word and the amplitude by the difference of two DACs (low and high voltage) configured with code-to-voltage words. 
		\textbf{b,} DAC code-to-voltage transfer function. 
		\textbf{c,} Pulse code-to-time transfer function. 
		\textbf{d,} Typical pulse duration to exchange angle qubit calibration curve.
	}
	\label{figS5}
\end{figure}

Figure \ref{figS4} summarizes the analog theory of operation and illustrates how the gate driver circuitry executes tune-up, SPAM, and exchange pulsing operations.
Each of the six gate driver sectors also features a multiplexing network to connect any constituent gate driver output to a dedicated output pin.
These six outputs in the analog layer are routed to room temperature for in-situ functionality screening and DAC code-to-voltage calibration.
Pulse generator codes are calibrated to specific qubit rotation angles.
This is accomplished by first acquiring a high-resolution code-to-pulse width transfer function from one of the three available exchange gate outputs in sector F not connected to the qubit device. 
That transfer function is then used to generate finely-tuned code-to-angle calibrations for all other pulse generators on the chip. 
Example code-to-voltage and code-to-time curves and how they are related to voltage pulse waveforms are shown in \figref{figS5}. 

\subsection{Analog performance}\label{analog-performance}

\begin{figure}
	\centering
	\includegraphics[width=\columnwidth]{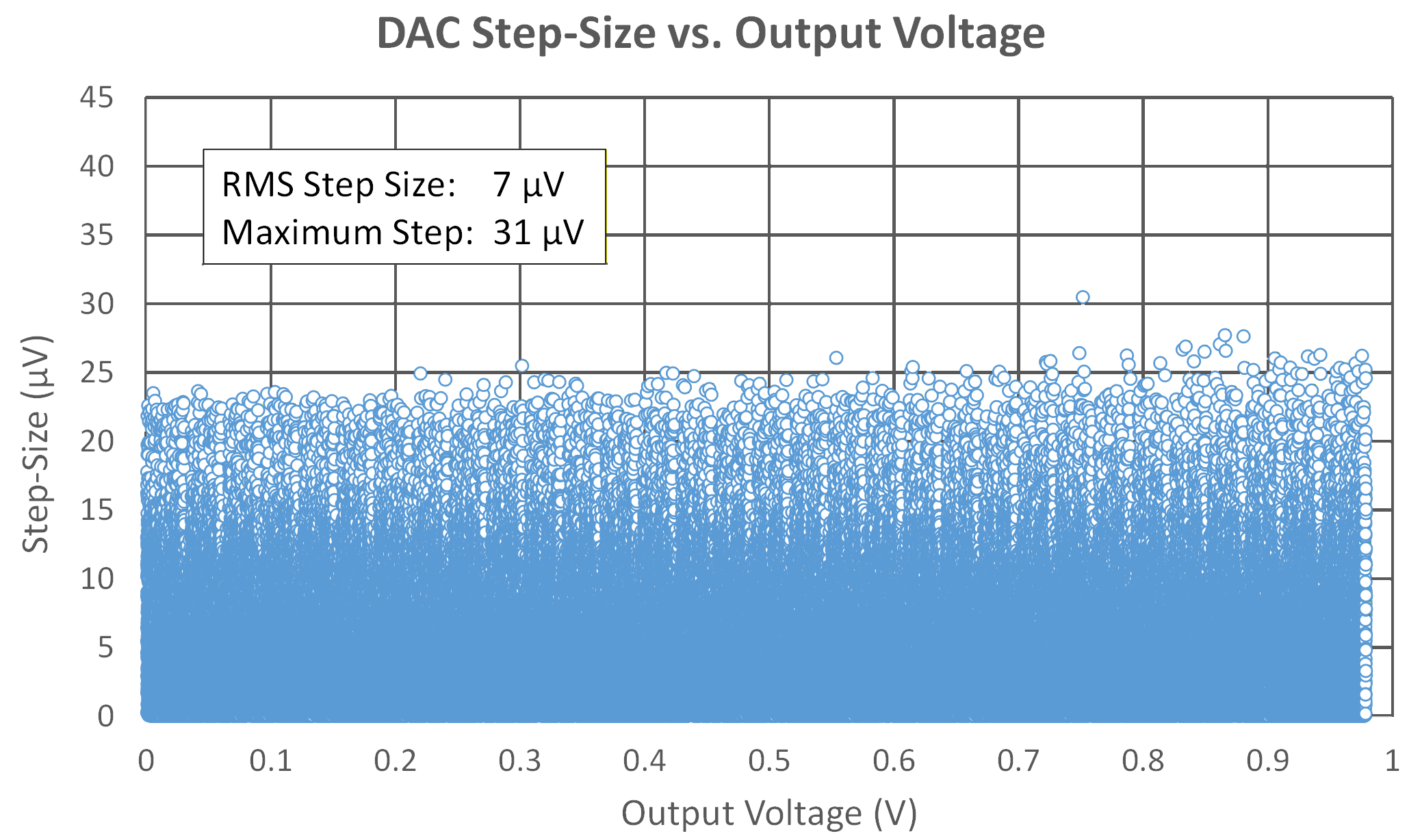} 
	\caption{
		\textbf{Data showing DACs are programmable from 0-1~V with an RMS step size \textless{} 10~$\mu$V.}
	}
	\label{figS6}
\end{figure}

\begin{figure}
	\centering	
	\smartincludegraphics[width=\columnwidth]{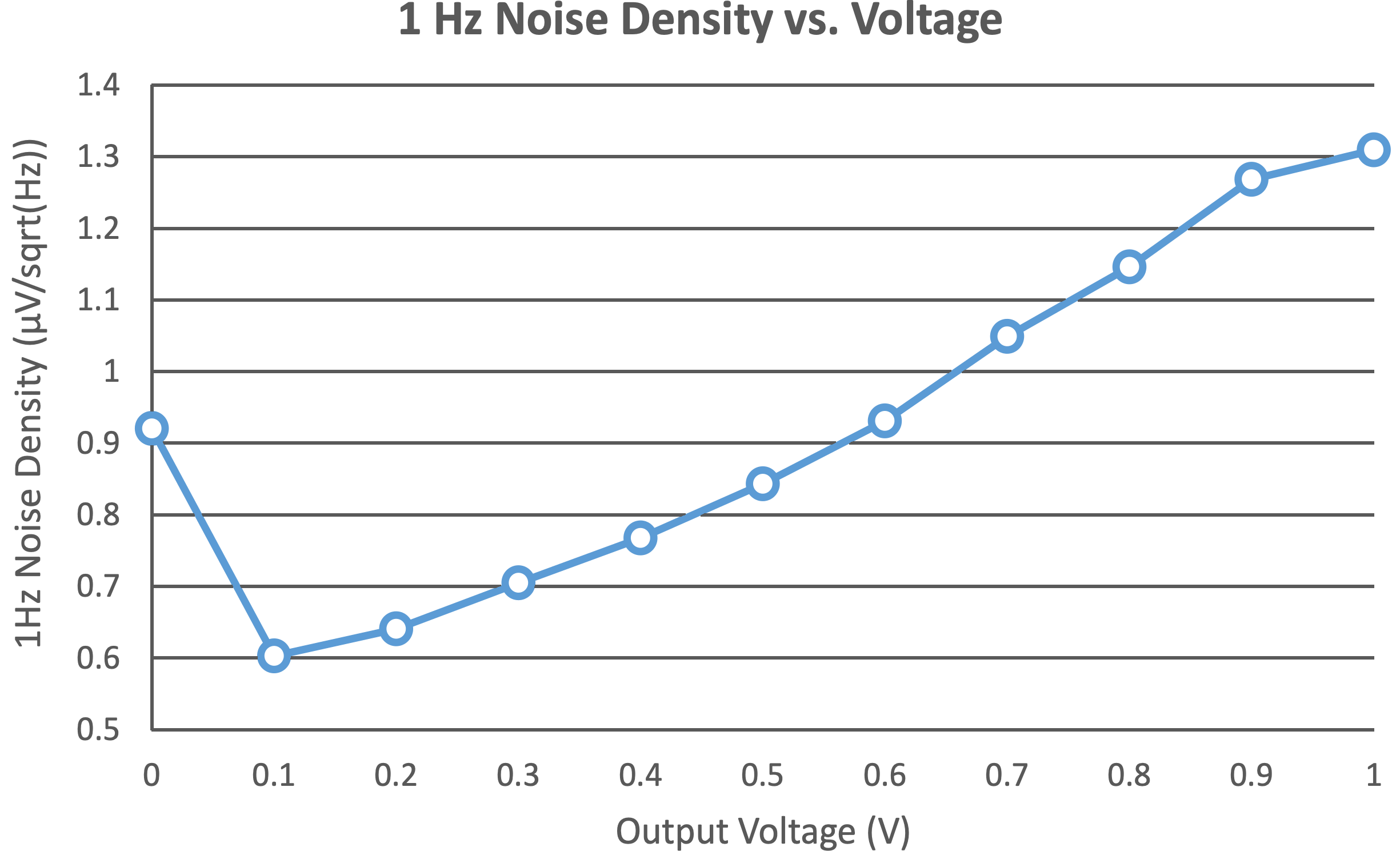} 
	\caption{
		\textbf{Data showing the typical driver's output noise density at 1~Hz is 
			\(< \mathbf{1.4}\ {\boldsymbol{\mu} \textbf{V}}/{\sqrt{\textbf{Hz}}}\).}
	}
	\label{figS7}
\end{figure}

DACs are programmable from 0-1~V with an RMS step size \textless{}~10~$\mu$V and are designed together with the buffer amplifiers to be low noise.
Low frequency ``flicker'' noise is typically a significant contributor in baseband circuits and becomes a significant contributor to qubit error if it is comparable to the intrinsic low frequency charge noise of the qubits. 
This section presents representative performance characterization data at 4~K from a typical gate driver output routed to room temperature instrumentation. 
Figure \ref{figS6} plots the step-size from a 200,000-point, sorted DAC-code-to-voltage transfer function spanning the full output voltage range.
Figure \ref{figS7} plots the noise density at 1~Hz as a function of set voltage and the results show it is 
\(< 1.4\ {\mu \textrm{V}}/{\sqrt{\textrm{Hz}}}\).

\begin{figure}
\centering
\includegraphics[width=\columnwidth]{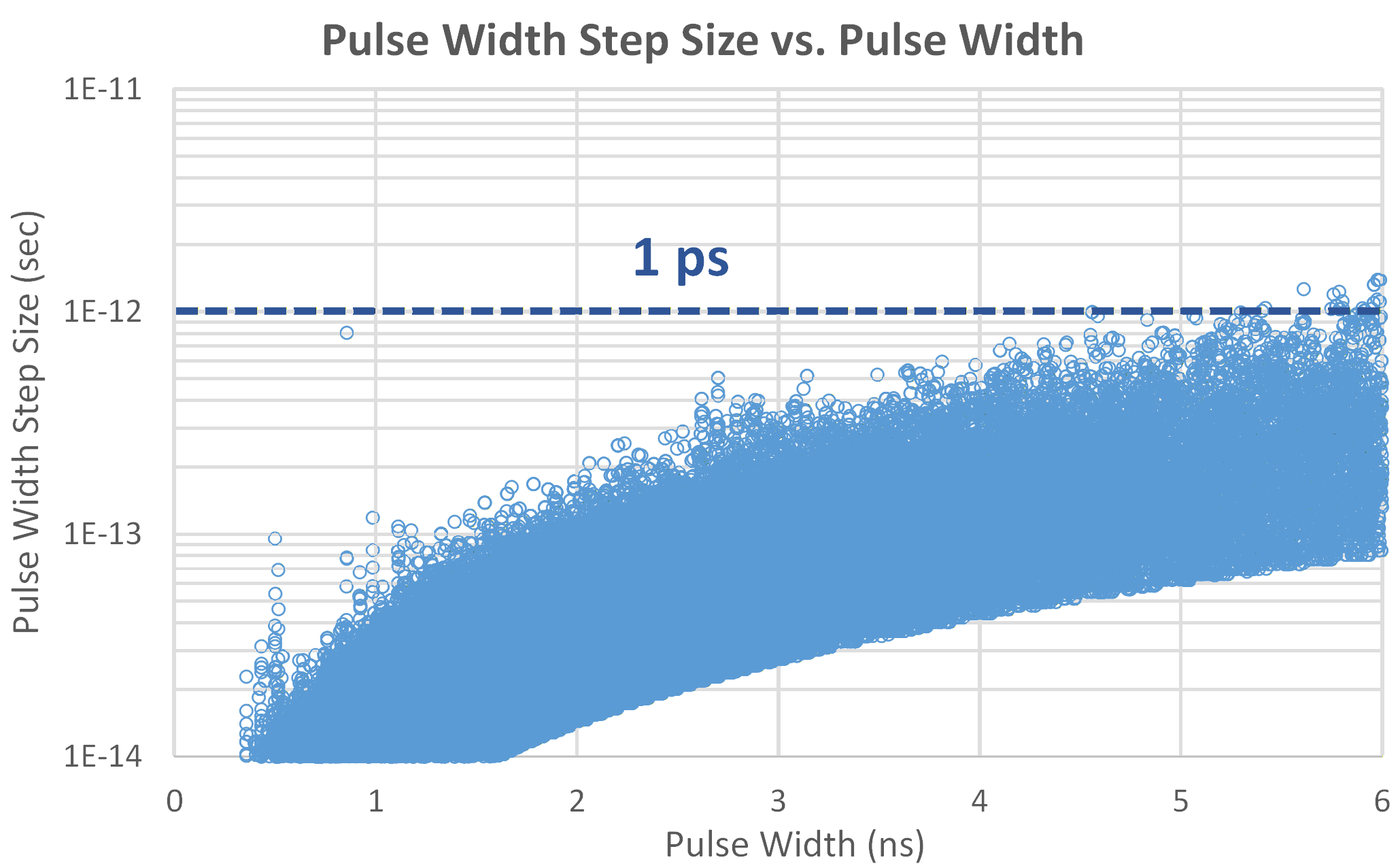} 
\caption{
	\textbf{Data showing the pulse generator step size is approximately ≤ 1~ps from 0.4 to 6~ns.}
}
\label{figS8}
\end{figure}

\begin{figure}
\centering
\smartincludegraphics[width=\columnwidth]{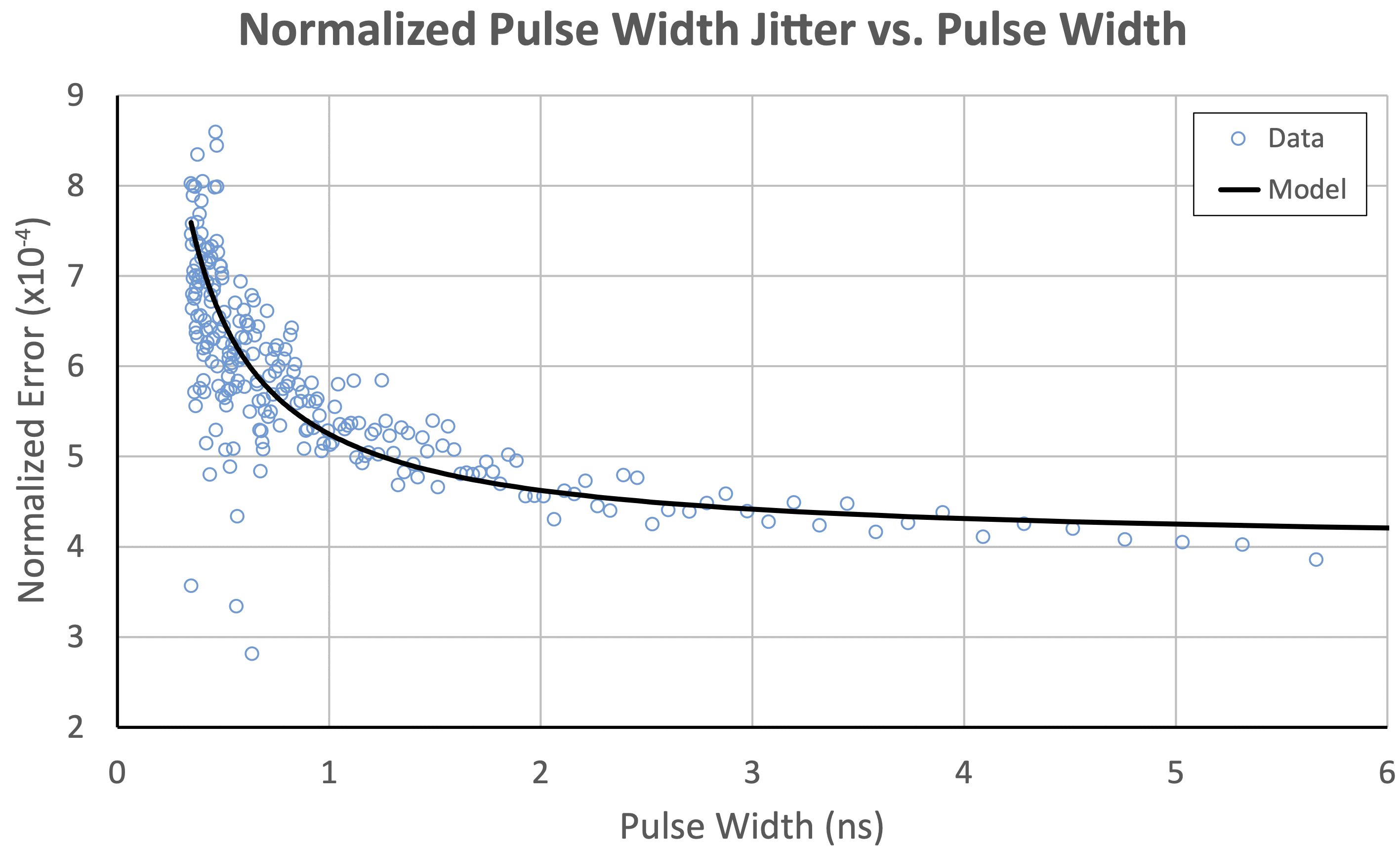} 
\caption{
	\textbf{Measurements and model fit of normalized pulse width jitter from 0.4 to 6~ns.}
	The model fit function is 125~fs/PW + 4\e{-4}.
}
\label{figS9}
\end{figure}

The pulse generator is designed to produce performant pulses at a maximum pulse repetition frequency (PRF) of 41.67~MHz. 
Pulse width is tunable from 0.4 to 6~ns with approximately picosecond resolution across the full range and with sub-picosecond resolution below 4~ns.
Figure~\ref{figS8} plots measurements of the output pulse width step-size as a function of set pulse width. 
Figure~\ref{figS9} plots the pulse width jitter expressed as a normalized error (jitter divided by pulse width) across the full pulse width range. 
Jitter error is 7.5\e{-4} for the shortest pulses and decreases to 4\e{-4} for longer pulses. 
These datasets were taken through 10~feet of high-bandwidth coaxial cabling into a 256~Gsps oscilloscope.

\subsection{Power consumption}\label{power-consumption}

\begin{figure*}
\centering
\smartincludegraphics[width=120mm]{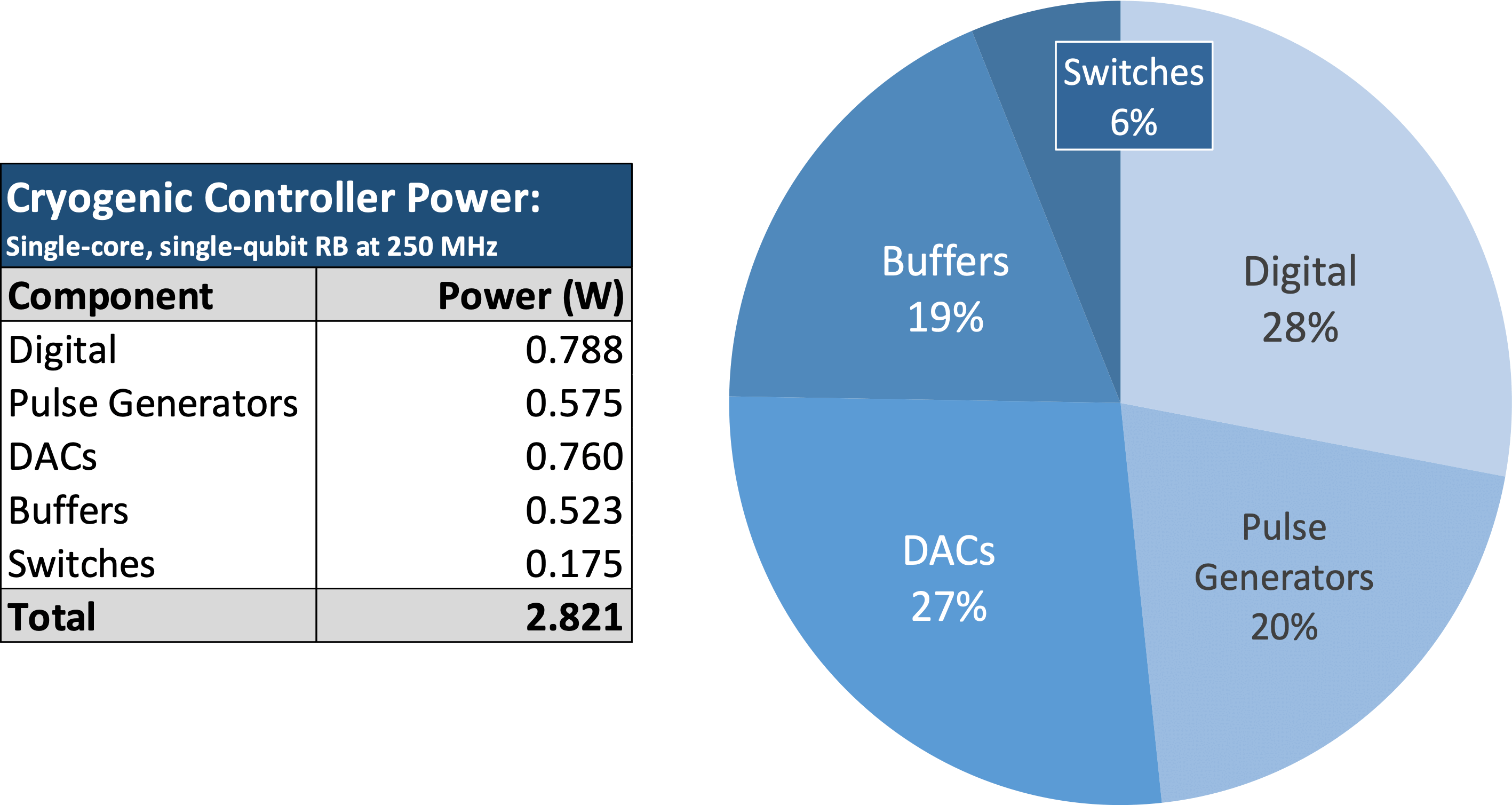}
\caption{
	\textbf{Detailed breakdown of cryo-controller power consumption during an optimized single-sequencer, single-qubit RB experiment at 250~MHz.}
}
\label{figS10}
\end{figure*}

The cryo-controller requires one digital power supply and six analog power supplies from room temperature. 
Power consumed by the digital architecture is dynamic power and is therefore a strong function of the operating frequency and the complexity of the program. 
The analog gate driver layer power consumption is mostly static because the typical duty cycle for pulsed outputs during an experiment is \textless~10\%.
Empirical data suggests randomized benchmarking consumes more digital power that the other kinds of experiments discussed in this publication.
For an optimized single-sequencer, single-qubit RB experiment at 250~MHz, the chip consumes approximately 2.8~W. 
A detailed power breakdown of this scenario is shown in \figref{figS10}. 
It is estimated that power consumption would increase up to 3.5~W if all six sequencers were independently running randomized benchmarking.

The total power consumption is within the typically-available thermalization budget of the 4~K stage in commercial dilution refrigerators. 
The controller features granular power control such that most digital assets and analog sub-circuits can be powered down when not in use. 
Approaches to substantially reducing the power-per-qubit through architecture and scaling have been identified, but were not required for this work.

\section{Superconducting interconnect}\label{superconducting-interconnect-sup}

\subsection{Fabrication}\label{fabrication}

\begin{figure}
\centering
\smartincludegraphics[width=\columnwidth]{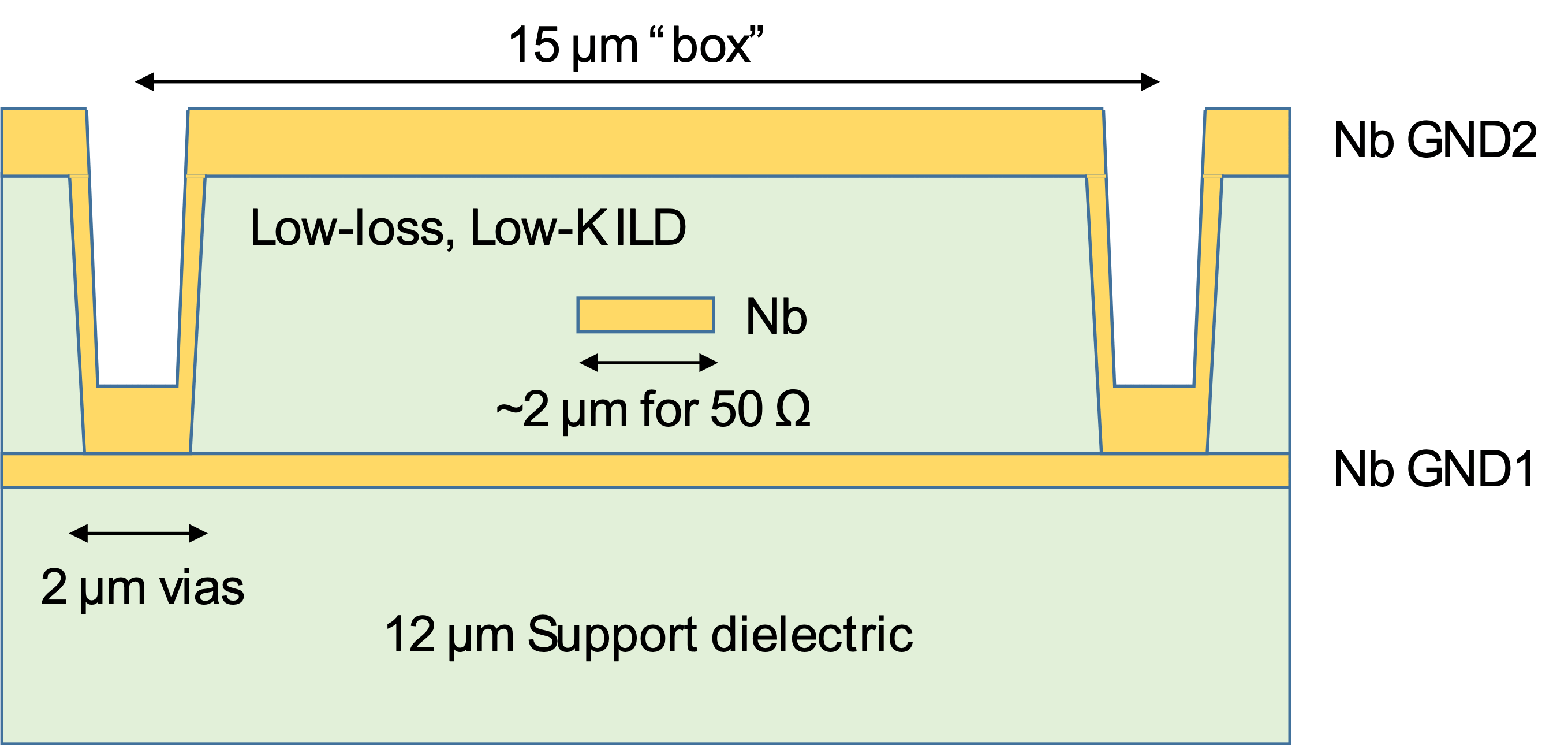} 
\caption{
	\textbf{Schematic cross-section of a single trace in the ribbon cable.}
	All metal layers are Nb and the ILDs are polyimide.
}
\label{figS11}
\end{figure}

The superconducting ribbon cables are fabricated on 150~mm silicon wafers using Nb for the superconducting layers and polyimide for the dielectric. 
Polyimide layers serve as both low-K interlayer dielectric (ILD) and in providing structural support for the ribbon cable. 
Three levels of Nb metal realize a lower ground plane, a signal layer, and an upper ground plane to create a fully-enclosed stripline environment. 
Via trenches are formed between each signal trace, electrically connecting the upper and lower ground planes. 
The resulting structure (\figref{figS11}) is a rectangular coaxial structure with micron-scale signal wires on a 15~$\mu$m pitch. 
Pads are finished with Au to support assembly operations. 
Typical cable designs are 6~cm long, approximately 1~cm wide, and implement 200 to 500 independent signal wires. 
High-yielding wafer-scale processes are able to routinely realize devices with 100\% interconnect yield for these 6~cm\textsuperscript{2} devices.

\subsection{Packaging}\label{packaging}

The ribbon cables are indium bonded to ten-layer laminate package substrates with length matching within related wire bundles. 
The interface ends of the ribbon are capillary-underfilled to enhance the mechanical strength of the bond and provide environmental protection to the joints. 
The package mates to the printed circuit board (PCB) using a land-grid spring-probe array with an 800~$\mu$m pitch. 
The pads on the package and PCB are gold plated. 
The package-to-PCB interface has undergone over 10 remate cycles on multiple test systems with no loss of functionality. 
A photograph of the assembly is in Extended Data Fig.~3.

\subsection{Electrical performance}\label{electrical-performance}

\begin{figure}
\centering
\smartincludegraphics[width=\columnwidth]{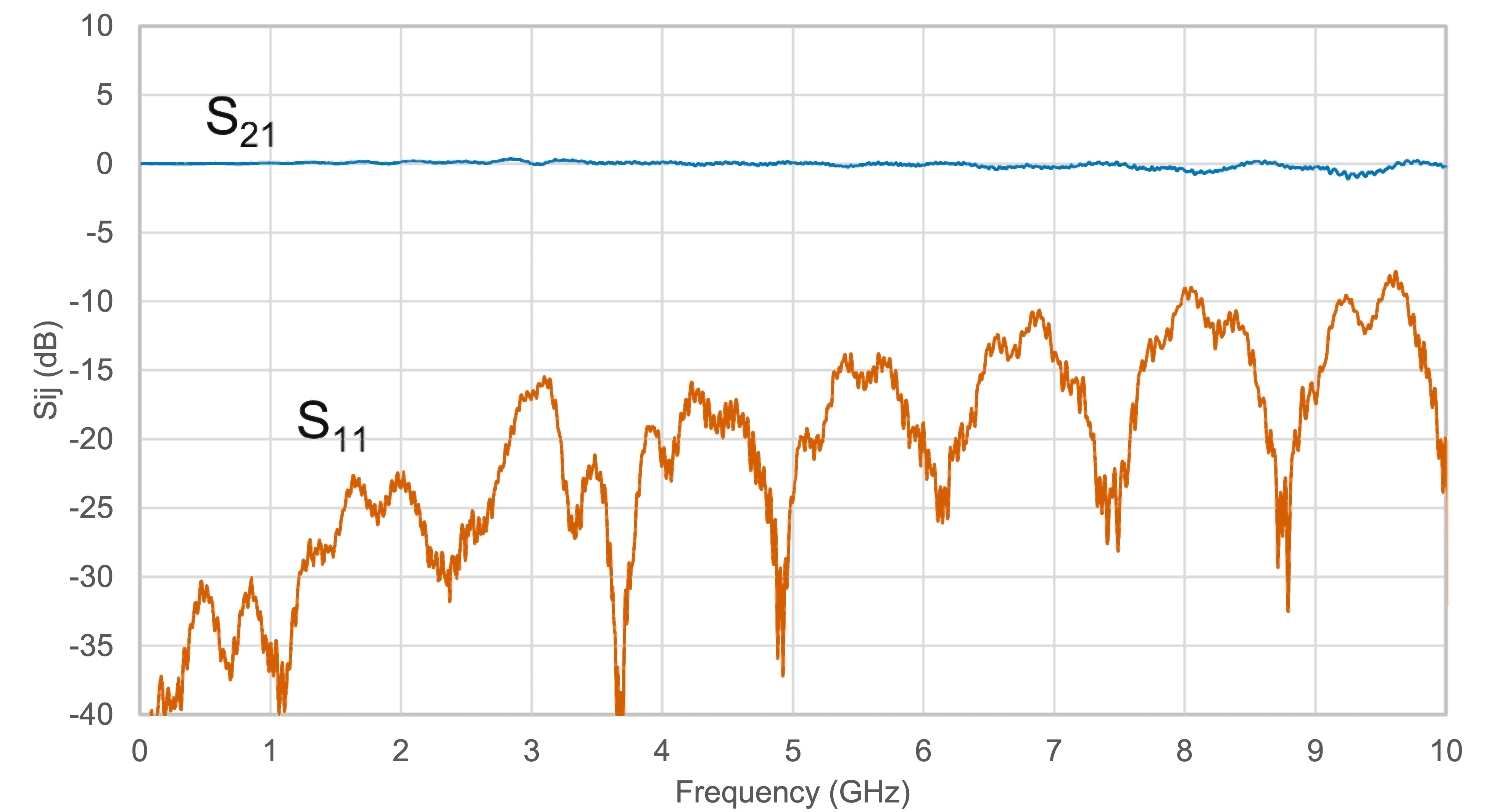} 
\caption{
	\textbf{6~cm ribbon cable insertion loss at 4~K.} 
	Actual ribbon cable insertion loss (S21) and return loss (S11) are de-embedded from the non-superconducting test fixturing using \textit{in situ} cryo calibration standards measured at the end of the cryo test fixturing. 
	Imperfections in the fixturing and de-embedding process leave some residual ringing which causes minor artificial non-passivity at some frequencies. 
	A fit of the de-embedded data yields a loss of approximately $-1.0$~dB/GHz/m.
}
\label{figS12}
\end{figure}

\begin{figure}
\centering
\smartincludegraphics[width=\columnwidth]{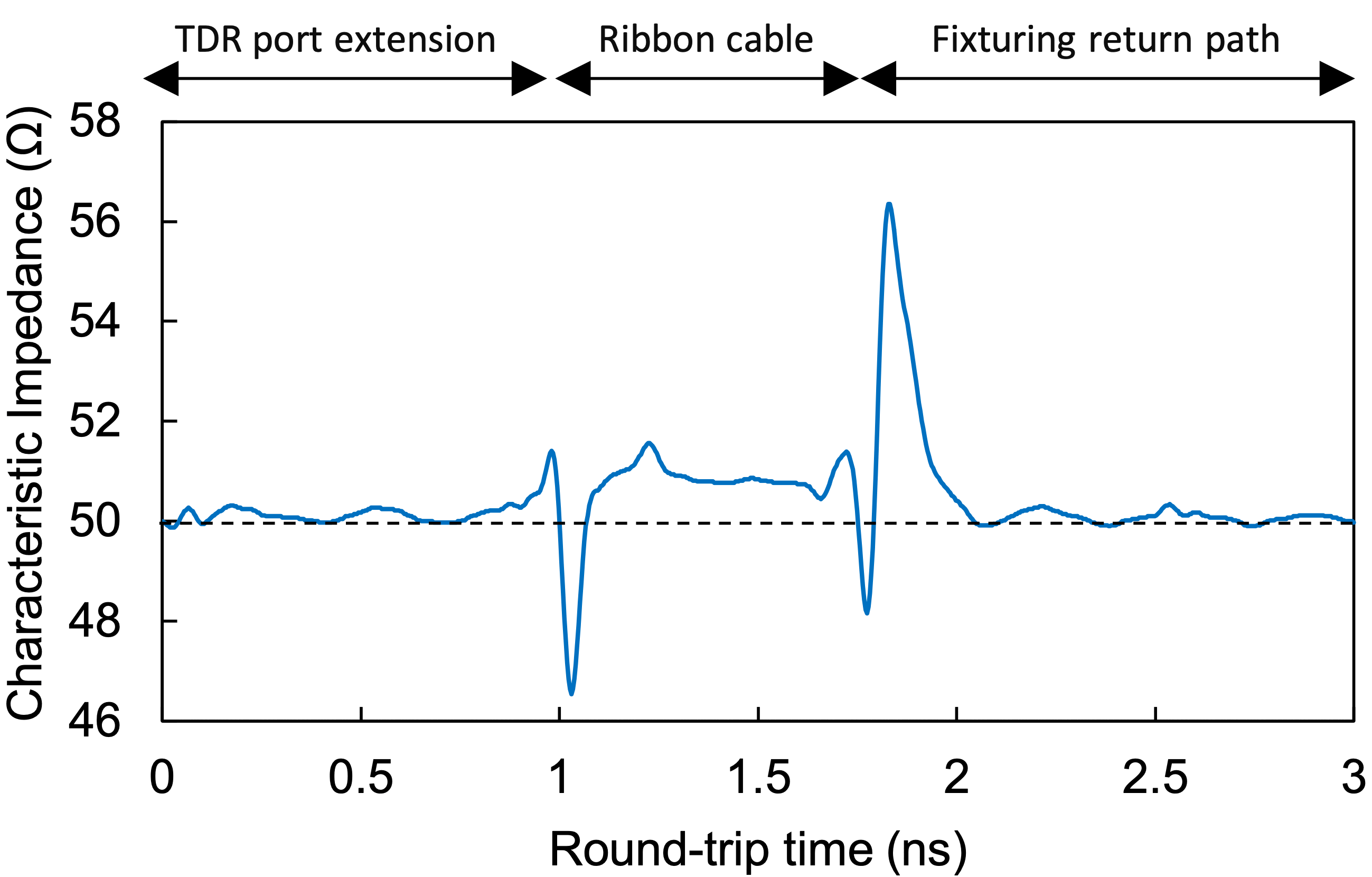} 
\caption{
	\textbf{Time-domain reflection response of 6~cm ribbon cable at 4~K.}
	De-embedded S11 data has been transformed to yield a characteristic impedance profile for the ribbon cable assuming a 50~ps rising edge.
	Between the capacitive drops at the interfaces on either end of the cable associated with the bond pads (at 1.03 and 1.76~ns), the characteristic impedance averages 50.8~$\Omega$. 
	The initial 1~ns is a time-domain reflectometer (TDR) port extension, an artificially calculated response prior to the device under test, and is associated with the time-domain transformation technique.
}
\label{figS13}
\end{figure}

The RF performance of the ribbon cable was measured at 4~K to determine insertion loss, impedance, and cross-talk.
Figure \ref{figS12} shows a typical insertion loss (S11) measurement for a single transmission line in the cable. 
The de-embedded cable loss is approximately $-1.0$~dB/GHz/m.
Figure \ref{figS13} shows a single transmission line impedance with 50~±~5~Ω being typical for all wires within a cable.
Cross-talk between nearest neighbor wires within the cable is $<-90$~dB at 1~GHz and $<-80$~dB at 10~GHz. 
Wire critical currents are approximately 150~mA. 
All cables are screened electrically inline on the wafer and at 300~K and 4~K after packaging.

\subsection{Thermal performance}\label{thermal-performance}

The heat leak through the cable from 4~K to about 50~mK is approximately 10~$\mu$W. 
The cables have survived over 20 thermal cycles across multiple test systems with no signs of fatigue or degradation.

\section{Environmental control}\label{environmental-control}

\subsection{Thermal management}\label{thermal-management}

High-fidelity state preparation and measurement of our exchange-only spin qubit processor requires operation at ultra-low cryogenic temperatures. 
To accomplish this, we utilize a commercial Bluefors XLD1000 dilution refrigerator, capable of rejecting 1~mW of heat at 100~mK. 
Cooling power at 4~K is provided by three Cryomech PT-425 pulse tube coolers: one dedicated for pre-cooling the circulating \textsuperscript{3}He/\textsuperscript{4}He mixture and two providing a nominal 5~W of cooling power to the 4~K plate of the cryostat. 
This 4-Kelvin cooling capacity supports the heat loads from cabling as well as the cryo-controller (\secref{power-consumption}). 
The CMOS-generated control signals are routed to the qubit chip via a 296-trace superconducting Nb/Polyimide ribbon cable (\secref{superconducting-interconnect}) that launches off a printed circuit board (PCB) which houses the cryo-controller at the 4~K stage and lands on a separate PCB that houses the qubit chip at the mK stage.

\begin{figure*}
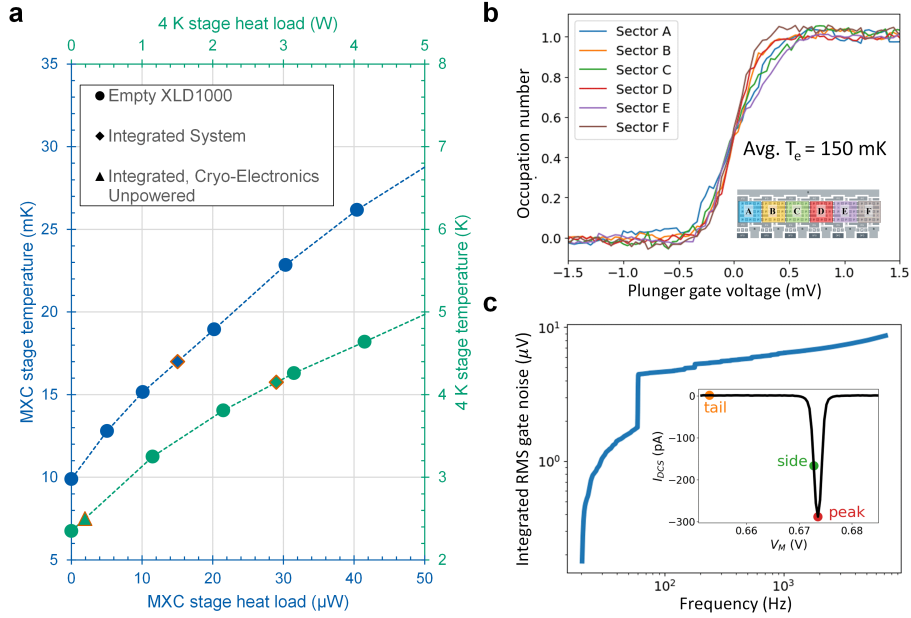

\centering
\smartincludegraphics[width=120mm]{./media/fig_s14} 
\caption{
	\textbf{QPU thermal and electrical environment.} 
	\textbf{a,} Cooling power data (circles) measured in two bare XLD1000 dilution refrigerator cryostats before integrating cabling and control electronics. 
	Markers (diamonds) denote the steady-state temperature of both stages after each cryostat is fully integrated with control hardware, and power is supplied to cryogenic electronics. 
	An additional marker (triangle) shows the 4~K stage temperature of an integrated system with cryogenic electronics deliberately unpowered. 
	Referring these temperatures to the cooling power data of the base cryostats implies \textless200~mW of conductive cabling heat load, and \textless3~W of cryo-control power dissipation at the 4~K stage. 
	\textbf{b,} Charge loading measurements for one quantum dot in each of the 6 sectors across a 54-dot qubit chip, indicating an average electron temperature of 150~mK. 
	All curves are normalized according to the DCS current measured deep in the \(N = 0\ \)and \(N = 1\) charge cells respectively. 
	The lower right inset shows the location of each qubit chip sector (``A''-``F'').
	\textbf{c,} The effective RMS gate voltage fluctuation, defined as the square root of \(\int S_{g}df\), measured with the same qubit chip. 
	The RMS fluctuation is found to be \textless10~$\mu$V over the integration range of 20~Hz to 7~kHz. 
	The inset shows the measured current through the DCS as a function of the M-dot plunger bias, as well as the three biases used to calculate \(S_{g}\) as discussed in the text.
}
\label{figS14}
\end{figure*}

The signals to support the qubit chip and cryo-controller are binned into several functional roles, outlined in Fig.~1 of the main text, each having their own electrical requirements. 
Digital and analog power rails for the cryo-controller, as well as readout amplifier power, are delivered with individual coaxial cables to minimize IR drops and quiescent power dissipation to the 4~K stage. 
The signals supporting the cryo-controller digital I/O and reference voltages (along with a select set of DC qubit chip biases) are routed from room temperature to the 4~K stage of the cryostat via two chains of custom, 100-trace flexible cryogenic cables (see  \secref{flex-cable}). 
The thermal loads at the 50~K and 4~K stages of the dilution refrigerator are minimized by appropriate choice of material according to the requisite signal bandwidth: DC references and qubit chip static biases are routed through a higher resistance CuNi/CuNi (signal/ground) cable chain with 10~MHz of bandwidth, while the high-speed digital I/O is routed through a hybrid Cu/CuNi-Cu/Cu cable chain with a combined bandwidth of 450~MHz.
We estimate the total conductive heat load to 4~K from all room temperature signals to be less than 200~mW, consistent with measured data shown in \figref{figS14}a.

The analog and digital power dissipation of the cryo-controller itself dominates the parasitic loads discussed above, and is estimated at \textless3.5~W (\secref{power-consumption}). 
Close proximity of the cryo-controller to the qubit chip in our architecture improves signal integrity but also necessitates effective thermal isolation of the two components. 
In order to avoid thermal gradients across the motherboard exceeding the superconducting critical temperature (8.5~K) of the mK interconnect, the chip die is directly heat sunk to the 4~K plate through a copper framework. 
Based on sweeps of directly applied 4~K heat while monitoring interconnect trace resistance, we estimate approximately 6~W of cooling power available for cryo-control at 4~K in the current design. 
The combined heat leak to the qubit chip from amplifier cabling and the Nb interconnect is measured to be less than 20~$\mu$W, as shown in \figref{figS14}a, enabling mixing chamber temperatures as low as \textless20~mK in a fully integrated system.

\subsection{Qubit temperatures}\label{qubit-temperatures}

In general, the phonon temperature of the cryostat represents only a lower bound on the effective qubit temperature relevant to a quantum circuit, which may be significantly higher due to thermal interface resistances and electrical noise contributions. 
We directly verify low qubit temperatures by measuring the width of a quantum dot charge number transition as a function of applied plunger gate voltage and fit the data to a characteristic form as in Ref.~\cite{s_dicarlo_2004}. 
With this method we find an average electron temperature of \(\overline{T_{e}} = \) 150~mK across an entire device based on the \(N = 0\ \)to \(N = 1\ \)transition of the quantum dot closest to the 2DEG bath gate in each sector, as shown in \figref{figS14}b. 
This compares favorably to electron temperatures reported from our previous system architecture utilizing all-room-temperature control \cite{s_blumoff_fast_2022}.

We also perform low-frequency charge noise spectroscopy on the same qubit chip by measuring the power spectral density of current fluctuations through a readout dot-charge-sensor (DCS) while biased to maximum conductance (peak), minimum conductance (tail), and maximum transconductance (side). 
Taking linear combinations of the resulting spectral densities \(S_{i}\), we can extract an ``effective plunger gate voltage noise'' \(\ts{S}{g}\):
\begin{equation}
\ts{S}{g} = \frac{1}{G_{m,\textrm{side}}^{2}}\left( \ts{S}{side} - \ts{S}{tail} 
- \left( \frac{\ts{G}{side}^{2}}{\ts{G}{peak}^{2}} \right)\left\{ \ts{S}{peak} - \ts{S}{tail} \right\} \right),
\end{equation}
where \(G_{x}\) are the DCS conductances, and \(G_{m}\) is the transconductance. 
The experimentally measured plunger-gate-referred RMS fluctuation, shown in \figref{figS14}c, is found to be \textless10~$\mu$V in the frequency range of 20~Hz to 7~kHz, demonstrating minimal coupling of 60~Hz mains or other low-frequency noise sources through the cryo-controller/qubit chip interconnect, though this width is still believed to be dominated by system noise rather than intrinsic noise.

\begin{figure*}
\centering
\includegraphics[width=6.49167in,height=0.95972in]{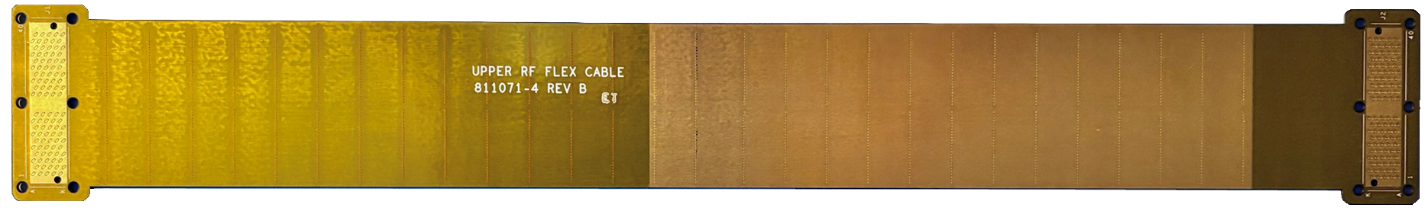} 
\caption{
	\textbf{One of the flex cables that together carry signals between room temperature and 4~K.}
}
\label{figS15}
\end{figure*}

\subsection{Flex cable}\label{flex-cable}

We implement our cryogenic interconnect from room temperature to 4~K using a series of flexible cables (\figref{figS15}) manufactured using standard flex PCB processes. 
The cables are comprised of six metal layers designed in a stacked stripline configuration, with the two stripline layers joined by an adhesive. 
Signal traces are etched from a metal layer and vias are used to form electrical connections to the signal traces, as well as to join the GND planes together to prevent unwanted resonant modes from forming during the transmission of high-speed electrical signals.

Using a combination of CuNi for ground planes and either Cu (for high speed) or CuNi (for low speed) for signal traces maximizes the signal bandwidth while minimizing the passive thermal load. 
We form a cable-to-board connection using Samtec ZRAY spring interposer arrays to form a mezzanine connection between an array of pads on the ends of the flex cables and a matching array of pads on the corresponding PCB. 
Cable thermalization to 50~K and 4~K stage plates are accomplished by means of clamping the cables with a flat plate onto areas of the cables with exposed GND metal. 
These are then clamped to an Au plated Cu cold finger attached to a stage plate within the dilution refrigerator.

For the vacuum feedthrough at the top of the fridge, we epoxy a PCB into a vacuum flange with an appropriately milled channel. 
The PCB has a ZRAY connection internal to the cryostat and uses Samtec HLCD microcoaxial cable on the external side to connect to instruments at room temperature by means of an appropriate breakout PCB. 
We externally verify the resistance and isolation of each trace of each cable using an automated wire harness tester. 
Once installed into the cryostat, we again verify the electrical continuity and isolation of the composite signal chain using an automated wire harness tester while the cryostat is both warm/vented and cold/evacuated using a loopback PCB at the end of the signal chain.

\begin{figure}
\centering
\smartincludegraphics[width=\columnwidth]{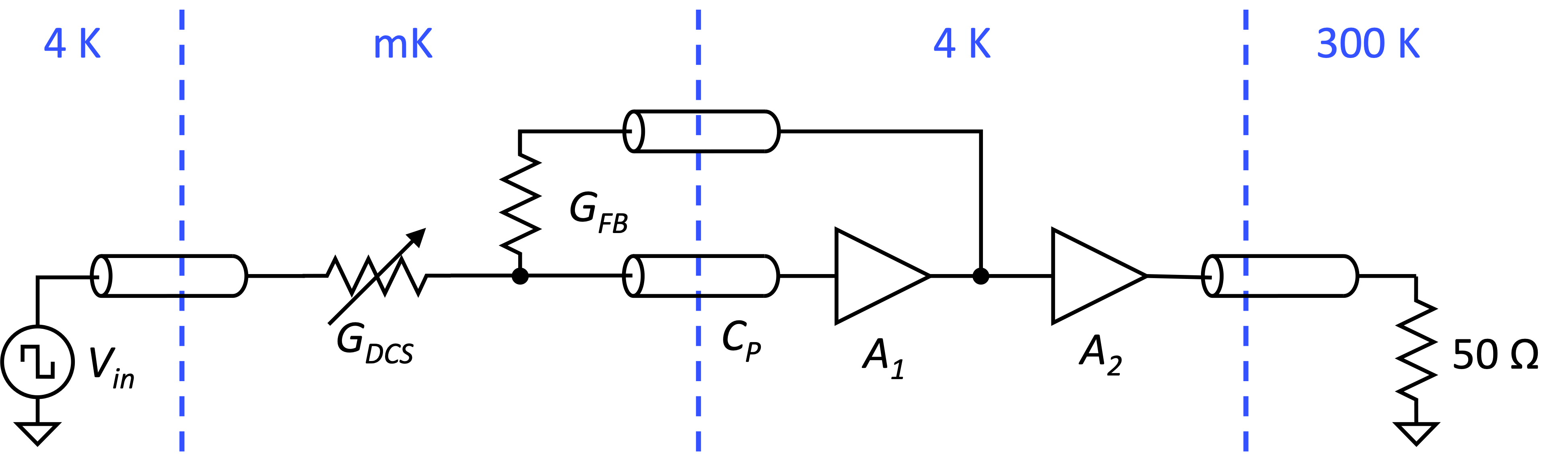} 
\caption{
	\textbf{Simplified electrical circuit of the custom 4~K TIA.} 
	Note this design requires two connections between mK and 4~K per channel.
}
\label{figS16}
\end{figure}

\begin{figure}
\centering
\includegraphics[width=\columnwidth]{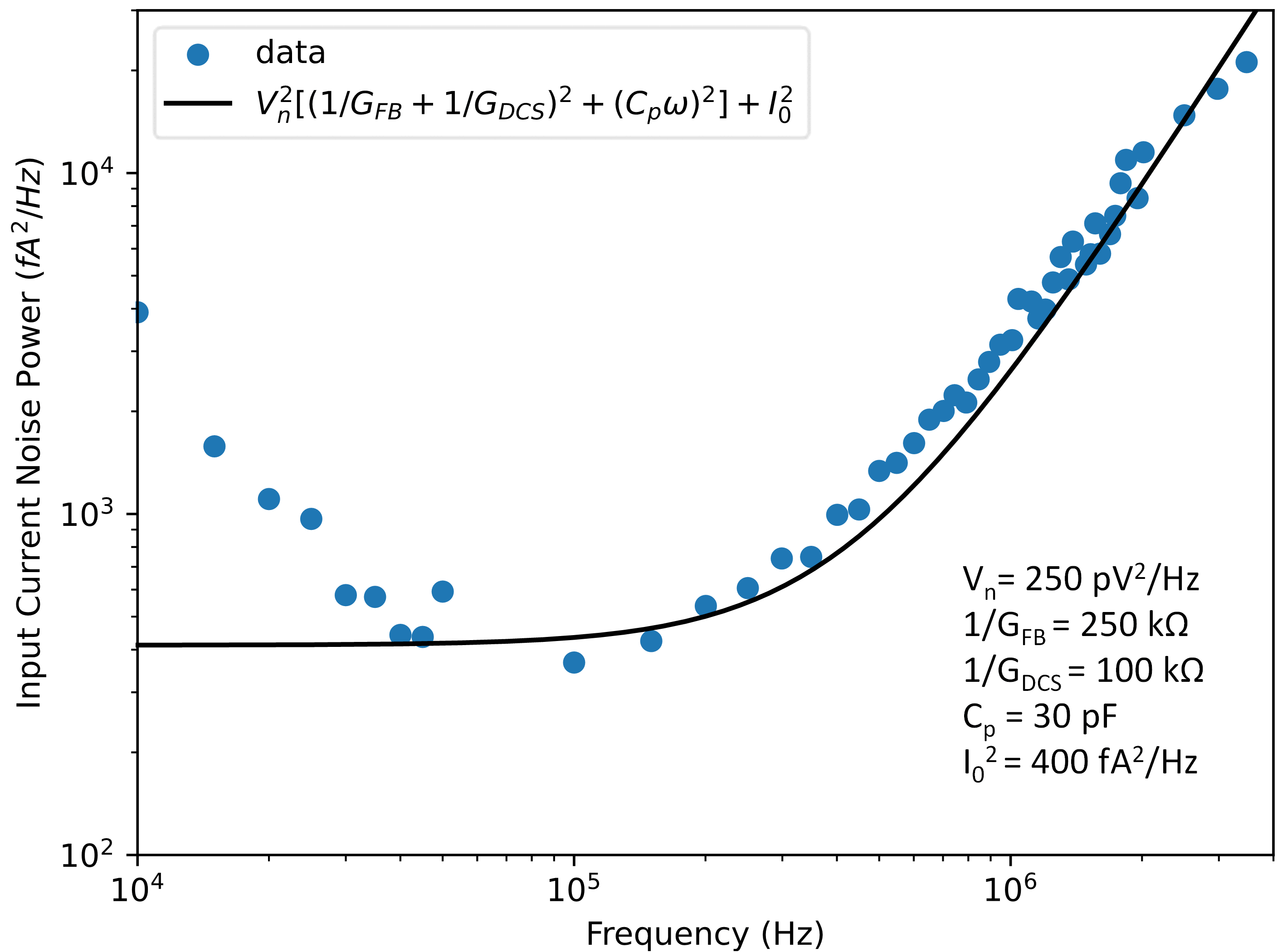} 
\caption{
	\textbf{Transimpedance amplifier input current noise power versus frequency.} 
	At high frequencies $f$ the $f^2$ term proportional to the parasitic input capacitance dominates the modeled noise spectrum (black line) and describes experimental data well (blue circles). 
	Independent estimates of $V$\textsubscript{n}, $G$\textsubscript{FB}, $G$\textsubscript{DCS}, and $C$\textsubscript{p} are shown in the bottom right, while the white noise term $I$\textsubscript{0}, set by the approximately 100~pA measured HEMT gate-current shot noise, resistor thermal noise, and potentially uninvestigated noise channels, was left a free parameter. 
	At low frequencies, higher-order noise channels not included in this model begin to contribute, e.g. series capacitances multiplying the HEMT gate current. 
	Maximum readout signal-to-noise ratio (SNR) is achieved by centering the fundamental modulation frequency around 100 to 200~kHz, where it matches the observed input noise minimum, though this is slower than our typical measurement cadence which must respond to other requirements.
}
\label{figS17}
\end{figure}

\subsection{4~K transimpedance amplifier}\label{transimpedance-amplifier}

Measurement is accomplished by applying a square wave voltage to a variable conductance dot-charge-sensor \(\ts{G}{DCS}\), as in Ref.~\cite{s_blumoff_fast_2022}, and measuring the associated current into a custom low-noise transimpedance amplifier (TIA) based on the model ATF-38143 high-electron-mobility transistor (HEMT). 
A simplified effective circuit of the TIA is shown in \figref{figS16}. 
The transimpedance architecture creates an amplifier input admittance for which $A_{1}\ts{G}{FB}$ in parallel with $j\omega C_{p}$ is much greater than $\ts{G}{DCS}$, where (per \figref{figS16}) $A_1$ is the gain of the first stage amplifier, $\ts{G}{FB}$ is the conductance of the feedback resistor, and \(C_{p}\) is the total parasitic capacitance to ground between the DCS and HEMT gate.   The resulting signal bandwidth and associated settle time is approximately 10~MHz.
We model the amplifier input current noise by including the effects of HEMT voltage noise, thermal Johnson noise, and shot noise according to
\begin{equation}\begin{split}
I_{n}^{2} \approx V_{n}^{2}\left( \left( \ts{G}{DCS} + \ts{G}{FB} \right)^{2} + \omega^{2}C_{p}^{2} \right) + \\ 4k\ts{T}{mK}\left( \ts{G}{DCS} + \ts{G}{FB} \right) + 2qI_{G},
\end{split}\end{equation}
where \(V_{n} \approx 250~\text{pV}/\sqrt{\text{Hz}}\) is the input referred voltage noise of the HEMT and \(I_{G}\) is the gate current of the HEMT.  
Figure \ref{figS17} shows a comparison of this model to measured amplifier current noise data. 
Further noise reduction and increased bandwidth can be accomplished by using a lower noise HEMT and reducing \(C_{p}\).

Overall qubit measurement performance of the integrated QPU is similar to that reported in Ref.~\cite{s_blumoff_fast_2022}.

\section{Software}\label{software}

All qubit experiments in this work were performed using our custom ``Quiver'' control software. 
Quiver has numerous features and integrations to support device operation including, but not limited to, data collection and pulse sequence compilation.

\subsection{Pulse sequence compilation}\label{pulse-sequence-compilation}

Qubit experiments require pulse sequences that are too complex for a human to generate manually. 
We use our custom domain-specific language, ``PulseScript," along with associated software tools to define, manipulate, and compile voltage-time waveforms, i.e. pulse sequences. 
PulseScript contains several high-level language features for programming pulse sequences, such as subroutines, loops, parallelism, synchronization, and native support for sweeping parameters at runtime. 
While PulseScript programs can be written directly, we typically use Python bindings for programmatically generating PulseScript programs for device tune-up, calibration, benchmarking, etc.

The PulseScript compiler targets multiple backends while maintaining strict timing of the input program. 
The primary backend targets the cryo-controller, with additional backends for the cryoMUX systems and to generate metadata for qubit simulation and debugging. 
Several optimization passes are performed to minimize the size of the compiled program, which is particularly important given the memory constraints of the cryo-controller (\secref{digital-architecture}).

\subsection{Circuit compilation}\label{circuit-compilation}

We use a custom structured format called ``Isthmus'', along with associated software tools, for defining and manipulating quantum circuit programs.
Isthmus contains several semantic constructs for programming quantum circuits such as qubits, gates, exchange angles (``exchangles''), initializations, measurements, subroutines, loops, etc. 
We programmatically generate parameterized circuit programs for the specific qubit layout which has been tuned up on the array, ensuring operations are parallelized when possible. 
Generated programs are then translated by expanding qubit gates into sequences of exchange pulses using a database of predefined constructions, preserving parallelism and SPAM operations. 
Dynamical decoupling exchange pulses are woven into gate definitions on any qubits not undergoing exchange rotations or SPAM. 
Care is taken with respect to spacing between pulses on same and adjacent axes to reduce the effects of inter-symbol interference. 
The resulting exchange pulse and SPAM program is translated to PulseScript by querying exchange and SPAM calibration parameters from the target system. 
Compilation continues from here utilizing the previously described pulse-sequence compilation process.

\subsection{Constructing qubit operations}\label{constructing-qubit-operations}

\begin{figure}
\centering
\smartincludegraphics[width=\columnwidth]{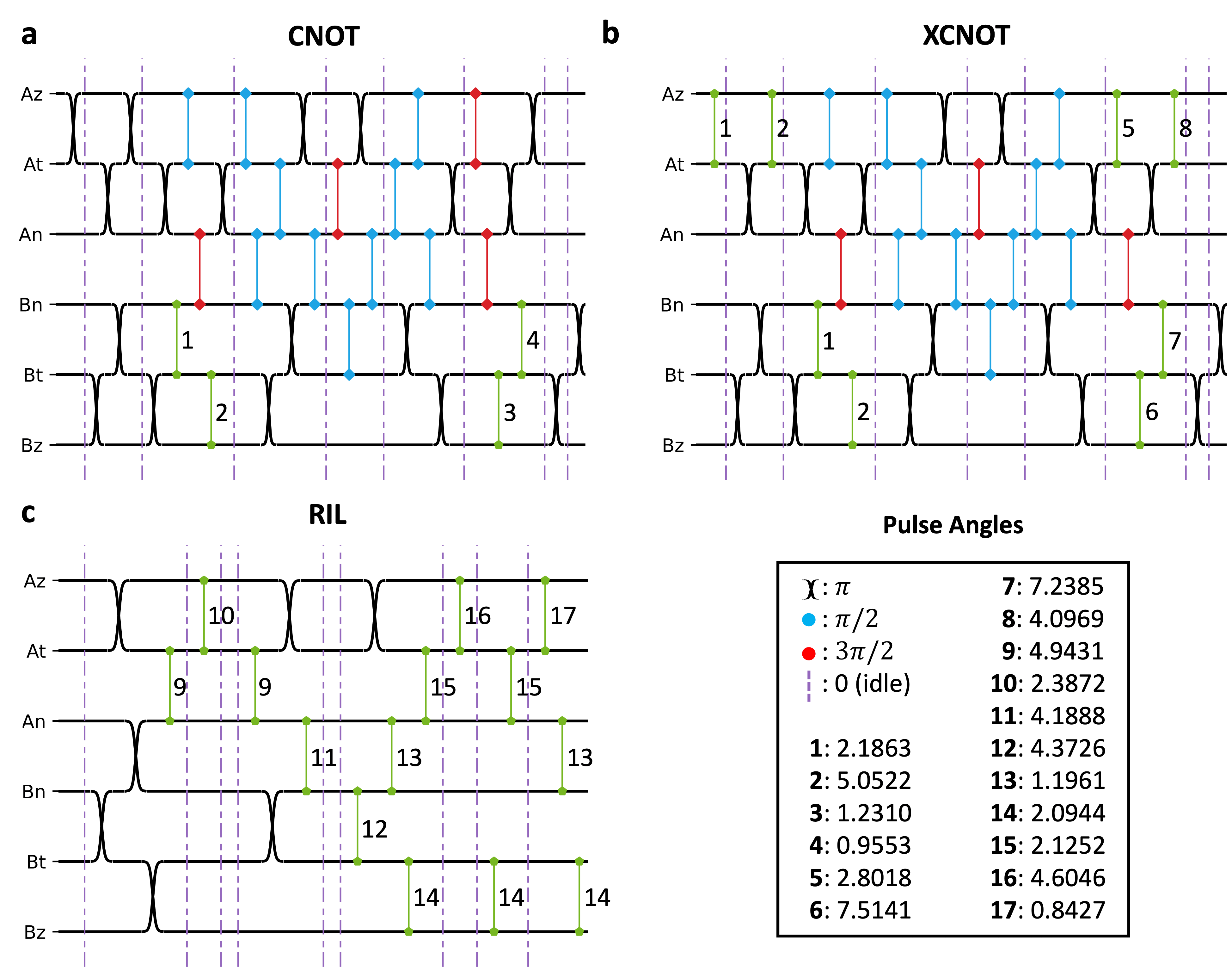} 
\caption{
	\textbf{Exchange-angle implementations of two-qubit operations.}  
	Each horizontal line represents a spin, with vertical connections representing exchange pulses. 
	\(\pi\)-pulses are denoted as crossing lines, to emphasize their equivalence to spin-level swaps.
	``A$x$'' and ``B$x$'' labels represent the spins in encoded qubits. 
	Within each qubit, $x$ can signify ``z'', the outermost spin in the initialization pair, ``t'' the center spin, or ``n'' the third spin. 
	Pulse angles of \(\pi/2\) (blue) and \(3\pi/2\) (red) are also common in entangling gates. 
	Other exchangles are shown in green, and their numeric values are given in the key.
	Idle timesteps (purple dashed lines) do not affect the functionality of the operation, but they enforce a regular cadence that mitigates some types of error. 
	These implementations of 
	\textbf{a,}~CNOT and 
	\textbf{b,}~XCNOT differ in only six places, because they were derived from the same Fong-Wandzura CZ; other sequences, such as 
	\textbf{c,}~RIL can show larger differences. 
	These implementations are for an n-n coupling between encoded qubits. 
	Slightly modified versions are used for z-z coupling.
}
\label{figS18}
\end{figure}

\begin{figure*}[h]
\centering
\smartincludegraphics[width=120mm]{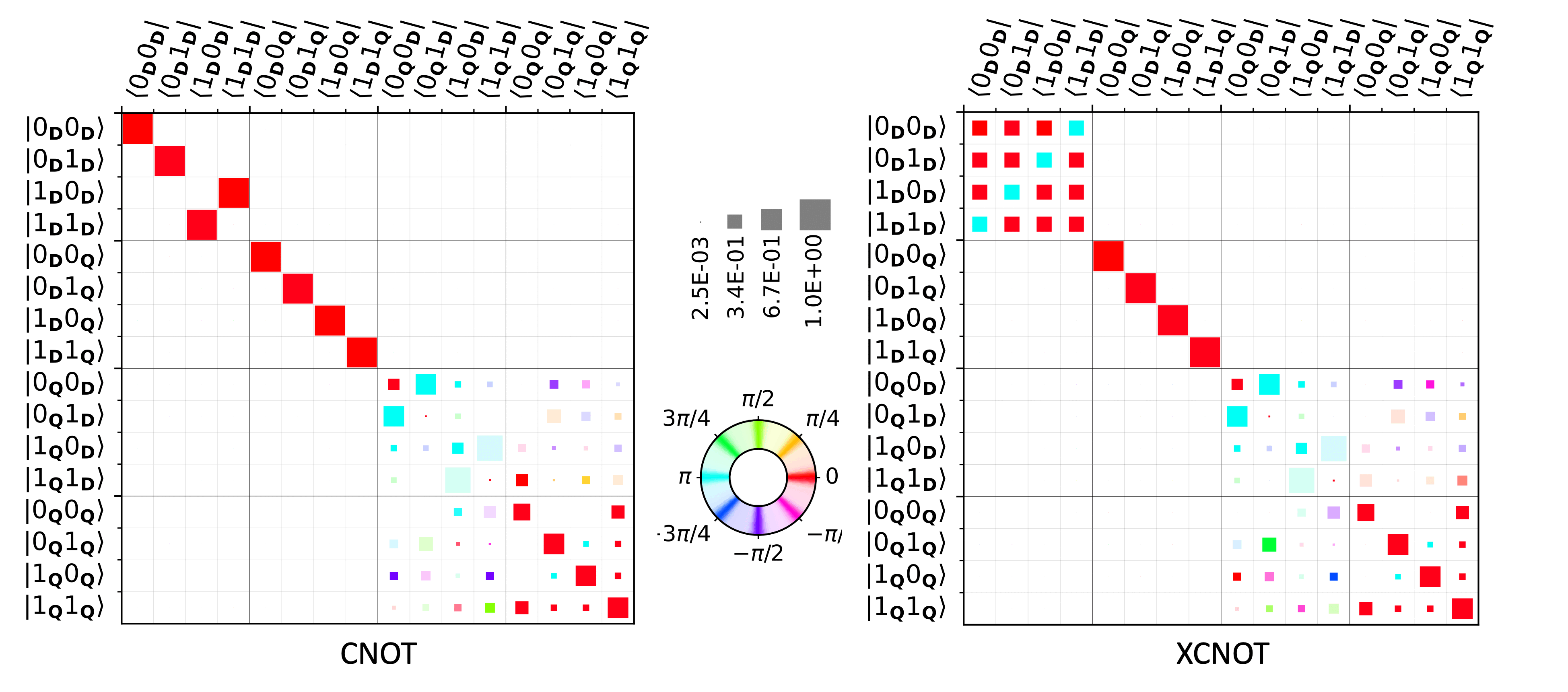} 
\caption{
	\textbf{Matrix representations of CNOT and XCNOT sequences in \figref{figS18}.} 
	After gauge-averaging we obtain GAQQMap (section \secref{gaqqmap}), which we represent in the Kraus operator-sum representation: \(\mathcal{E}(\rho) = \sum_{i}^{}{K_{i}\rho K_{i}^{\dagger}}\ \)acting on a ququart state \(\rho\) with Kraus operators \(K_{i}\). 
	Here we show the average of the Kraus operators \(|K_{i}|\) that summarizes the sequence dynamics. 
	The matrix is organized by the encoded (D) and leaked (Q) property of each of the two DFS qubits. 
	For example, the block ``DQ'' represents ``first qubit encoded, second qubit leaked.''
	The size of each square represents the magnitude of the matrix element, while the color represents the complex phase.
	The top left of each plot shows the action of the sequence on the encoded state, showing either the familiar structure of CNOT (left) or the less-familiar structure of the XCNOT (right). 
	Both the CNOT and the XCNOT are only directionally leakage-controlled, as evidenced by the large block of non-zero elements in the lower right of both process matrices. 
	This observation, contrasting the leakage-controlled CNOT sequence demonstrated in Ref.~\cite{s_weinstein_universal_2023}, highlights the fact that if the first qubit is in the leaked (Q) state, the second qubit is not guaranteed to remain encoded under either sequence.
}
\label{figS19}
\end{figure*}

\begin{figure*}
\centering
\includegraphics[width=165mm]{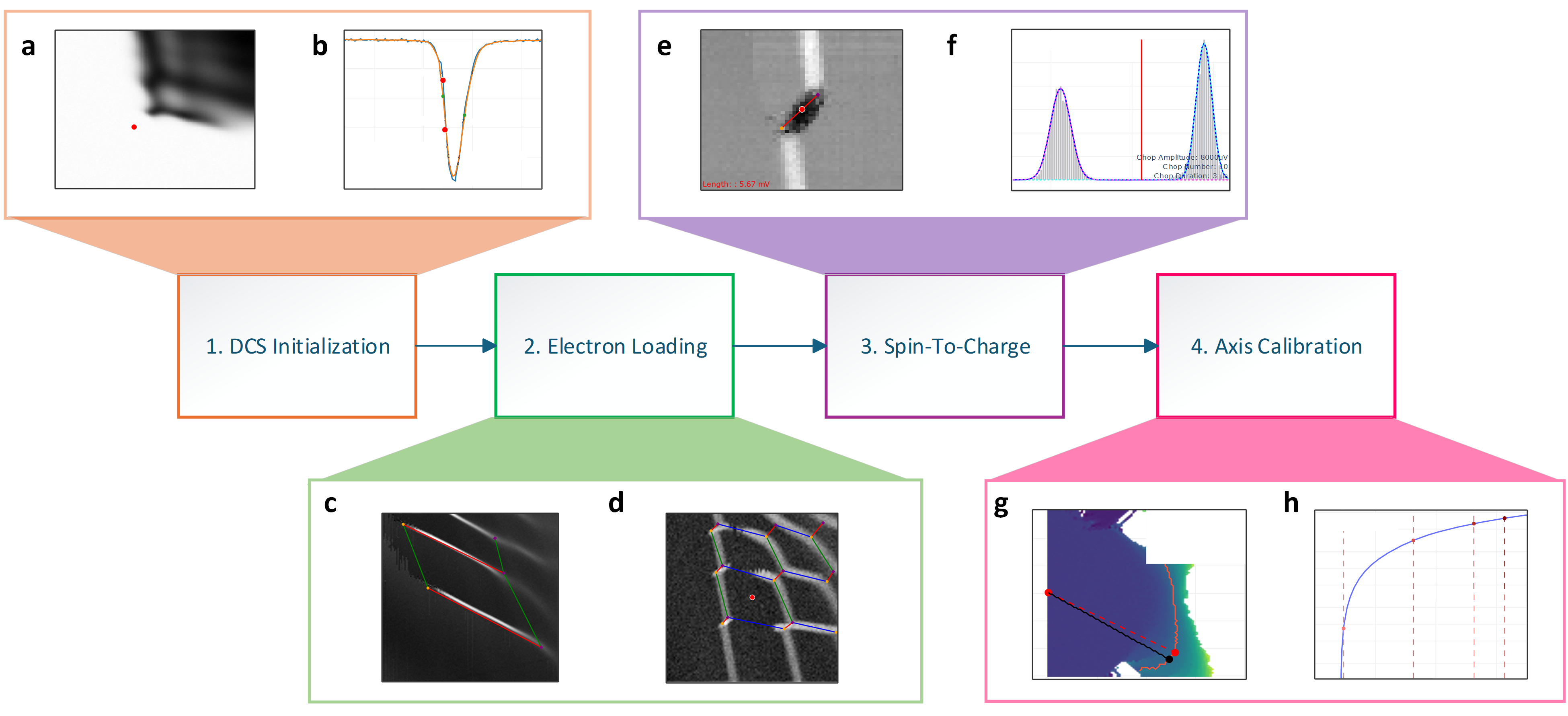} 
\caption{
	\textbf{Tune}-\textbf{up flow for an exchange-only spin qubit which proceeds via dot charge sensor (DCS) initialization, electron loading, spin-to-charge conversion, and coherent axis calibration.} Selected steps of each phase are shown in subfigures.
	\textbf{a,} A Z-gate pinch-off experiment identifies Z-gate voltages
	where current between the source and drain is pinched off. 
	\textbf{b,} A tune-DCS experiment sets M-gate voltage to the point of peak sensitivity. 
	\textbf{c,} A plunger-versus-tunnel experiment (PvT) displays electron loading lines as identified by a CNN. 
	\textbf{d,} A plunger-versus-plunger experiment (PvP) reveals charge stability cells as identified by a CNN. The idle operating voltage setpoint is	automatically updated to the center of the (1,1) charge cell.
	\textbf{e,} A ``tiebar'' PvP evaluated with a CNN provides a rough measure of interdot tunnel coupling and an initial guess of the spin-to-charge measurement coordinates. 
	\textbf{f,} A histogram of measured currents from a spin-to-charge experiment reveals Gaussian peaks associated with the singlet and triplet spin states, a threshold mcurrent value, and measurement SNR. 
	\textbf{g,} For a fixed exchange pulse duration, the mapping of phase as a function of pulse voltages identifies the symmetric axis \& maximizes charge-noise insensitivity \cite{s_reichardt_422_2025}. For time-domain control we then choose a single bias point along that axis at the target exchange frequency $J/h$. \textbf{h,} The exchange pulse duration is swept and a fit provides an invertible pulse duration to exchange angle mapping.
}	
\label{figS20}
\end{figure*}

The above compilation process relies on a pre-defined database of operations, mapping qubit-level instructions such as $H$ or CNOT into their implementations as sequences of exchange pulses. 
Some sequences, such as SWAP or single-qubit Cliffords, are simple enough for implementations to be analytically derived. 
Other two-qubit operations are generally more complicated, and we employ a variety of analytic and numeric techniques to add new sequences to the library.

The entangling operations used in this work were derived from the Fong-Wandzura CZ sequence~\cite{s_fong_universal_2011}. Adding Hadamards before and after the CZ ($Z$-controlled-$Z$), either to the target qubit or to both qubits, gives us a CNOT ($Z$-controlled-$X$) or an XCNOT ($X$-controlled-$X$), respectively. 
From here, the sequence can be manipulated in several ways: inserting pairs of \(\pi\) pulses that resolve to identity, commuting $Z$-rotations with the entire CZ, or permuting pulses allowed by the commutation or Coxeter relations of the exchange algebra [e.g. \(E_{ij}(\pi)E_{jk}(\phi)E_{ij}(\pi) = E_{jk}(\pi)E_{ij}(\phi)E_{jk}(\pi)\), for exchange unitary $E_{jk}(\theta)$ executing exchangle $\theta$ between spins $j$ and $k$].
Using these techniques, we manipulate the sequences into a more structured cadence, where pulses on any given spin pair are evenly spaced in time and no spin is ever touched by two consecutive pulses.
This cadence roughly doubles the number of pulses required per sequence, but the added structure suppresses error due to pulse-to-pulse crosstalk and imparts partial magnetic noise decoupling. 
The end result is a set of 45-timestep CNOT and XCNOT implementations (\figref{figS18}a and b) for all encoded qubit pair layouts on our devices. 
These implementations, while longer than the original CZ sequences, have preferable intersymbol interference and magnetic decoupling properties.

Reset-if-leaked (RIL) operations were found using a custom exchange sequence search tool, dubbed SpEQuLator (Spin Exchange Quantum Logic Gate Cooker). 
SpEQuLator uses a replica-exchange-Monte-Carlo algorithm~\cite{s_hukushima_exchange_1996} to search for pulse sequences that approximately implement target unitary operations, then switches to a GRAPE (GRadient Ascent Pulse Engineering) algorithm~\cite{s_khaneja_optimal_2005}  to refine the sequences to high precision. 
For RIL specifically, the optimization target is a unitary that (1) acts as an identity when the ancilla is in \(\left. |0 \right\rangle\) and the data qubit is unleaked, and (2) deterministically maps a \(\left. |0 \right\rangle\) ancilla with a leaked data qubit to a \(\left. |1 \right\rangle\) ancilla with the data qubit in some unleaked computational state. 
The precise state of the data qubit after leakage reduction, or the gate action with an input ancilla in the \(\left. |1 \right\rangle\) state, are unspecified. 
Once we had a SpEQuLator-derived RIL sequence, we applied the same transformation techniques as above to produce a 30-timestep RIL with a regular cadence (\figref{figS18}c), which we used in the experiments presented here. 
Other unitary constraints have allowed SpEQuLator to generate operations including CS, iSWAP, CCZ, and leakage-controlled variants of CZ and CNOT.

\subsection{GAQQMap}\label{gaqqmap}

Considering the notation \(|S_{12},S;m\rangle\) described in the \hyperref[methods-section]{Methods section}, recall that encoded qubits live in the manifold $S=1/2$ and $m=\pm 1/2$. 
We refer to these four states (two gauge states for each of two qubit states) as $D$, referring to the gauge ``Doublet.'' 
However, errors cause leakage into \emph{S}=3/2 and its Quadruplet of $m$ values, a manifold we call $Q$. 
In the presence of arbitrary errors, relative populations and coherences of all eight states could in principle be important, but the lack of spin polarization at small magnetic field renders a symmetry enabling us to neglect the sign of $m$. We therefore model a gate as acting on states maximally mixed in sign($m$), and (in simulation as well as in physical measurement) ignore (trace over) this degree of freedom after each operation. 
With only this binary degree of freedom traced, our resulting state space for each triple-dot is the ququart of states $\ket{S_{12},S;|m|}  = \ket{0,1/2;1/2}, \ket{1,1/2;1/2}, \ket{1,3/2;1/2},$ and $\ket{1,3/2;3/2},$ abbreviated as $\ket{0\textsubscript{D}}, \ket{1\textsubscript{D}}, \ket{0\textsubscript{Q}},$ and $\ket{1\textsubscript{Q}},$ respectively. 
Hence this is a two-dimensional encoded (\emph{D}) qubit and a two-dimensional leakage (\emph{Q}) space. 
This averaged manifold is referred to as a Gauge-Averaged QuQuart (GAQQ), and logical gates, with or without error, can be described by a map between double-Pauli operators on these states, which we call a GAQQMap. 
We may represent the map in the Kraus operator-sum representation, \(\mathcal{E}(\rho) = \sum_{i}^{}{K_{i}\rho K_{i}^{\dagger}}\) where \(K_{i}\) are the Kraus operators corresponding to each traced gauge state sign($m$). 
The Kraus operators then intuitively represent the dependence on sign($m$).  
While non-physical, for visualization of our generated sequences, we compute the mean of the Kraus operators of the GAQQMap and organize them by the encoded and leakage spaces of each qubit. 
See \figref{figS19}\ for examples of this technique applied to CNOT and XCNOT two-qubit gates. 
For modeling gate error via simulation, which we discuss further in \secref{qubit-characterization}, GAQQMaps including simulated errors from full spin dynamics are created and used to populate event-based simulations.

\section{Tune-up and calibration}\label{tune-up-and-calibration}

Our Quiver control software includes a large suite of available measurements, analyses, and automated routines that enables users to efficiently tune-up, calibrate, and characterize the qubit device and controller with a high degree of flexibility.

Due to material and fabrication non-uniformity, both within and between devices, device tune-up is exploratory and non-deterministic.
Figure \ref{figS20} shows an overview of the basic tune up workflow divided in to four phases. 
Effective optimization within and between phases often requires iterative trial, evaluation, and adjustment. 
Dead ends require returning to known-good checkpoints where alternative paths can subsequently be explored. 
The software is designed to be flexible for users with loops, backtracking, conditional branching, and adaptive heuristics supported as first-class concepts. 
In addition to providing all the components required to enable tune-up, we have begun capturing proven workflows that combine the components into automated routines such that tune-up can proceed with increasingly less human interaction.

\begin{figure*}
\centering
\smartincludegraphics[width=165mm]{./media/fig_s21}
\caption{
	\textbf{Baseline CNN model architecture}. 
	A relational network generates high-dimensional embeddings of the information in the image, to which voltage coordinate information is added so the network is aware of voltage space distributions of the features. 
	An attention transformer decoder refines the embeddings which are used by multi-layer perceptrons (MLPs) and a relational graph convolutional network (RGCN) together to predict ``keypoint'' locations in voltage space as well as the links/relationships between them, which respectively form the nodes and edges of a graph.
}
\label{figS21}
\end{figure*}

\begin{figure*}
\centering
\includegraphics[width=140mm]{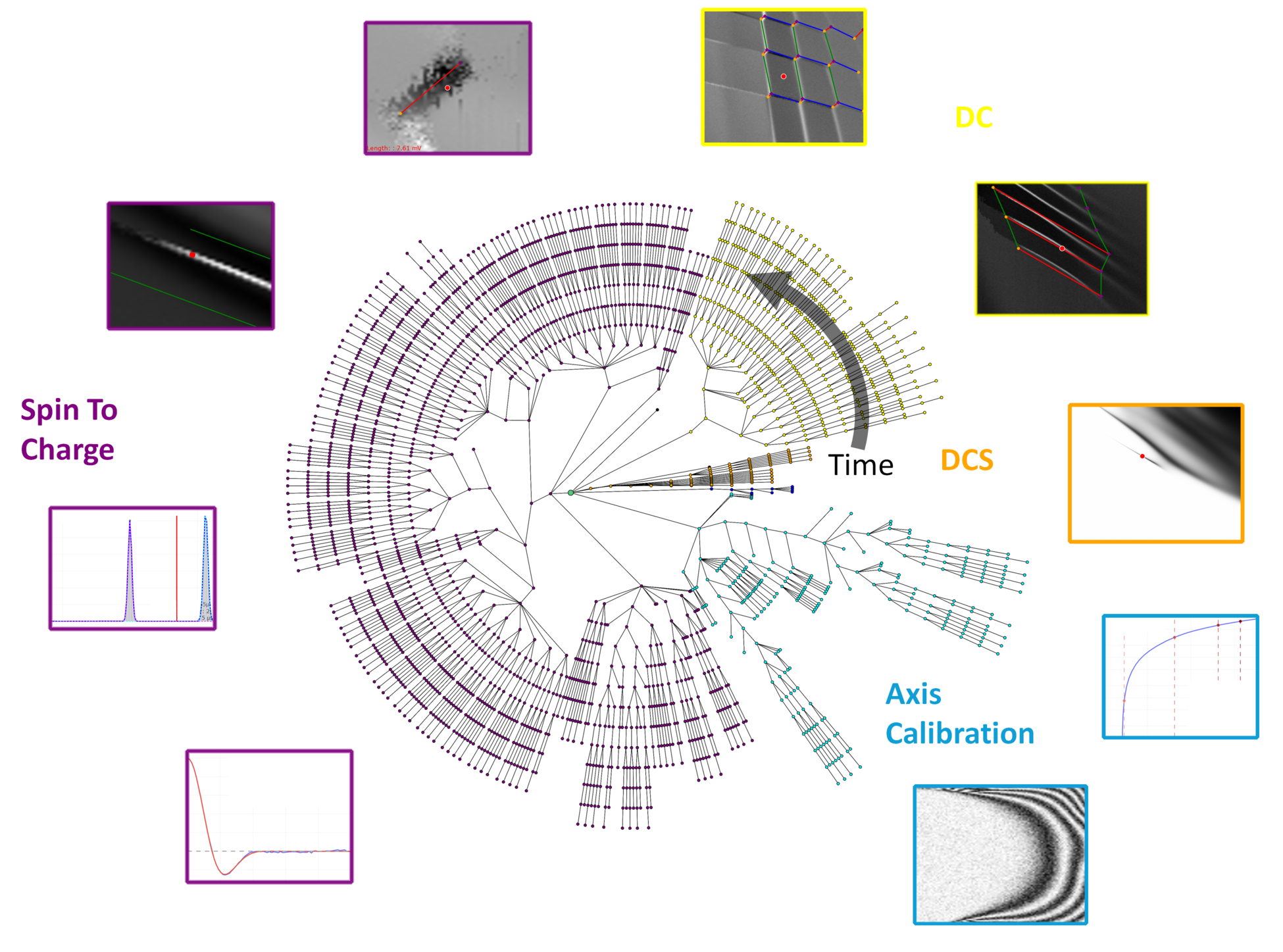} 
\caption{
	\textbf{An example automation tree capturing the sequence of fully automated actions taken during tune up of a single qubit.} 
	The green central node represents the start, and a depth-first traversal indicates the total ordering. 
	The sequence of actions is dynamic, with each action having the ability to insert additional actions (child nodes) or yielding to the parent node. 
	An example of the dynamic nature is when analysis of an experiment (e.g. PvP) indicates unsuccessful determination of the desired operational parameters and further data collection is inserted into the tree.
}
\label{figS22}
\end{figure*}

\subsection{Parallelism}\label{parallelism}

Our software can take multiple routines for tune-up on separate device locations and merge experimental steps to create a routine that executes the multi-location experiment in parallel. 
Subsequent analysis and automated control-flow logic then occur sequentially. 
This parallelization technique enables a speedup of up to a factor of 4 to 5 when each DCS is used in the multi-location experiment. 
We primarily use this technique during the electron loading phase of tune-up, where many charge stability experiments are run at each location to converge on the target location in the high dimensional voltage space. 
Tune-up and calibration of larger devices is expected to be embarrassingly parallel.

\subsection{Neural nets for charge stability inference}\label{neural-nets-for-charge-stability-inference}

We use custom machine learning models to analyze and extract features from charge-stability data, enabling automated loading of one electron in each plunger gate and the tuning of tunnel couplings. 
We have separate convolutional neural net (CNN) models for PvT, PvP and ``tiebar'' experiments (tiebar is a PvP experiment zoomed-in on the (1,1)-(2,0) charge transition). 
Each CNN model is trained on thousands of human-labeled data sets. 
The CNN models produce a graph of ``keypoints'' which an automated routine can use to adjust gate voltages to values of physical significance for the state of the device, e.g. electron occupancies or tunneling rates. 
The baseline model architecture is shown in \figref{figS21}.

\subsection{Automation}\label{automation}

While a simplified tune-up workflow is illustrated in \figref{figS21}, in practice a tree data structure records the complete sequence of actions performed during a fully automated tune-up run. 
Each node in this automation tree is an action taken (i.e.  data collection, analysis, decision logic, etc.) and a depth-first traversal indicates the total ordering of actions. 
Each node is automatically evaluated at the time it occurs and given a grade that enables quick assessment of status and later post-hoc interpretability.
Such a record is critical after long unattended operation over several hours or days. 
Furthermore, we leverage trees from multiple runs to perform systematic analysis of success rates and failure modes, enabling iterative improvement to automated routines.
Figure \ref{figS22} is an example of an automation tree for a successful tune-up of a single qubit.

\section{Qubit characterization}\label{qubit-characterization}

We use a variety of techniques to characterize the qubit chip and its performance. 
We highlight some of them here.

\subsection{Initialization window widths}\label{initialization-window-widths}

\begin{figure}
\centering
\smartincludegraphics[width=\columnwidth]{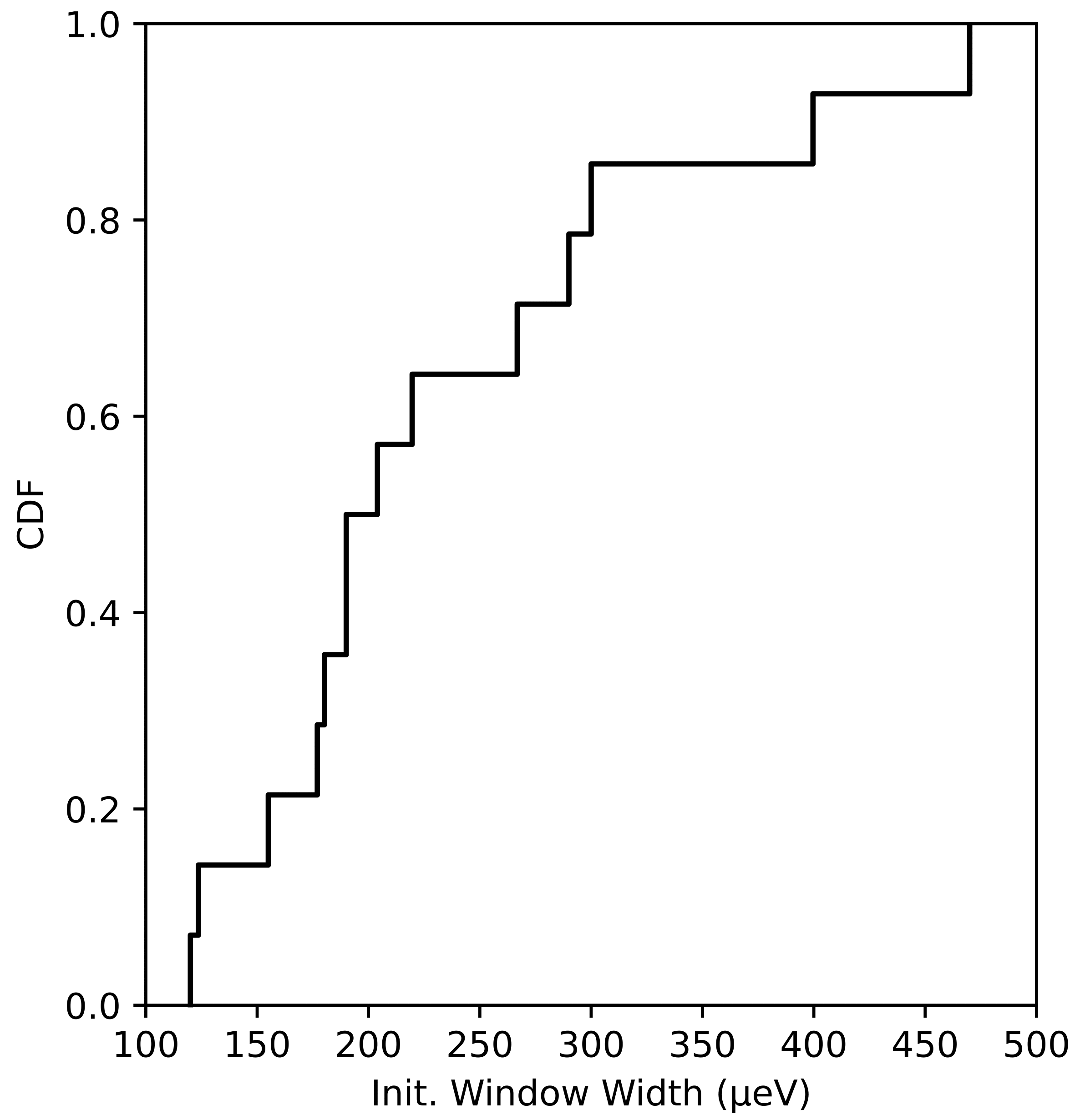}
\caption{
	\textbf{Measured width of the (2,0)-(3,0) initialization window, scaled to $\boldsymbol{\mu}$eV using a lever arm, sampled from nine different 54-dot devices.}
}
\label{figS23}
\end{figure}

Across nine devices, we estimate two-electron excited-state splittings from the width of the initialization bias and find a median energy splitting of roughly 200~$\mu$eV (\figref{figS23}). 
This is consistent with FCI simulations of orbital-limited excitations in these devices. 
At 150~mK effective temperature in the device, using this energy splitting in a Boltzmann distribution \cite{s_blumoff_fast_2022} yields an initialization infidelity of approximately 5\e{-9}, a negligible contribution to SPAM infidelity and consistent with our observations of it typically being challenging to differentiate from measurement error.

\subsection{Device voltage state}\label{device-voltage-state}

\begin{figure*}
\centering
\smartincludegraphics[width=165mm]{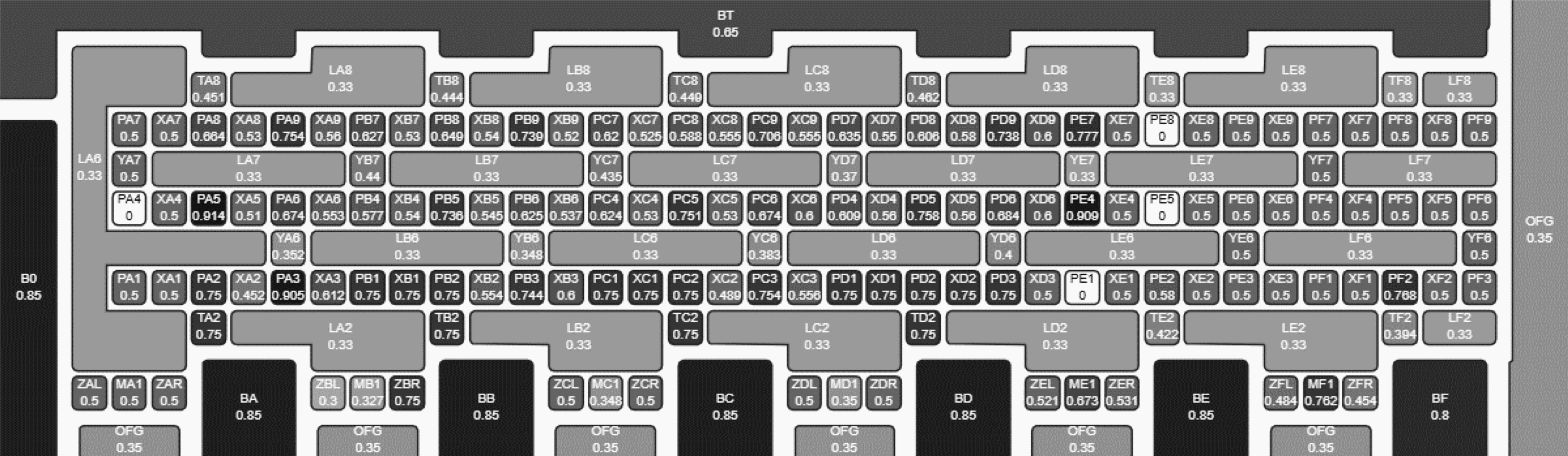}
\caption{
	\textbf{Biases applied to each gate electrode, in the configuration utilized for the distance-5 repetition code.}
}
\label{figS24}
\end{figure*}

The precise biases applied to each electrode in the configuration employed for the distance-5 repetition code is shown in \figref{figS24}. 
All biases applied to each gate electrode fall below 1.0~V, a requirement of the cryo-controller. 
A few gate electrodes have 0~V applied due to issues in the specific cryo-controller assembly used here that shorted some of the voltages, prohibiting them from being utilized.

\begin{figure*}
\centering
\smartincludegraphics[width=165mm]{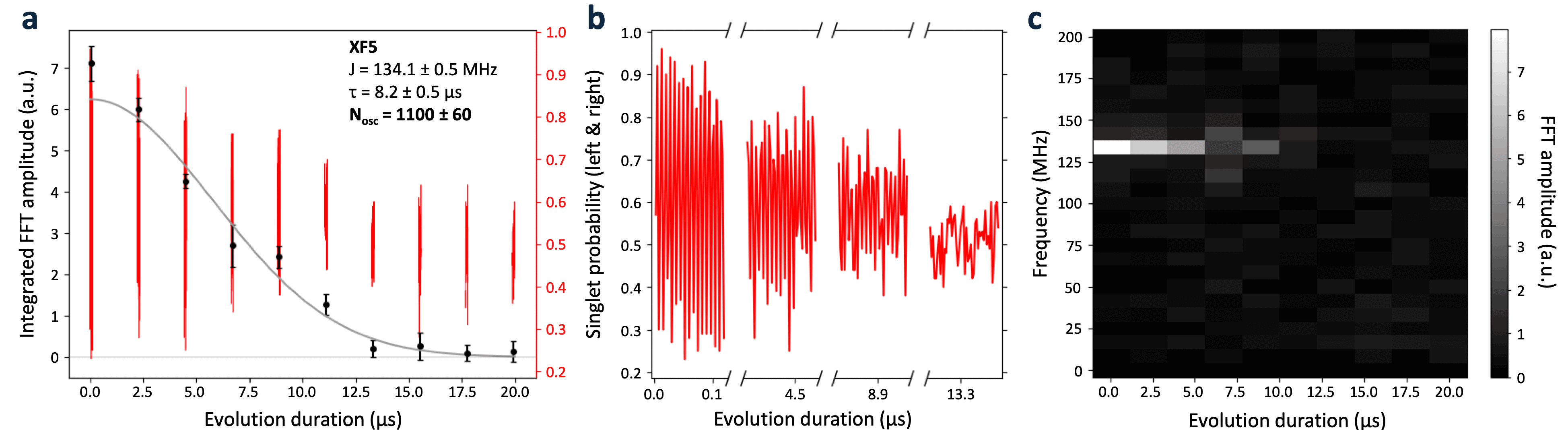}
\caption{
	\textbf{Stroboscopic} \textbf{\emph{N}\textsubscript{osc} methodology.}
	\textbf{a,} Example of stroboscopic exchange oscillation dataset from a sister 54-dot device, showing segmented singlet probability oscillations (red) with corresponding integrated FFT amplitudes (black datapoints) and fitted gaussian decay curve (black line). 
	The extracted decay time of 8.2~$\mu$s and measured exchange rate of 134~MHz yield a \Nosc\  of 1100. 
	\textbf{b,} Expanded view of coherent oscillations in selected segments of the dataset in the left panel, showing stronger coherence at earlier evolution durations. 
	\textbf{c,} Grayscale intensity map of the FFT amplitude vs. FFT frequency and evolution duration.
}
\label{figS25}
\end{figure*}

\begin{figure*}
\centering
\smartincludegraphics[width=165mm]{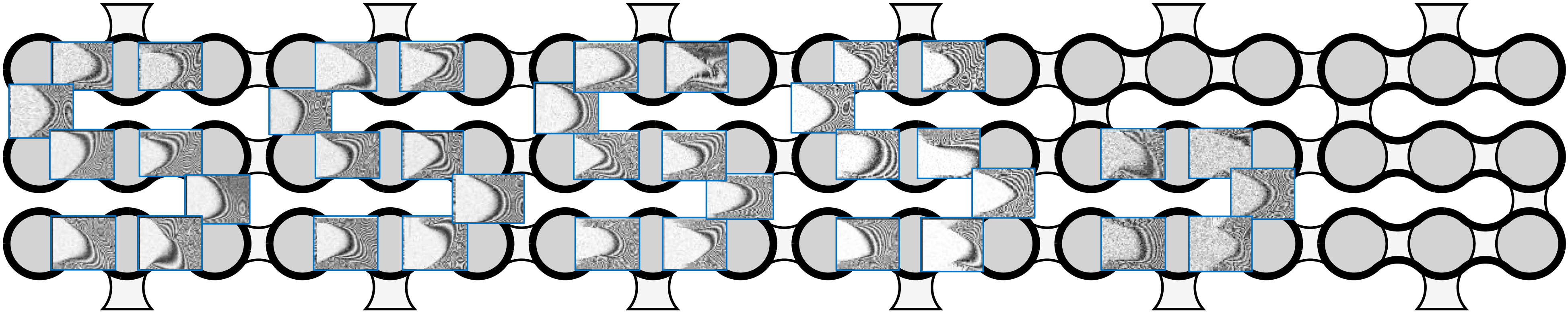}
\caption{
	\textbf{Representative device fingerprints.}
	Overlay of exchange axis ``fingerprints'' characterized (in a MUX system) on a qubit chip of the same generation as that used for the data in the main text. 
	Missing fingerprints in the tested sectors are not because of finite yield, but rather are a detail of with how MUX switching works, and were not attempted due to time constraints.
}
\label{figS26}
\end{figure*}

\begin{figure}
\centering
\includegraphics[width=\columnwidth]{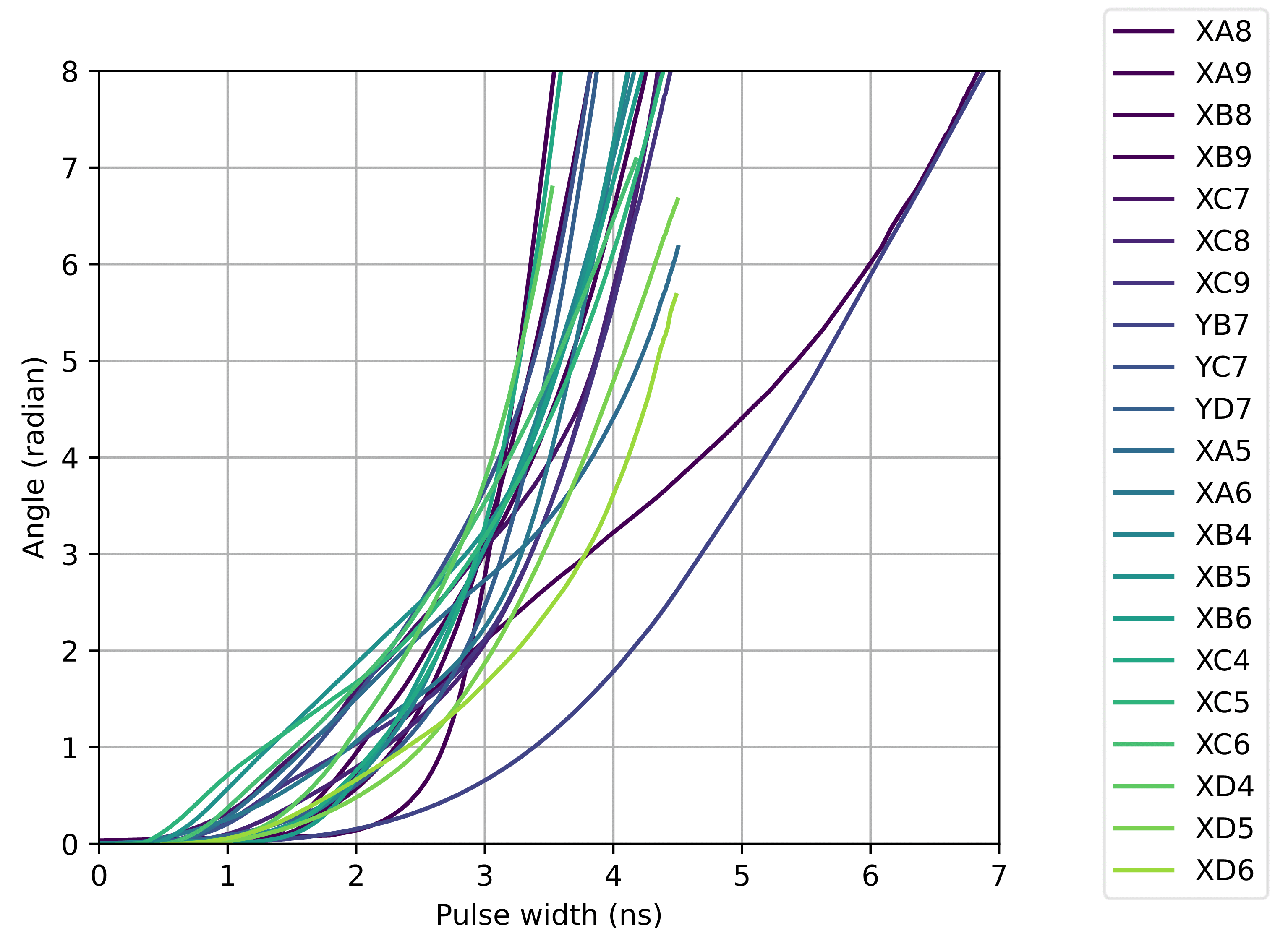} 
\caption{
	\textbf{Rotation angle to pulse-width mapping for all exchange axes utilized in the \mbox{distance-5} repetition code}. 
	Some axes (e.g. XB7, XA9) suffer from non-idealities in the daughterboard, and consequently targeted a lower exchange rate leading their calibration curves to fall outside the distribution here. 
	Nonetheless, all utilized rotation angles fall within the timing range {[}0.3~ns, 7~ns{]}.
}
\label{figS27}
\end{figure}

\begin{figure}
\centering
\smartincludegraphics[width=\columnwidth]{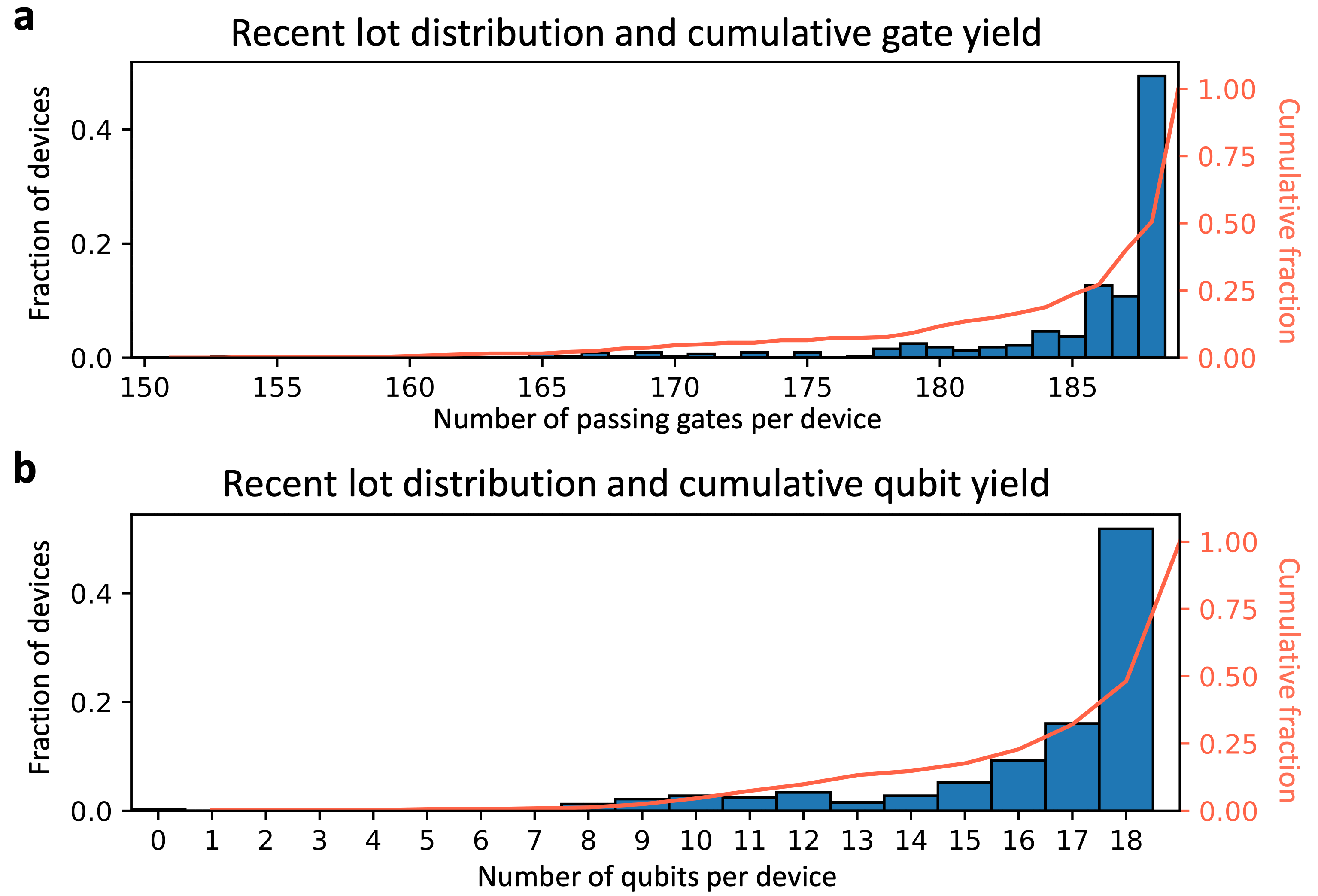} 
\caption{
	\textbf{Wire functional yield from a single device lot, obtained from room temperature wafer probe prior to die singulation.}
	This representative lot was fabricated more recently than the devices utilized in the main text.
	\textbf{a,} Fraction of devices exhibiting the indicated number of lines passing wafer probing, out of a total possible number of 188~lines (blue, left axis) and corresponding cumulative fraction of devices (orange, right axis). 
	\textbf{b,} Fraction of devices exhibiting the listed number of triple-dot qubits with all lines passing wafer probing, as defined from an assumed rectilinear layout of qubits on the device (blue) and corresponding cumulative fraction of devices (orange).
}
\label{figS28}
\end{figure}

\subsection{N\textsubscript{osc} methodology}\label{nosc-methodology}

To evaluate intrinsic charge noise, exchange oscillations are typically measured using room temperature electronics, with or without a cryogenic demultiplexing circuit (referred to here as cryoMUX, or simply MUX). 
The latter is an in-house, custom cryogenic demultiplexing chip, more similar in concept to those in \cite{s_paquelet_wuetz_multiplexed_2020, s_pauka_characterizing_2020} than to on-chip \cite{s_eastoe_method_2025, s_thomas_rapid_2025, s_tosato_crossbar_2026, s_wolfe_on-chip_2024, s_puddy_multiplexed_2015}, enabling measurement from a single qubit following control-signals delivered to every qubit in the array. 
Whether cryoMUX is employed or not, we measure exchange oscillations as a function of evolution duration at a frequency \(J/h \approx 100\ \)~MHz, where oscillations can be fully resolved and charge noise is typically the dominant source of decoherence. 
The observed \(1/e\) time \(\tau\) of the Gaussian decay envelope can be used to calculate the effective number of exchange oscillations \(\ts{N}{osc} = J\tau/h\). 
In the high-\(\ts{N}{osc}\) regime, hardware memory constraints can limit the maximum program length uploaded to a room temperature AWG, preventing rapid sampling of the entire decay curve with full time resolution.
While portions of the sweep could be compiled and measured sequentially, each portion would sample the local charge noise at different times with inconsistent averaging. 
When measuring \(\ts{N}{osc}\) with a cryo-controller we increment evolution duration with sweep pulse repetition count rather than pulse duration, however the stroboscopic sampling technique remains applicable.

We address these limitations by using a stroboscopic time-domain exchange oscillation measurement, sampling the oscillations at full time resolution in multiple equal-length segments along the decay curve. 
An FFT transform is applied to each segment, where a Gaussian fit is used to extract instantaneous frequency \(J/h\) and integrated amplitude. 
The FFT amplitude of each segment is then fit to a Gaussian decay profile to determine the remaining decoherence parameters \(\tau\) and \(\Nosc\).

Figure \ref{figS25} shows \(\Nosc\) measurement from a 54-dot qubit chip using the stroboscopic methodology. 
The left panel shows raw data acquired in 10 ``chunks.'' 
Within each chunk, the results are sampled on a 2.5~ns timescale to fully resolve coherent oscillations. 
Black data shows the magnitude of the FFT of each chunk, which is a metric for the degree of coherence within that sampling time. 
A Gaussian fit to the black data yields a decay timescale of 8.2~$\mu$s. 
This quantity multiplied by the exchange frequency 134~MHz, gives a coherence Q-factor \(\ts{N}{osc}\) of 1100.

The middle panel of \figref{figS25} shows sampling of chunks from the first panel, highlighting the coherent oscillations present at early dephasing times, and decohered noisy variation at the latest times sampled. 
The right panel shows the FFT of each chunk of acquired data, indicating the operational frequency of  about 134~MHz and the decay in coherence with increasing evolution duration.

Note that the \(\Nosc\) data presented in Fig.~3 of the main text was taken for \({J}/{h} \approx 100\)~MHz. 
For reasons described in Ref.~\cite{s_reed_reduced_2016}, insensitivity and therefore \(\ts{N}{osc}\) typically increases with \(J\). 
Based on the calibration curves in \figref{figS27} and accounting for finite pulse rise times, \(J/h\) is likely significantly higher than 100~MHz during the QPU demonstrations shown in the main text.
Though insensitivity was not characterized in this regime, the reported values of \(\ts{N}{osc}\ \)are therefore likely an underestimate.

\subsection{Fingerprint compilation}\label{fingerprint-compilation}

To illustrate the ``quantum functional yield'' of the 54-dot devices, we examined the extent to which a ``fingerprint'' of coherent exchange could be attained on every available exchange axis \cite{s_reichardt_422_2025}, shown in \figref{figS26}.
This work was pursued in a cryoMUX system with room temperature DACs and AWGs and a cryogenic demultiplexer. 
This system is limited to individual controllability to nine dots or three qubits worth of channels (each qubit must also be on a distinct row of the device) at a time. 
This work commenced on the left-most side of the device and progressed to the right, but was halted during tune-up of the penultimate device sector for a non-technical reason. 
Additionally, the labeled X gate electrodes between sectors that do not have fingerprints superimposed on them are missing here due to the specifics of which device segments were chosen when switching the MUX state, and were not yet attempted at the time this experiment ended.

\subsection{Exchange calibration uniformity}\label{exchange-calibration-uniformity}

The spread of calibration times necessary for all exchange axes utilized in the distance-5 repetition code are shown in \figref{figS27}. 
There we plot the mapping of rotation angle (in radians) to calibrate voltage pulse width for each calibrated axis. 
Two mappings stand out from the more clustered group of calibrations -- YB7 and XA9. 
Both of these exchange axes utilize gate electrode PB7, whose voltage impulse response suffered from non-idealities of the daughterboard in the system. 
This subsequently motivated utilizing a lower exchange rate to partially mitigate the effect, reflected in their calibrations. 
The rest of the curves lie in a relatively tight distribution.

\subsection{Yield of devices}\label{yield-of-devices}

To capture the manufacturability of the 54-dot qubit chips, we present in \figref{figS28} the yield statistics of all such devices from a recent fabrication lot. 
The figure shows the distribution of yield of individual lines on the qubit chip leading up to the gate electrodes (top) and the corresponding number of anticipated operable qubits expected assuming a rectilinear layout of qubits on the device (bottom).  
This lot exhibits higher yield rates than the lots that sourced the devices reported in the main text, but is typical of recent runs.

\subsection{Valley splitting}\label{valley-splitting}

\begin{figure*}
\centering
\smartincludegraphics[width=165mm]{./media/fig_s29}
\caption{
	\textbf{DAPS measurement} \textbf{of valley splitting and orbital energies}. 
	\textbf{a,d}, DAPS differential current measurement vs. detuning energy and neighboring exchange gate bias. 
	Varying the neighboring exchange gate bias changes the orbital confinement. 
	It is therefore expected that the orbital energy will change more strongly than the VS, which allows us to identify each. 
	\textbf{b,e}, 1-D line cuts of a,d along the dashed line at zero exchange gate bias.
	\textbf{c,f}, Relative peak position along detuning tracked vs. the y-axis of a,d. 
	Items a-c show an example measurement where the ground orbital excited valley peak (G-V) is easily distinguishable and lower in energy from the first excited orbital ground valley peak (G-O). 
	Items c-f show an example where the valley splitting is larger in energy than the orbital energy, identified as such by DAPS peak response to orbital confinement changes.
}
\label{figS29}
\end{figure*}

\begin{figure*}
\centering
\includegraphics[width=165mm]{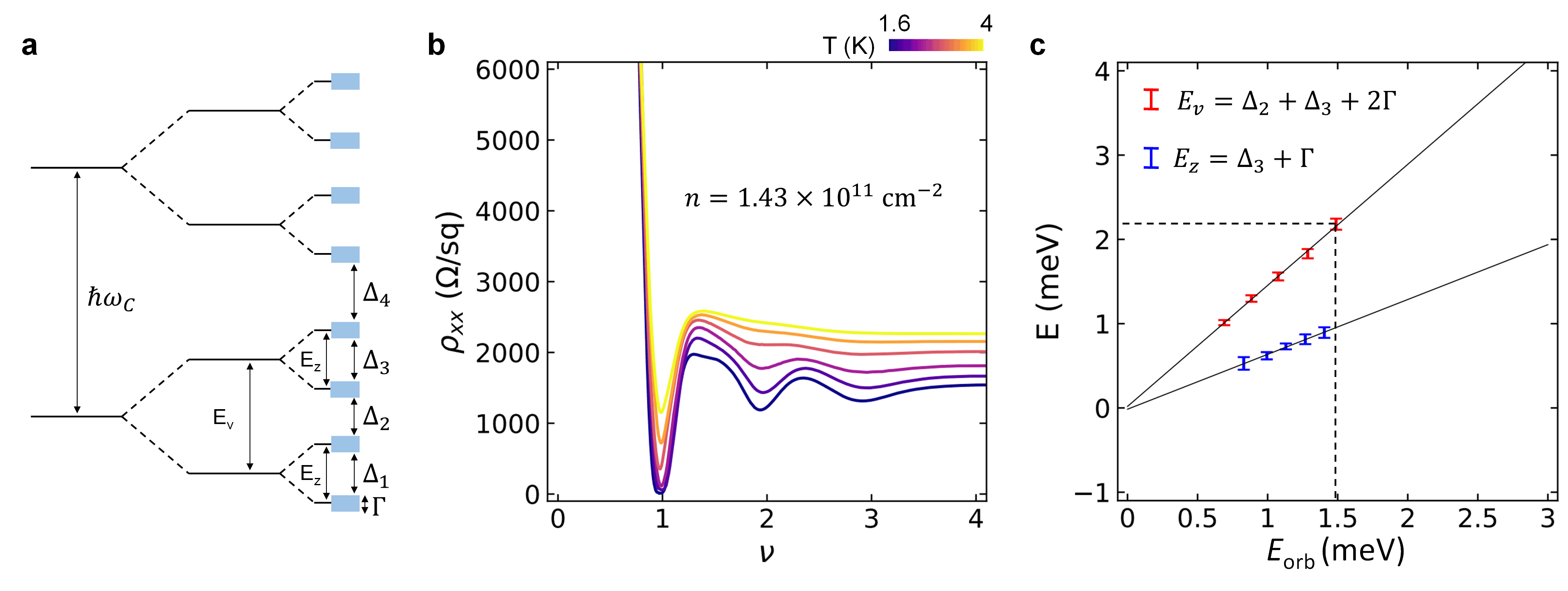} 
\caption{
	\textbf{Valley splitting estimation via the quantum Hall effect}. 
	\textbf{a}, Landau level energy ladder in the quantum Hall regime with large valley splitting, including disorder broadening \(\Gamma\). 
	Landau, valley, and spin levels are split by the cyclotron energy \(\hslash\omega_{c}\), valley splitting \(E_{v}\), and Zeeman splitting \(g\mu_{B}B\), respectively. 
	\textbf{b}, Longitudinal resistivity (\(\rho_{xx}\)) as a function of integer filling factor (\(\nu = nh/eB)\) and temperature (T). 
	\textbf{c}, Valley splitting energy (\(E_{v} = \Delta_{2} + \Delta_{3} + 2\Gamma\)) as a function of orbital energy (\(\ts{E}{orb} = \hslash^{2}/2l_{B}^{2}m^{*}\)).
}
\label{figS30}
\end{figure*}

Large and uniform valley splitting (VS) is a crucial ingredient for any scalable Si quantum dot technology. 
Because valley mixing is very sensitive to the local microscopic structure of the quantum well, VS measurements generally show large dot-to-dot variability (over distances spanning a few hundred nm to microns) \cite{s_eposti_vs_2024, s_marcks_valley_2025}. 
Our devices use an internally-conceived and optimized quantum well design that consistently yields meV-scale valley splitting energies across entire wafers.
This energy range is significantly higher than previously demonstrated for Si/SiGe quantum wells (where typical VS span tens to a few hundred~$\mu$eV) and is comparable to or larger than single-electron orbital excitations, which are usually around 1~to~2~meV under conventional device biasing.

At these large splittings, we use a few different techniques to probe and validate our process. 
Detuning axis pulsed spectroscopy \cite{s_chen_detuning_2021} is an effective method to extract valley and orbital energies for quantum dots. 
In \figref{figS29}, we show representative examples of DAPS measurements taken from devices using our process. 
To assign orbital and valley states, we track excited states as we change neighboring gate biases to alter the lateral confinement. 
Energies that change significantly with bias are assumed to be orbital-like, whereas the bias-invariant states are assumed to be valley. 
In the instances shown in \figref{figS29}, we use this method to extract valley splittings of 1.04~meV and 1.81~meV. 
Even if these assignments are incorrect, the lowest excited states visible in our measurements are greater than 1~meV, setting a lower bound on the valley splitting in these devices.

Nonetheless, it is often difficult to unambiguously differentiate between valley and orbital states using dot-level spectroscopy. 
As supporting evidence, we can also infer valley splitting from transport measurements of Hall bars in the quantum Hall regime. 
By measuring the dependence of longitudinal resistivity at low filling factors \(\nu\) as a function of temperature, the mobility gaps of the Zeeman, valley, and Landau levels can be extracted and identified based on their magnetic field dependences (\figref{figS30}a). 
As illustrated in \figref{figS30}c, we identify the valley splitting of about 2~meV with the summed mobility gap in the form \(E_{v} = \Delta_{2} + \Delta_{3} + 2\Gamma\) \cite{s_wuetz_vs_2020, s_stehouwer_engineering_2025}.
(The higher Zeeman split-level \(\Delta_{3}\) was utilized to extract \(E_{z}\) in lieu of \(\Delta_{1}\) due to magnetic-field constraints.) 
The large energy splittings in our structures make it possible to perform these measurements above 1.6~K as both \(E_{z}\) and \(E_{v}\) are well above the Landau level broadening (\(\Gamma \approx\) 0.36~meV) at relevant Landau orbital energy values (\(E_{orb} \approx\) 1.5~meV). 
Further evidence that valley splitting does not limit device operation comes from the qubit initialization and measurement operating windows (\secref{initialization-window-widths}). 
Typically, we find that the singlet-triplet splittings that define these windows can be large (up to several hundred $\mu$eV) and change predictably with neighboring gate biases in agreement with numerical full configuration interaction (FCI) simulations, indicating we are limited by two-electron orbital rather than valley excitations.

\subsection{Blind randomized benchmarking (BRB)}\label{blind-randomized-benchmarking-brb}

To characterize the performance of our encoded qubit operations, we relied on blind randomized benchmarking (BRB), as presented in Ref.~\cite{s_andrews_quantifying_2019} but generalized to multiple qubits. 
Like traditional randomized benchmarking (RB), BRB homogenizes many possible error channels into a simple metric of error-per-operation that can be fit to an exponential decay without knowledge of the underlying noise model.
The key difference is that BRB does so in a leakage-aware way, so instead of one error metric it produces two: a total error rate \(\varepsilon\) and a leakage rate \(\Gamma\).

\begin{figure}[ht]
\centering
\smartincludegraphics[width=\columnwidth]{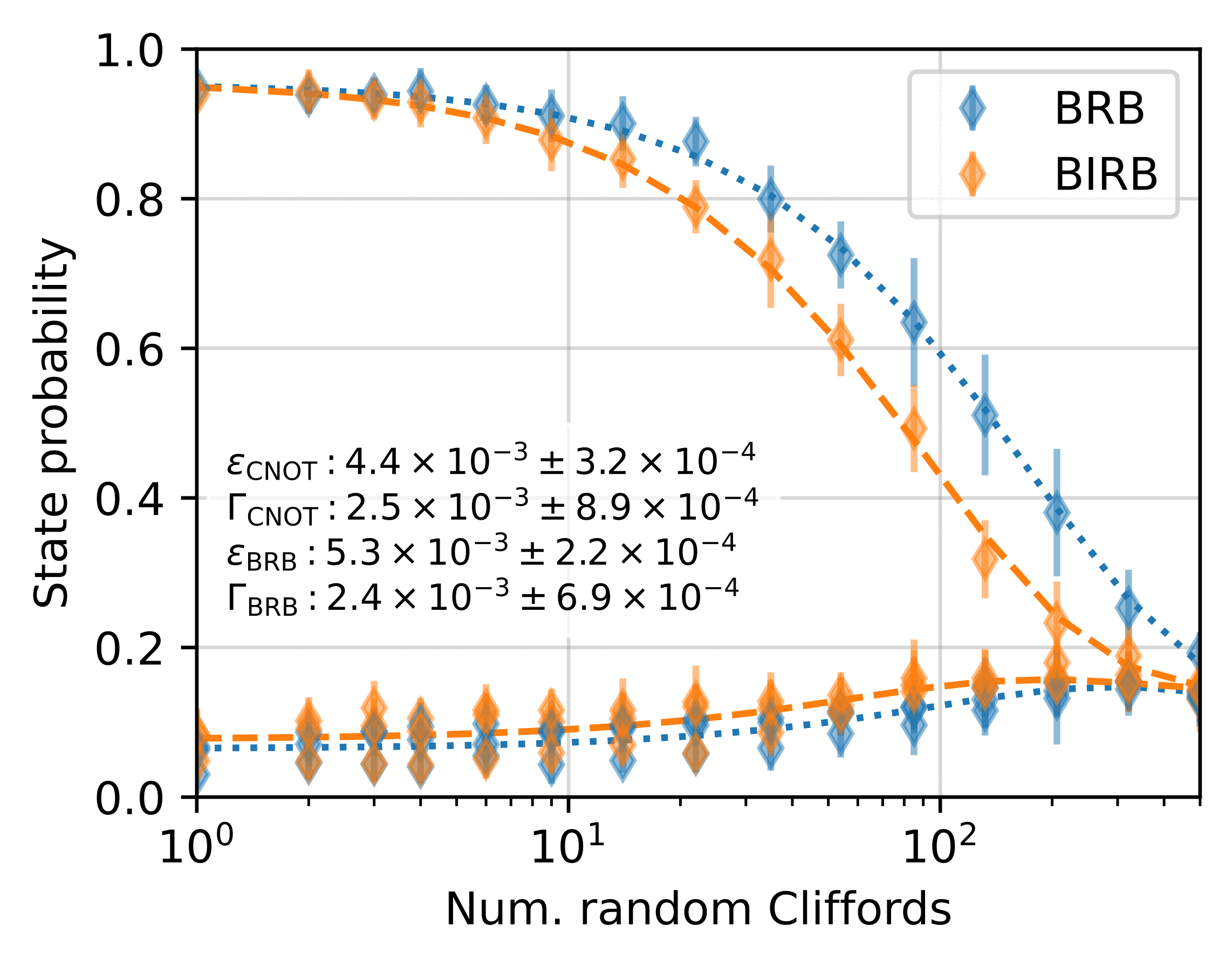} 
\caption{
	\textbf{Interleaved BRB characterizing CNOT error on a pair of qubits used in the {[}{[}4,2,2{]}{]} experiment.} 
	Dark blue (orange) circles show average singlet probability for the two branches of BRB (BIRB).
	Dashed lines show the best fit to equations \ref{eq:birb_fit1} and \ref{eq:birb_fit2}.
	Dotted lines show the equivalent fit for non-interleaved ``reference'' BRB.
}
\label{figS31}
\end{figure}

\begin{figure}[h]
\centering
\smartincludegraphics[width=\columnwidth]{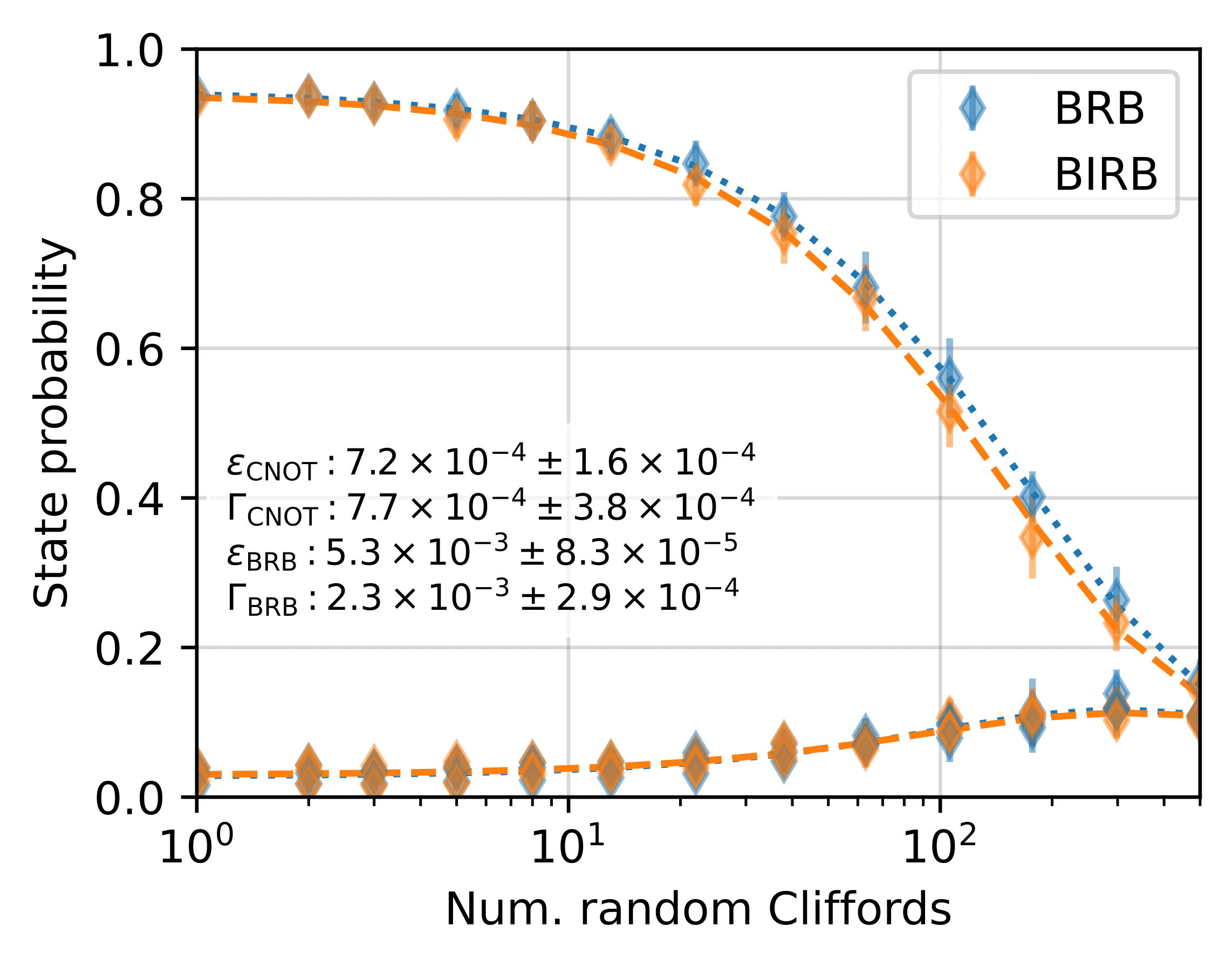} 
\caption{
	\textbf{Lowest-error BRB data to date.} 
	Note that with \(\ts\varepsilon{CNOT} < 0.2\ts\varepsilon{BRB}\), this CNOT error is not quantitatively reliable; we can also see that interleaved BIRB is failing because \(\ts\Gamma{CNOT} > \ts\varepsilon{CNOT}\), which is not physical. 
	Still, we can draw two significant conclusions. 
	First, CNOT error is substantially smaller than the BRB error of approximately 5\e{-3}. 
	Second, this behavior is consistent with pulse-to-pulse crosstalk being a dominant error source, since
	random Cliffords irregularly mixing one- and two-qubit operations. 
	Interleaving high-fidelity CNOT operations may increase the regularity of the sequence as a whole, offsetting some of the error from the CNOT itself.
}
\label{figS32}
\end{figure}

As with traditional RB, BRB admits an ``interleaved'' variant (BIRB), where an operation of interest is interleaved into the BRB sequence to characterize its performance specifically.  
We used BIRB to characterize CNOT, XCNOT, and RIL performance on connected qubit pairs as relevant to the experiment. 
Figure \ref{figS31} shows one such experiment, measuring CNOT error on a representative qubit pair. 
We also show in \figref{figS32} an example of an exceptionally low-error CNOT, where \(\ts\varepsilon{CNOT}\) is so small relative to \(\ts\varepsilon{BRB}\) that its uncertainty is greater than its estimated value.

\subsubsection{BRB implementation and analysis}\label{brb-implementation-and-analysis}

The \emph{N}-qubit BRB experiment proceeds as follows:

\begin{enumerate}
\def\labelenumi{\arabic{enumi}.}
\item
\emph{Initialize the system in}
\(\left. 
\ |0 \right\rangle^{\otimes N}\)
\item
\emph{Apply $n$ random 2-qubit Clifford operations,}
\(C_{1},\ C_{2},\ \ldots C_{n}\)
\item
\emph{Choose a target Clifford} \(\ts{C}{tot}\) \emph{from among all
	combinations of single-qubit identity or Pauli-$X$ operations. 
	E.g. 
	a
	one-qubit} \(\ts{C}{tot}\) \emph{could be either $I$ or $X$, while a
	two-qubit} \(\ts{C}{tot}\) \emph{could be any of $II, IX, XI,$ or $XX$.}
\item
\emph{Apply an inverting Clifford} \(\ts{C}{inv}\)\emph{, chosen such
	that} \(\ts{C}{inv} \cdot C_{n}\ldots \cdot C_{2} \cdot C_{1} = \ts{C}{tot}\)
\item
\emph{Measure the final singlet population for each qubit}
\item
\emph{Repeat across many different choices of random Clifford
	realization} \(C_{i}\)\emph{, overall Clifford} \(\ts{C}{tot}\)\emph{, and
	Clifford count $n$}.
	\end{enumerate}
	With no error, \(\ts{P}{singlet} = 1\) for the \(\ts{C}{tot} = I^{\otimes N}\) branch, and 0 for all other branches. 
	In the limit of large encoded-space error, the system approaches a fully depolarized state, and \(\ts{P}{singlet} \rightarrow \frac{1}{d}\) for all branches, where \(d = 2^{N}\) is the dimensionality of the qubit space. 
	By contrast, any leaked DFS qubit measures as a triplet, so in the limit of large \emph{leakage} error \(\ts{P}{singlet} \rightarrow 0\).
	
	To analyze BRB data, we assume that the system undergoes two exponential decays as a function of \emph{n}: one within the encoded space, and another into a larger, partially leaked space. 
	This is only strictly accurate for a single qubit with a sufficiently strong external B field, but numeric and empirical evidence suggests that it is a good approximation for a wide range of one- and two-qubit error models. 
	Under this approximation and neglecting SPAM errors, the probability \(P_{I}\) of measuring all singlets in the \(I^{\otimes N}\) branch is
	\begin{equation}
P_{I} = \frac{1 - L}{d} + \frac{d - 1}{d}(1 - p)^{n} + \frac{L}{d}(1 - q)^{n},
\end{equation}
where \emph{p} and \emph{q} are the encoded and leakage decay rates, respectively, and \emph{L} is the asymptotic leakage probability. 
The remaining \(d - 1\) branches---collectively termed the ``X-type'' branches---behave very similarly to each other, starting at \(\ts{P}{singlet} = 0\) and decaying up to some asymptote \(\leq {1}/{d}\). 
It is convenient to consider the mean singlet probability \(P_{X}\) of all such branches:
\begin{equation}
P_{X} = \frac{1 - L}{d} - \frac{1}{d}(1 - p)^{n} + \frac{L}{d}(1 - q)^{n},
\end{equation}
The sum and difference of these terms can be fit to single-exponential ansatz:
\begin{equation}\label{eq:birb_fit1}
P_{I} - P_{X} = 2B(1 - p)^{n}, 
\end{equation}
\begin{equation}\label{eq:birb_fit2}
\frac{1}{d}P_{I} + \frac{d - 1}{d}P_{X} = A + C(1 - q)^{n}.
\end{equation}
Finally, we derive error rates accounting for SPAM error following the algebra in \cite{s_andrews_quantifying_2019}:
\begin{align}
\Gamma &= \frac{C}{2B}q, \\
\varepsilon &= \frac{d - 1}{d}p + \frac{1}{d}\Gamma.
\end{align}

\subsubsection{Interleaved BRB}\label{interleaved-brb}

A BRB experiment on \emph{N} given qubits tells us the average \(\Gamma\) and \(\varepsilon\) over all possible Cliffords on those qubits. 
To characterize a specific Clifford of interest \emph{U}, we modify the BRB experiment to interleave \emph{U} after each random Clifford, making the full sequence \([C_{1},U,C_{2},U,\ \ldots C_{n},U,\ts{C}{inv}]\). 
If errors on \emph{U} are not strongly correlated with those on the random Cliffords, then the errors will add incoherently in the interleaved BIRB experiment, and we can approximate the errors due to \emph{U} alone by subtraction:
\begin{equation}
\varepsilon_{U} \approx \ts\varepsilon{BIRB} - \ts\varepsilon{BRB},
\end{equation}
and likewise for \(\Gamma\).

CNOT and XCNOT are Clifford operations with error rates comparable to the random Clifford error rate, so they are usually well-suited to characterization via interleaved BRB (although \figref{figS32} shows a striking counterexample).
RIL, by contrast, is not a Clifford gate or even an operation confined to the computational space: with general two-qubit inputs (including ancilla), it can produce leaked states. 
However, if the ancilla qubit is guaranteed to be in singlet \(\ket{0}\), RIL acts as an identity on any (unleaked) data qubit state. 
The process of initializing the ancilla and then applying RIL thus forms a non-unitary, leakage-reducing gadget that can be interleaved into \emph{one}-qubit RB, with random Cliffords applied to the data qubit.

Finally, although SWAP is also a critical component of our experiment sequences, its error rate is much lower than the random Clifford error rate, and its periodic structure can lead to destructive interference with random Clifford error channels. 
This means that BRB provides unreliable measurements of SWAP errors. 
While SWAP error is included in our spin-level simulations and therefore in our error budgeting for multiqubit circuits, we do not rely on experimental interleaved BRB measurements of SWAP error for model calibration.

\subsubsection{Random Clifford sampling}\label{random-clifford-sampling}

To efficiently represent long BRB sequences in limited cryo-controller memory, we rely on the on-chip PRNG to sample random Cliffords, as described in \secref{digital-architecture}. 
For one qubit BRB, we supply the controller with unique sequence definitions for all 24 one-qubit Cliffords and randomly choose among them. 
For two qubits, we add definitions for SWAP and an entangler (e.g. CNOT), then construct uniformly random two-qubit Cliffords with the following algorithm:

\begin{enumerate}
\def\labelenumi{\arabic{enumi}.}
\item
\emph{With probability 0.5, apply a SWAP.}
\item
\emph{Apply random one-qubit Cliffords to both qubits.}
\item
\emph{With probability 0.9, apply the entangler.}
\item
\emph{If the entangler was applied in step 3, apply random one-qubit
	Cliffords to both qubits.}
	\end{enumerate}
	
	For some datasets we have applied the entangler with unit probability (rather than 90\%), which results in an overly pessimistic estimate for two-qubit Clifford error.
	
	\section{Simulation}\label{simulation}
	
	\begin{figure*}[ht]
\centering
\smartincludegraphics[width=165mm]{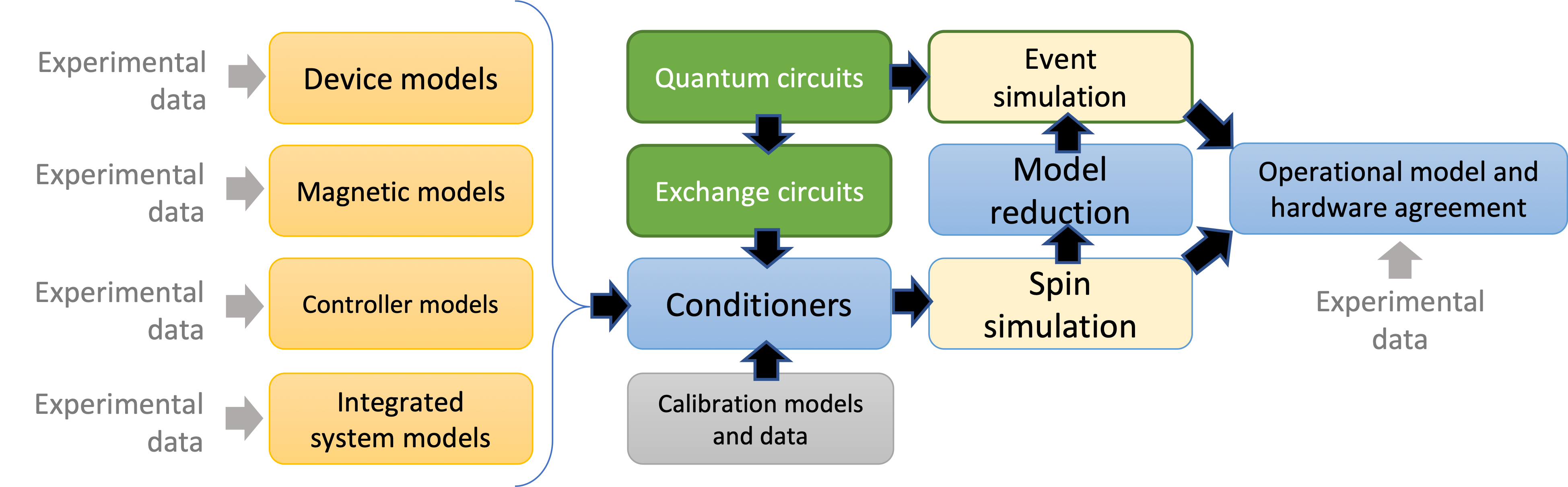}
\caption{
	\textbf{Schematic summary of models and simulations.} 
	Experimental data feeds semiconductor device models, including charge noise and uniformity; models of noisy magnetic fields including dephasing; models of cryogenic controllers including white and flicker noise; and models of the integrated system including interconnect and impacts on signal fidelity. 
	These models are colored orange. 
	The pulse sequences for experiments originate as quantum circuits, and become compiled into exchange circuits (green). 
	Noise models, as well as models and data from calibration, feed into ``conditioners'', which apply those models to exchange-pulse sequences. 
	These feed into coherent spin simulations tracking the effects of noisy exchange sequences and magnetic fields on individual spins.  
	Results of spin simulations then populate reduced models, which track Pauli and leakage error rates at the encoded logic gate level.
	These may be compared with the original quantum circuit in event simulators. 
	Finally, the expected measurement statistics of event simulators, spin simulators, and experimental data are compared to validate the model and indicate need for adjustment at previous steps.
}
\label{figS33}
\end{figure*}

The combination of qubit characterization data and qubit models with system-level simulations is summarized in \figref{figS33}.
This figure depicts a flow by which data informs models of noisy exchange gates and magnetic fields, which ultimately become simulated via the coherent spin simulation methods to be described in \secref{coherent-spin-simulation}. 
Although charge noise and magnetic noise are important parts of the model, the dominant error in the present demonstration arises from signal integrity issues which require integrated system models to capture; we describe these in \secref{interconnect-simulation}. 
These models enable population of errors treated as events at gate-level, including due to contextual miscalibration from signal integrity imperfections; via the process of model reductions, these errors can be included in event simulators as described in \secref{event-simulation}. 
The basis for comparing the result of spin and event simulations with experiment is in detector error models (DEMs), methodologies for which we describe in \secref{detector-error-models-dems}. 
For the present demonstration, operational model and hardware agreement requires an ersatz parameter determination, in which the results of simpler experiments such as randomized benchmarking populate a reduced model for miscalibration, and this parameter is subsequently fed to model multiqubit experiments, a process elaborated in \secref{quasi-static-miscalibration-model}.
The actual results of such modeling and simulation for the experiments described in the main text are elaborated in \secref{multiqubit-experiments}.

\subsection{Coherent spin simulation}\label{coherent-spin-simulation}

Our qubit architecture can be modeled as a collection of spins subjected to magnetic fields and interacting with each other via the exchange interaction. 
This is described by the following Hamiltonian:

\begin{equation}\label{eq:pyquest_hamiltonian}
H(t) = \ \sum_{j}^{}{\gamma_{j}{\vec{B}_{j}}(t) \cdot {\vec{S}}_{j}} + \sum_{\left\langle j,k \right\rangle}^{}{J_{jk}(t){\vec{S}}_{j} \cdot {\vec{S}}_{k}}.
\end{equation}
Here, \(\gamma_{j}\) is the gyromagnetic ratio for the electron in dot \(j\), \({\vec{S}}_{j}\, = \,{\sigma}_{j}\text{/}2\) the spin operator for the electron in dot \(i\), and \(\left\langle j,k \right\rangle\) denotes nearest neighbors. 
The magnetic fields \({\vec{B}}_{j}(t)\) for each dot are a sum of the global applied magnetic field, any static gradients present in the device, along with fluctuating effective fields of magnitudes consistent with those originating from hyperfine interactions with stray spinful nuclei \cite{s_kerckhoff_magnetic_2021}. 
The time-dependent exchange couplings \(J_{jk}(t)\) encode both the applied pulses (which we approximate as square pulses), as well as fluctuations due to charge noise.

We solve the time-dependent Schrödinger equation under the Hamiltonian in Eq. \ref{eq:pyquest_hamiltonian} using a custom software package called the Python Quantum Exchange Simulator Tool Suite (pyquest).
Simulations of our qubit architecture are run in pyquest following these steps:

\begin{enumerate}
\def\labelenumi{\arabic{enumi}.}
\item
The user passes a \emph{resource topology}, which provides the device graph, whose nodes are the spins and whose edges are the possible exchange couplings. 
The topology also may or may not contain an \emph{encoding}, which describes how spins are grouped to form DFS qubits. 
We use the encoding for GAQQMap (\secref{gaqqmap}) to track spin labels belong to DFS, but in general circuit simulation we only rely on the collection of spin and their connectivity.
\item
The user passes a \emph{flatseq} object, which is a data structure that contains the sequence of pulse operations and state preparation and measurement (SPAM), operations to apply to the spins.

\begin{enumerate}
	\def\labelenumii{\alph{enumii}.}
	\item
	This is typically encoded as a matrix, where each row is a timestep, each column a particular exchange axis, and each entry either zero or the desired pulse angle.
	\item
	A non-zero pulse exchangle \(\theta\) for an axis between \(i\) and \(j\) gives the time-integral of the exchange coupling \(J_{ij}(t)\) over the duration of the pulse. 
	Explicitly, we have: \emph{\hfill\break}\[\theta = \int_{t_{1}}^{t_{2}}{J_{ij}\left( t' \right) dt' }.\]
	The start and end times \(t_{2}\) and \(t_{1}\) for each time step are inferred from user-supplied data.
	
	\item
	State preparation (i.e., flushing to a singlet) and measurement are indicated by an additional list of flags specifying at which timesteps the operation(s) occur, and on which exchange axis.
\end{enumerate}

\item
The user specifies noise parameters for the system, such as \(\ts{N}{osc}\) for each axis or \(T_{2}^{*}\) for each dot, the (static) magnetic field experienced by each spin, and static pulse miscalibration.  Note that in 
pyquest, \(T_{2}^{*}\) is a property of each \emph{dot}, rather than each \emph{exchange axis}. 
For a DFS qubit with three spins
\(\left\{ z,t,n \right\}\), with pairwise \(T_{2}^{*}\) values
\(\left\{ T_{2}^{*}(z,t),T_{2}^{*}\ (z,n),T_{2}^{*}\ (t,n)\  \right\}\),
the effective \(T_{2}^{*}\) inputted to simulation for all three spins is
\[T_{2}^{*}\ \  = \ \left\lbrack \frac{1}{3}\left( \frac{1}{{T_{2}^{*}\ (z,t)}^{2}} + \frac{1}{{T_{2}^{*}\ (z,n)}^{2}} + \frac{1}{{T_{2}^{*}\ (t,n)}^{2}} \right) \right\rbrack^{- 1/2}.\]
\item
Pyquest then generates a \emph{noise trajectory} (i.e., a particular sampling of fluctuating magnetic fields or fluctuating exchange couplings) and simulates application of the sequence with the sampled noise trajectory.
This process can be repeated for a desired number of shots, with pyquest automatically re-sampling noise for each shot.
\end{enumerate}

Fluctuating magnetic fields at each dot, along with fluctuating exchange couplings, are generated via the Voss algorithm \cite{s_voss_efficient_2022}. 
When dealing with time-dependent $({\vec{B}}_{j})$ or \(J_{jk}\), we Trotterize the applied operators; e.g., for a single-spin magnetic field term we do:
\begin{equation}\begin{split}
\mathcal{T}\left\{ \exp\left( - \frac{i\gamma_{j}}{\hslash}\int_{t_{1}}^{t_{2}}{{\vec{B}}_{j}\left( t' \right)} \cdot {\vec{S}}_{j}\,\text{d}t' \right) \right\}\ \  \mapsto \\ \ \ \prod_{n = 0}^{N}{\exp\left\lbrack - \frac{i\gamma_{j}}{\hslash}{\vec{B}}_{j}\left( t_{1} + n\frac{t_{2} - t_{1}}{N} \right) \cdot {\vec{S}}_{j}\frac{t_{2} - t_{1}}{N} \right\rbrack},
\end{split}\end{equation}
where \(\mathcal{T}\) is the time-ordering operator, \(N = \left\lceil \left( t_{2} - t_{1} \right)\text{/}\tau \right\rceil\), and \(\tau\) is the splitting time. 
Exchange, i.e. \(\ts{N}{osc}\), noise is Trotterized similarly. 
For the simulations presented in this manuscript, we take the splitting time to be \(\tau = \ T_{2}^{*}\ \text{/}20\), where \(T_{2}^{*}\) is the \emph{shortest} \(T_{2}^{*}\) across all spins involved in the operator being Trotterized.

Pyquest has two different backends for performing simulations: one using dense state vector methods, and the other using tensor network methods (more specifically, matrix product state methods). 
We provide high-level overviews of these simulation backends below.

\subsubsection{Spin state vector (SSV)}\label{spin-state-vector-ssv}

The spin state vector (SSV) backend stores the state of the \(N\)-spin system in a dense state vector with \(2^{N}\) entries. 
State vectors in the computational basis are mapped on to indices in the state vector in the standard way: namely, the state
\(\left| \sigma_{0}\sigma_{1}\ldots\sigma_{N - 1} \right\rangle\), with \(\sigma_{i} \in \text{\{}0,1\text{\}}\), corresponds to index \(\sum_{j = 0}^{N - 1}{\sigma_{N - 1 - j}2^{j}}\) in the state vector (see, e.g., Ref.~\cite{s_lin_exact_1993} for details).

Single-spin magnetic field terms on dot $j$, $\vec{B}_{j} \cdot \sigma_{j}$, and exchange terms $\vec{S}_{j} \cdot \vec{S}_{k}$, are constructed respectively as $2 \times 2$ and $4 \times 4$ matrices in Pauli basis.
%
%
Single-spin terms are exponentiated directly using \(\exp\left( i\vec{B} \cdot \sigma \right) = \cos\left( \left| \vec{B} \right| \right)I + i\sin\left( \left| \vec{B} \right| \right)\vec{B} \cdot \sigma \text{/}\left| \vec{B} \right|\), where \(\vec{B}\) includes the stepwise integrated field(s) (after Trotterizing as described above).
Matrices for multi-spin terms (e.g. exchange with fields or multiple overlapping exchanges) are either directly diagonalized and exponentiated (if involving three spins or fewer) or Trotterized to consecutive single or two-site terms (if involving more than three spins). 
These operators are then applied to the state vector by iterating over all possible configurations of the spins not involved in the operation.

To implement a SPAM operation on a given spin pair, we first construct the projectors onto the four possible angular momentum states \(\left\{ S,T_{0},T_{+},T_{-} \right\}\) for the pair in question. 
In the computational basis, these read:
\begin{align}
	P_{S} &= \frac{1}{2}\ \begin{pmatrix}
		0 & 0 & 0 & 0 \\
		0 & 1 & - 1 & 0 \\
		0 & - 1 & 1 & 0 \\
		0 & 0 & 0 & 0
	\end{pmatrix},&
	P_{T_{+}} &= \ \begin{pmatrix}
		1 & 0 & 0 & 0 \\
		0 & 0 & 0 & 0 \\
		0 & 0 & 0 & 0 \\
		0 & 0 & 0 & 0
	\end{pmatrix},&
	\notag\\
	P_{T_{0}} &= \frac{1}{2}\ \begin{pmatrix}
		0 & 0 & 0 & 0 \\
		0 & 1 & 1 & 0 \\
		0 & 1 & 1 & 0 \\
		0 & 0 & 0 & 0
	\end{pmatrix},&
	\ P_{T_{-}} &= \ \begin{pmatrix}
		0 & 0 & 0 & 0 \\
		0 & 0 & 0 & 0 \\
		0 & 0 & 0 & 0 \\
		0 & 0 & 0 & 1
	\end{pmatrix}.
\end{align}
Next, the probabilities \(p_{i} = \ \left\langle \psi \middle| P_{i} \middle| \psi \right\rangle\), for \(i \in \left\{ S,T_{0},T_{+},T_{-} \right\}\), of each angular momentum state are computed. 
Then one outcome is sampled (weighted by the probabilities \(p_{i}\)) and the appropriate projection operator is applied. 
For measurement, we apply either \(P_{S}\) (if the sampled result is \(S\)) or \(P_{T_{0}} + P_{T_{+}} + P_{T_{+}}\) (if the result is one of the triplet states) to the state and re-normalize. 
In either case, the measurement result is recorded.  
For initialization, we project to the same state that was sampled, and then apply a correction:

\begin{itemize}
	\item
	For \(T_{0}\), we apply a rotation by \(\pi\) about the \(z\) axis, \(\exp\left( - i\pi\sigma_{z}\text{/}2 \right)\), to the first spin.
	\item
	For \(T_{+}\) and \(T_{-}\), we apply a rotation by \(\pi\) about the \(x\) axis, \(\exp\left( - i\pi\sigma_{x}\text{/}2 \right)\), to the first spin, and then apply the operator \(P_{S}\) and re-normalize.
\end{itemize}

In either case, the result of these corrections is that the system is in the singlet (\(S\)) state; the purpose of doing initialization in this manner is to correctly simulate the back-action of the initialization on the rest of the system. 
For initialization, we can also simulate errors via an affine transformation of the probabilities
\begin{equation}
	\begin{pmatrix}
		p_{S} \\
		p_{T_{0}} \\
		p_{T_{+}} \\
		p_{T_{-}}
	\end{pmatrix} \mapsto A\begin{pmatrix}
		p_{S} \\
		p_{T_{0}} \\
		p_{T_{+}} \\
		p_{T_{-}}
	\end{pmatrix} + \,\left| B \right\rangle,
\end{equation}
with \(A\) a \(4 \times 4\) matrix and \(B\) a 4-element vector parameterizing the initialization error. 
The outputted state can then be sampled from these new probabilities, and constructed by reversing the corrections performed above. 
Measurement errors can be simulated after the fact by probabilistically altering the recorded measurement results.

The advantage of SSV is that, when there are no overlapping exchange operators applied at the same time, all the operators applied to the state are constant size (\(2 \times 2\) or \(4 \times 4\)). 
Applying gates or SPAM operations to the state is then \(\mathcal{O}\left( 2^{N} \right)\) in both space and time, rather than \(\mathcal{O}\left( 4^{N} \right)\) when constructing the full matrices.
The looping over configurations of uninvolved spins can also be done very efficiently using bit-twiddling techniques and pointer arithmetic.
Consequently, the SSV backend has very good performance for small (\(\lesssim 7\) DFS qubits, i.e. \(< 21\) spins) systems. 
This is particularly advantageous for simulating one and two qubit RB, or single and two qubit gates.

\begin{figure*}[ht]
	\centering
	\smartincludegraphics[width=125mm]{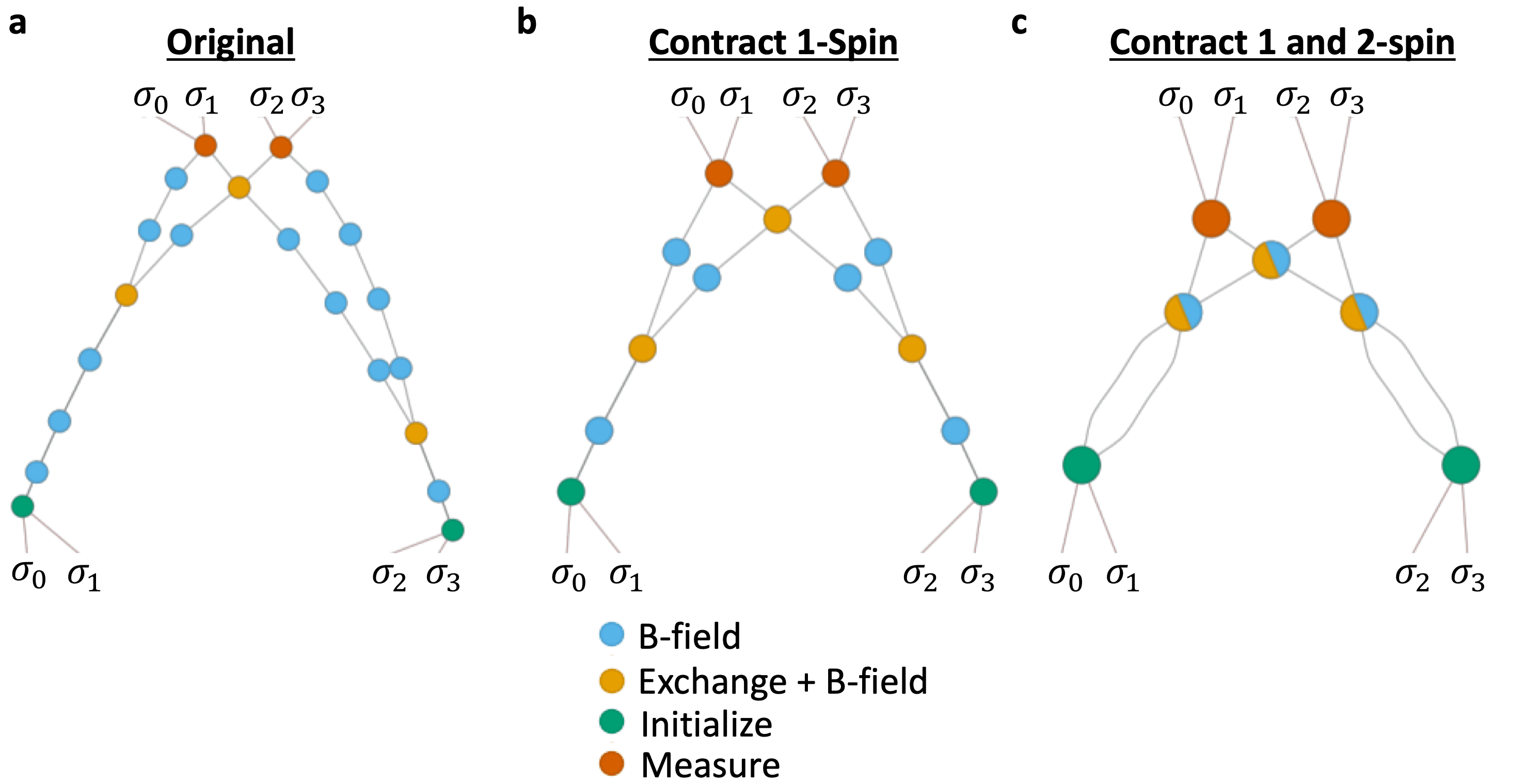}
	\caption{
		\textbf{Example of contracting the tensor network representing evolution under a given $\mathbf{H(t)}$.} 
		In this case, we have a four-spin system, where two pairs of spins are initialized into singlet pairs, then exchange operations are performed on three pairs of spins, and then singlet/triplet measurements are performed. 
		Between all these operations, the system idles in the presence of a magnetic field.
		\textbf{a,} Initial tensor network, with every single and two-spin operator represented by a single node. 
		\textbf{b,} Tensor network after contracting all single-spin (i.e., magnetic field evolution) terms together. 
		\textbf{c,} Tensor network after contracting all single-spin operators into nearby two-spin operators. 
		Note the dramatic reduction in the number of nodes in the network; this contraction scheme represents a significant savings in memory compared to the initial network. 
		At this point, one could also contract two-spin operators together as well, to reduce the number of SVDs required; this could be potentially useful for networks that have a deep network of nodes acting an isolated subset of the spins (as is the case, e.g., when performing a multi-qubit gate on a subset of spins in a large system).
	}
	\label{figS34}
\end{figure*}

\subsubsection{Matrix product state (MPS)}\label{matrix-product-state-mps}

The matrix product state (MPS) backend stores the state of the \(N\)-spin system as an MPS with a maximal bond dimension \(\chi_{\max}\). 
Each spin can be assigned to a single MPS site, or multiple spins can be grouped onto a single MPS site. 
Operators for single and multi-spin terms are constructed analogously as in SSV and applied to the \emph{physical} indices of the MPS. 
Singular value decompositions (SVDs) are then performed as needed, along with truncation of the bond dimension. 
SPAM operations are implemented analogously as in SSV.

We use a modified version of the open-source quimb package \cite{s_gray_quimb_2018} which implements methods for tensor network operations (including contractions). 
Importantly, this package represents the evolution under \(H(t)\) for the whole experiment as a tensor network, which allows the software to efficiently contract certain types of operations \cite{s_gray_hyper_2021}. 
For example, consecutive operations occurring on the same MPS site (e.g., consecutive single-spin terms on the same spin, or exchanges acting on a pair of spins that are grouped in the same MPS site), and even operations occurring on neighboring MPS sites, can be contracted into a single operator.  See \figref{figS34} for an example.
This latter case is important, as fusing together multiple operations between the same pair(s) of MPS sites can reduce the number of SVDs required and improve performance.

Our MPS simulation backend can also be run on one or more GPUs via a user-passed argument. 
This is achieved by using the autoray \cite{s_gray_autoray_2026} library to dispatch tensor operations to the appropriate backend. 
To run tensor operations on GPUs, we use cupy \cite{s_okuta_cupy_2017} as the backend. 
For large systems and/or deep networks, this allows for significant improvement in simulation wall time; indeed, the simulations of larger circuits ({[}5,1,5{]} and {[}{[}4,2,2{]}{]}) presented in this manuscript make use of this feature. 
The network contraction described above is a vital feature when running the MPS backend on GPUs, as it ensures the network does not exceed the amount of memory available to the GPU.

Due to the overhead involved with storing the state as an MPS and contracting the tensor network, MPS is generally slower than SSV for systems with \(\lesssim 7\) DFS qubits. 
Above this scale, the memory savings from truncating the bond dimension (\(\chi_{\max}\)) leads to better performance relative to SSV (especially when run on a GPU). 
Our MPS simulator tracks the norms of the MPS after SVD truncation, allowing us to assess the reliability of the MPS approximation at a given \(\chi_{\max}\). 
A crucial consideration when setting up an MPS simulation is how to group the spins into MPS sites: putting more spins onto each MPS site can reduce the number of required SVDs, at the cost of larger MPS matrices (for the same max bond dimension). 
We find empirically that grouping the three spins of each DFS qubit into their own MPS site yields good performance, and seems to optimize (or nearly optimize) this tradeoff. 
We use this grouping for the simulations presented in this manuscript.

\subsection{Interconnect simulation}\label{interconnect-simulation}

In the previous section, we explained how the dot-level magnetic and charge noise parameters extracted from our tune-up and calibration experiments are added to a device simulation. 
However, there are also control errors that can occur in an experiment which are not caused by local noise near the qubits in the device. 
To that end, we have developed a framework to more accurately model how individual sources of control error propagate through our quantum simulation framework via ``conditioners''. 
In general, ``conditioners'' are software objects that consume a sequence of ideal exchange angles and return a modified sequences whose pulses have been altered in some way. 
For this specific use-case, we have designed a conditioner that captures how pulses generated by the controller propagate through the entire signal chain to the dot(s), which is implemented via linear \& time-invariant (LTI) techniques in conjunction with direct time-domain simulations. 
Using this conditioner framework, we capture both voltage-domain pulse errors from realistically-generated pulse shapes and unintended overlap of subsequent pulses (inter-symbol interference or ISI) and cross-talk (XTK) between pulses on separate control lines.

\subsubsection{Interconnect circuit model}\label{interconnect-circuit-model}

\begin{figure*}
	\centering
	\smartincludegraphics[width=165mm]{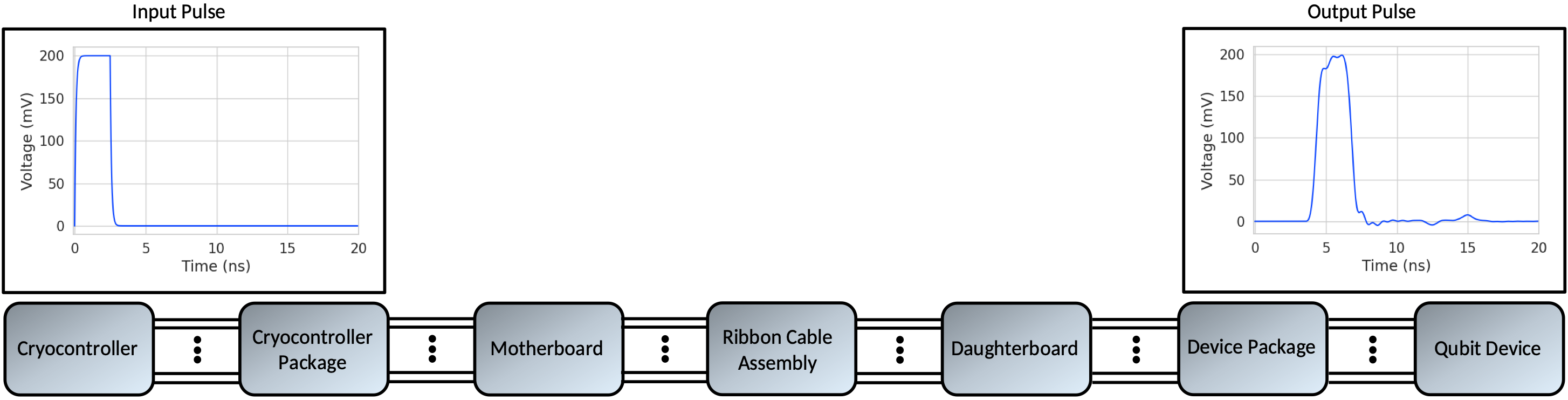}
	\caption{
		\textbf{A high-level block diagram of the end-to-end Simulation-Program-with-Integrated-Circuit-Emphasis (SPICE) circuit model of the system interconnect.}
		Each block in the signal chain represents a	discrete SPICE component model that has been developed to model the specific electrical characteristics of that component. 
		These component	models are obtained by applying a variety computational electromagnetic	techniques and lumped-element equivalent models. 
		A custom-developed python interface allows the users to easily specific the nets of interest, generate SPICE netlists for each component, and chain the components together to simulate the end-to-end integrated system.
	}
	\label{figS35}
\end{figure*}

We model voltage propagation through our entire analog signal chain assembly, consisting of a cryo-controller, various interconnect sub-assemblies and the qubit chip. 
Voltage pulses are simulated in the time-domain using industry-standard SPICE simulation software. 
ISI and XTK, driven by component design as well as unavoidable impedance mismatches, result in complex and unpredictable pulse reflections whose errors can only be fully-accounted via SPICE simulation of the full-scale, end-to-end interconnect of the integrated system. 
Such a circuit model is constructed by chaining together a series of (sub)circuit multiconductor models corresponding to individual components of the assembly (\figref{figS35}). 
These components span several orders of magnitude in physical size and embody a range of electrical complexity that merits the application of a unique modeling approach for each element in the end-to-end model. 
We have developed methods for generating high-fidelity models based on cross-sectional 2D TEM analysis, 3D full-wave electromagnetic simulations as well as lumped element models where appropriate. 
The end-to-end circuit model is used to generate the multi-input multi-output (MIMO) impulse response by performing a series of transient SPICE simulations, which is then convolved into sampled controller waveforms and mapped into the singlet-triplet exchange domain, to add the expected angle rotation errors to the ideal exchange sequence.

\subsubsection{V-J device model}\label{v-j-device-model}

\begin{figure*}
	\centering
	\includegraphics[width=165mm]{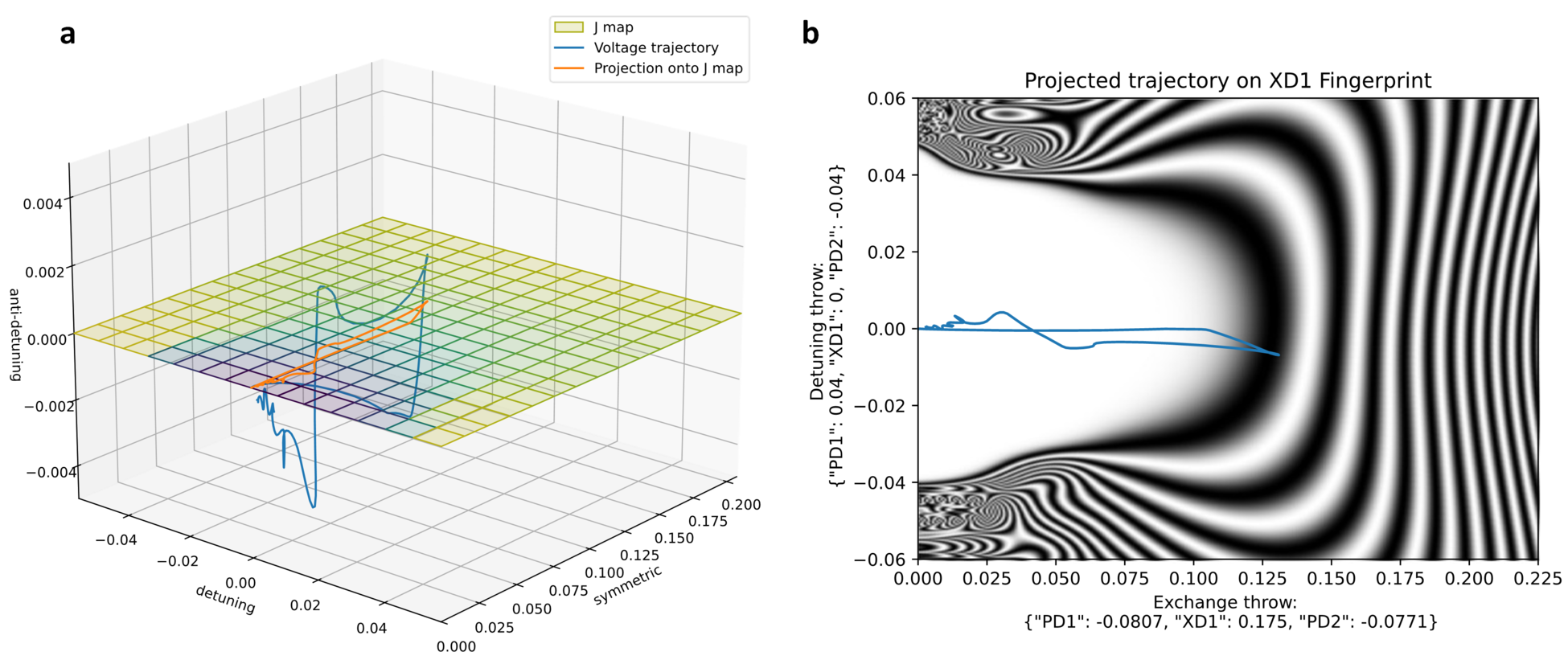} 
	\caption{
		\textbf{Pulse trajectory simulation.}
		\textbf{a,} Projection of 3-dimensional voltage trajectory of a 2.5~ns pulse onto a transformed-basis plane. 
		\textbf{b,} Detuning and symmetric axes are defined by the detuning and exchange throws of the simulated fingerprint, respectively.
	}
	\label{figS36}
\end{figure*}

Unwanted cross-capacitive coupling between metal gates and electrons is a significant source of control error. 
This type of ``front-end-of-line'' (FEOL) error is distinct from the ``back-end-of-line'' sources discussed in the previous section. 
It also will not be observed during single-axis procedures, and therefore is not accounted for in a local noise term like \(\Nosc\). 
To account for FEOL crosstalk, we require a voltage-to-exchange mapping that turns the applied voltage \(V\) on each gate into an effective exchange coupling \(J\) on each pair of dot-confined electrons. 
We generate this V-J map by combining a spatial lever-arm function and individual axis fingerprints from a single simulated device tune-up. 
Time-domain voltage signals obtained via impulse response convolution are projected into a transformed basis (``detuning,'' ``anti-detuning,'' and ``symmetric'' axes) which more accurately captures the operational modes of an axis between two dot-confined spins. 
Then, time-domain exchange energies per axis are estimated using rectangular spline interpolation of transformed-basis points on a log-normalized $J$ map. 
An example of this procedure is shown in \figref{figS36}a. 
In summary, this procedure converts a complicated, time-dependent voltage trajectory on multiple gates into a single number which predicts the total exchange applied to two dot-confined spins.

\subsubsection{Gate error predictions for ISI and crosstalk}\label{gate-error-predictions-for-isi-and-crosstalk}

\begin{figure*}
	\centering
	\includegraphics[width=165mm]{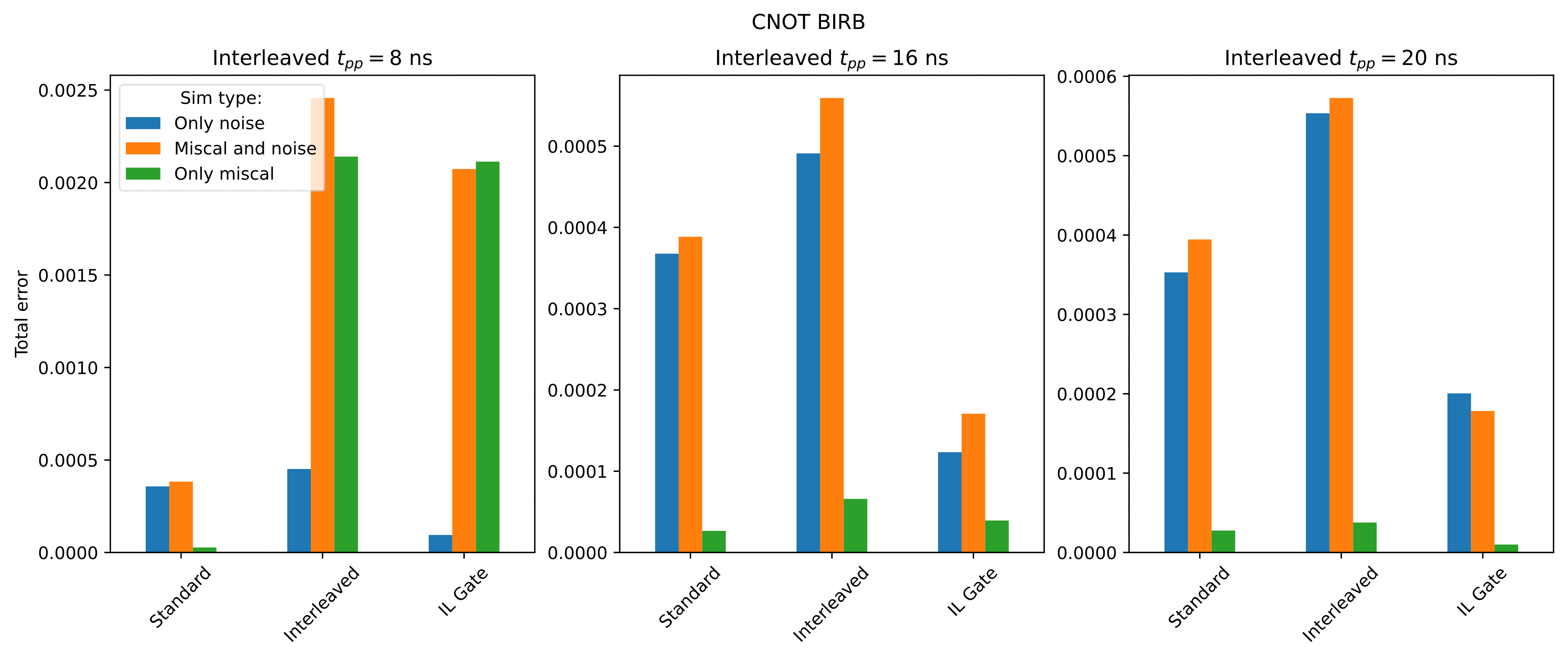} 
	\caption{
		\textbf{Coherent spin-simulations of a two-qubit BIRB sequence with an interleaved CNOT, varying the amount of time between the rise of adjacent pulses (pulse-to-pulse time, or} \(\boldsymbol{t}_{\mathbf{pp}}\)\textbf{).} 
		The three groups of bars represent errors from standard RB error (left), total interleaved RB error (middle), and the inferred interleaved gate error (right). 
		Bar colors represent simulations under an ideal sequence without noise (blue), a sequence with both control error and microscopic noise (orange), and a sequence with control error but without microscopic noise (green). 
		The random Clifford gates used in the BIRB sequence have \(\tpp = \) 20~ns, and single-axis exchanges were calibrated with \(\tpp = \) 24~ns.
	}
	\label{figS37}
\end{figure*}

\begin{figure*}
	\centering
	\includegraphics[width=165mm]{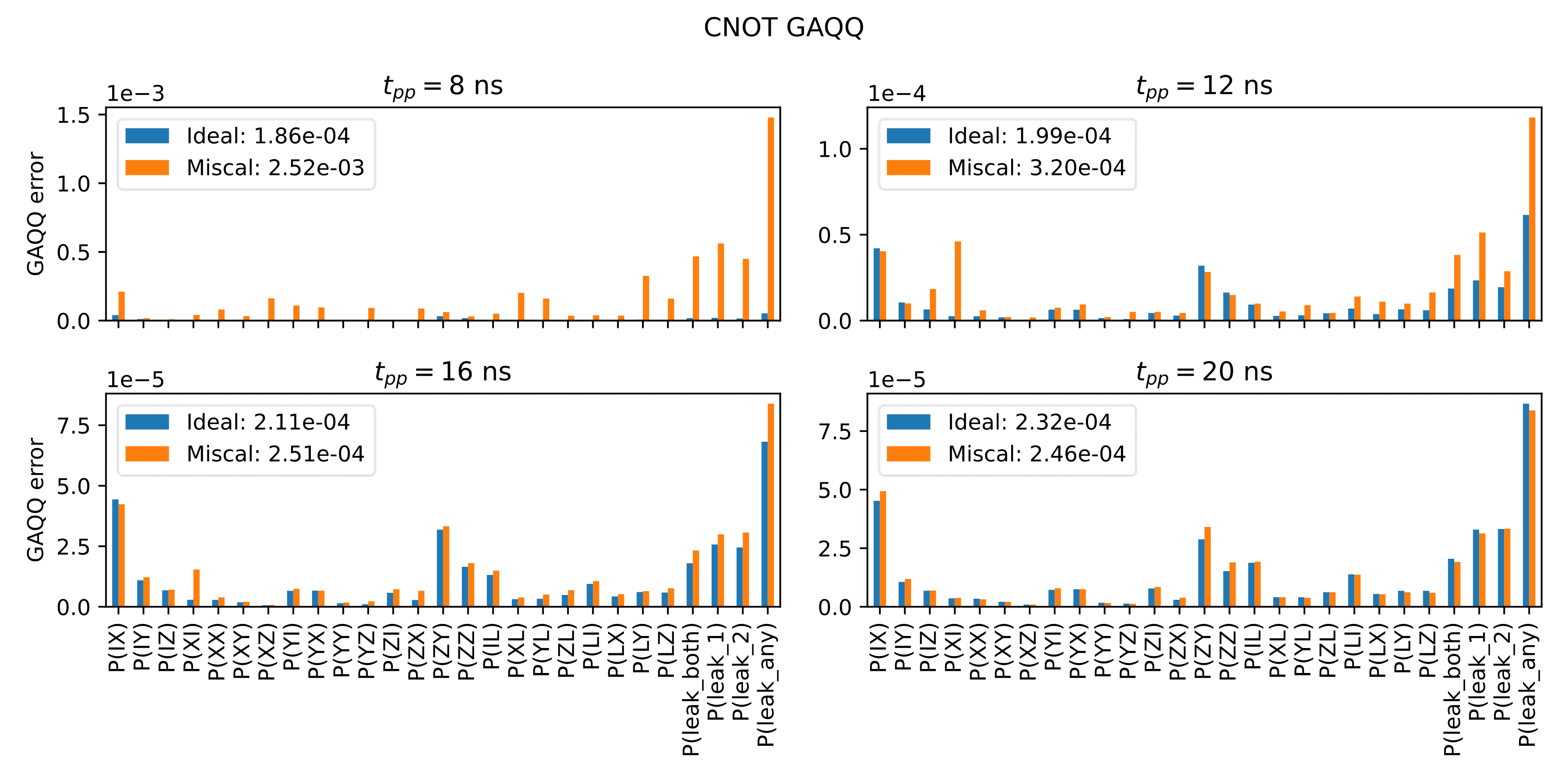} 
	\caption{
		\textbf{Error channels, as computed by GAQQMap (\secref{gaqqmap}) of the CNOT gate errors, with and without control error, at varying pulse cadences} \(\boldsymbol{t}_{\mathbf{pp}}\)\textbf{.} 
		Single-axis exchanges were calibrated with \(\tpp = \) 24~ns, and we used microscopic noise parameters of \(\Nosc = 100\), \({T_{2}}^{*} = \) 50~$\mu$s.
	}
	\label{figS38}
\end{figure*}

Using the interconnect circuit model and the V-J model, we can now make predictions of how the electrostatics acting on the control lines and FEOL affect qubit performance. 
Figure~\ref{figS37} compares a simulation of a two-qubit BIRB sequence with only microscopic noise (``noise''), one with only control error (``miscal''), and one with both (``miscal+noise''). 
In \figref{figS38} we also compare a noise-free (``ideal'') gate to one with control errors (``miscal'') along specific Pauli-error channels, as computed by GAQQmap (see \secref{gaqqmap}). 
By looking at the differences between each model's total gate error, we can isolate the contribution of ISI and XT to CNOT infidelity (see \secref{blind-randomized-benchmarking-brb}). 
It is informative to study how the relative amount of control error changes as one modifies the pulse cadence, or ``time-per-pulse'' \(\tpp\). 
We find that at a fast cadence (\(\tpp = 8\)~ns), the control error is almost five times larger than the microscopic noise sources. 
However, at the cadence used in our multiqubit demonstrations (\(\tpp = 20\)~ns), the control error is almost ten times \emph{smaller} than the microscopic noise source.

The strong dependence of gate error on pulse cadence is not surprising.
At short \(\tpp\) (\textless{}~16~ns), Same-Wire Intersymbol Interference (SWISI) from pulse tails is large, and pulses that are nearby spatially and/or temporally will experience a large integrated angle error. 
As we move toward the standard cadence of \(\tpp = \ \)20~ns, differences in the estimated CNOT error with and without miscal become much smaller, within the uncertainty of Clifford and noise sampling in the simulation. 
This indicates that SWISI and electronic crosstalk are not large contributors to the total error seen in experiments with typical operating parameters (calibration, pulse cadence, voltage operating points, etc.) and with well-behaved sequence structure (no same-axis consecutive pulses, etc.).

It is important to note that these simulations do not include internal cryo-controller nonidealities, and the device model does not account for differences between real device tune ups and those generated by our simulation. 
In a following section (\secref{quasi-static-miscalibration-model}), we use this finding to argue that cryo-controller errors are the likely source for a majority of the observed gate and circuit infidelity.

\subsection{Event simulation}\label{event-simulation}

Coherent simulation of each electron captures many microscopic noise phenomena, but the computational cost of the simulation grows exponentially with the number of spins. 
For this reason, we turn to qubit-level event simulation, which is a common approach for evaluating quantum circuits in the presence of noise-induced errors. 
All sources of microscopic noise are combined into an effective probability of a single encoded qubit error (explained in the GAQQmap procedure in \secref{constructing-qubit-operations}), and the quantum device is simulated as a series of expected (logical) and unexpected (noise) gates.

Our internal software package which handles event simulation of circuits is called Squeaky. 
Squeaky simulates quantum programs acting on DFS qubits made of three spins, and relies on an abstract syntax tree (AST) data structure. 
To more reliably simulate our exchange-only qubit architecture, leakage out of the encoded subspace is tracked by a leakage bit assigned to each qubit. 
When a leakage error occurs, this leakage bit is flipped to true, and the leakage is then propagated through the circuit according to the leakage properties of the applied gates. 
Leakage is only removed in this simulator when a qubit is reset to the encoded \(|0\rangle\) state.

\begin{figure*}
	\centering
	\smartincludegraphics[width=165mm]{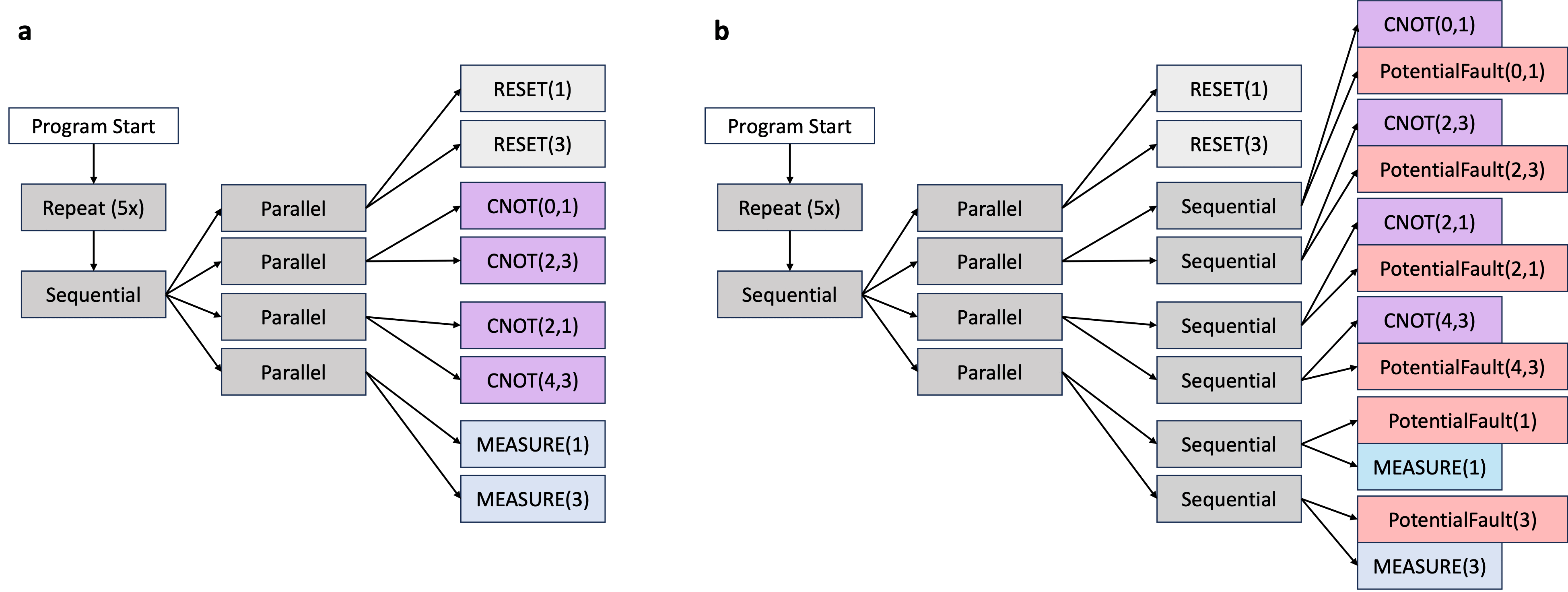}
	\caption{
		\textbf{Simple program example.} 
		\textbf{a,} Example program represented in Squeaky as an AST before errors are inserted. 
		\textbf{b,} The same program as in a but with potential faults inserted due to noise injection.
	}
	\label{figS39}
\end{figure*}

Noise is injected into an ideal circuit via the circuit's AST representation, which enables efficient calculation of when potential faults can occur and what their error distributions are. 
After this analysis, potential faults are inserted into the tree.
Figure \ref{figS39} provides an example of a simple program before and after noise injection. 
After this injection of potential faults, one then samples the circuit many times (usually in parallel, with use of MPI), stochastically replacing potential faults with realized faults before each shot.

We have used two different methods to evaluate the noisy circuits described above. 
The first is a coherent-qubit evaluator, where each qubit is represented by a single spin-1/2 degree of freedom (used for the {[}3,1,3{]} simulations). 
Note that this is distinct from the coherent-spin simulator of \secref{coherent-spin-simulation}, where each qubit is represented by \emph{three} spin-1/2 electrons. 
The other evaluator is a Pauli frame tracker, with an additional bit attached to each qubit to indicate whether it has leaked or not (used for the {[}{[}4,2,2{]}{]} simulations).

\subsection{Detector error models (DEMs)}\label{detector-error-models-dems}

Detector error models (DEM) can be estimated from experimental and simulated data to better understand both the qubit systems and simulators. 
A DEM is a list of detector error events \(E_{i}\) and their associated probabilities \(p_{i}\) for occurring. 
The \(E_{i}\) can be represented as length-\(N\) bitstrings that add together modulo 2. 
A sample \(x\) can be generated from a DEM of \(L\) events of size \(N\) via the equation \(x = E \cdot q\) where \(E\) is the \(N \times L\) matrix whose \(i\)th column is \(E_{i}\) and \(q\) is an \(L\)-bit binary vector generated from the probabilities whose \(i\)th entry is 1 with probability \(p_{i}\) and 0 otherwise.

A benefit of DEMs is that they allow for insights into higher weight events that objects like \(p_{ij}\) matrices are ill-equipped to provide. 
Recent progress summarized by Young and \mbox{Blume-Kohout} \cite{s_blume-kohout_estimating_2025} describe how to estimate a DEM that fits the detector data from many shots of a quantum error correcting code. 
Their codebase is available in pyGSTi~\cite{s_pygsti_DEM}, and we have used that to perform our DEM analysis. 
However, the cost of estimating DEMs scales exponentially with the number of detectors \(N\) and so is computationally intensive for \(N > 16\). 
This limits us to estimating DEMs for small and short codes. 
Nonetheless, this still provides a way to probe high weight error events for such cases and diagnose the ways in which our simulations agree and disagree with experiment.

Our workflow for doing this DEM comparison is as follows. 
First, we run the experiment or simulation and save the detector data to a universal format. 
Then, we use pyGSTi to estimate a DEM from each dataset.
Finally, we compare between experiment and simulation results in two ways. 
First, we compare probabilities event-by-event to probe how specific errors behave. 
Second, we coarse-grain an event into a weight class and compare the total probability for each weight class of error events. 
This second approach allows us to evaluate whether the probability of high-weight events are similar between two datasets even when the probability of each individual event is small compared to the total shot count.

\subsection{Quasi-static miscalibration model}\label{quasi-static-miscalibration-model}

A quantitative model of the pulse-angle contextual miscalibration errors (e.g. pulse interference effects) from the cryo-controller, interconnect, and package is difficult to achieve \emph{a priori}. 
The system models discussed in \secref{interconnect-simulation} have good qualitative and generic quantitative capture of such errors, and they certainly occur: when comparing to experimental results (see \secref{multiqubit-experiments}), we see that both the two-qubit gate errors and circuit errors are higher than would be expected when considering only noise sources described up to this point (e.g. \(T_{2}^{*}\), \(\Nosc\), SPAM, and interconnect). 
Futhermore, the nature of the ``speckle'' in random sequences suggests that this error is fundamentally coherent (i.e. a result of repeatable miscalibration).
We therefore claim that calibration error is a \emph{dominant} error source in the performed multi-qubit experiments, and have added a single-parameter miscalibration (``miscal'') error term to our model.

Arguably the simplest miscalibration model would be to add white noise on to each exchange pulse. 
However, we know from the pulsed exchange oscillation decay experiments (and the associated noise parameter \(\Nosc\)) that the white noise of our pulses is very low. 
Instead, we introduce a quasi-static model of miscalibration. 
To implement it, we pick a global value miscalibration magnitude, parameterized as ``\(\Delta\theta/\theta\)'', a fractional error for each exchange. 
For every shot of the coherent spin simulation, we sample a random value from a Gaussian distribution centered at zero with standard deviation given by \(\Delta\theta/\theta\) for each exchange axis \(a_{i}\), i.e. a unitless number \(\delta_{m}(i) = \mathcal{N}(0,{\Delta\theta}/{\theta})\). 
This collection of random numbers is then used to linearly transform every exchange angle in the circuit:
\begin{equation}
	\theta_{j}(i) \rightarrow {(1 + \delta}_{m}(i)) \times \theta_{j}(i),
\end{equation}
where \(j\ \)indices time. 
In other words, to generate a single shot of the quasi-static miscalibration model, every pulse along a given axis will be scaled by the same relative magnitude, with that magnitude varying axis-to-axis. 
The quasi-static miscalibration model is time-agnostic and strongly axis-dependent.

In the following section, we will see that although the quasi-static miscal model is able to reproduce the experimentally observed error rates of two-qubit gates, syndrome detectors, and logical codes, it leads to much larger probabilities of high-weight correlations in DEMs when compared to experiment. 
Further work is needed to explore calibration models or to optimize circuits to avoid these flavors of calibration errors.

\section{Multiqubit experiments}\label{multiqubit-experiments}

\subsection{Device layout and qubit orientation}\label{device-layout-and-qubit-orientation}

\newcommand{\swapsymbol}{$\times$\!\!---\!\!$\times$}
\begin{figure*}
	\centering
	\smartincludegraphics[width=\textwidth]{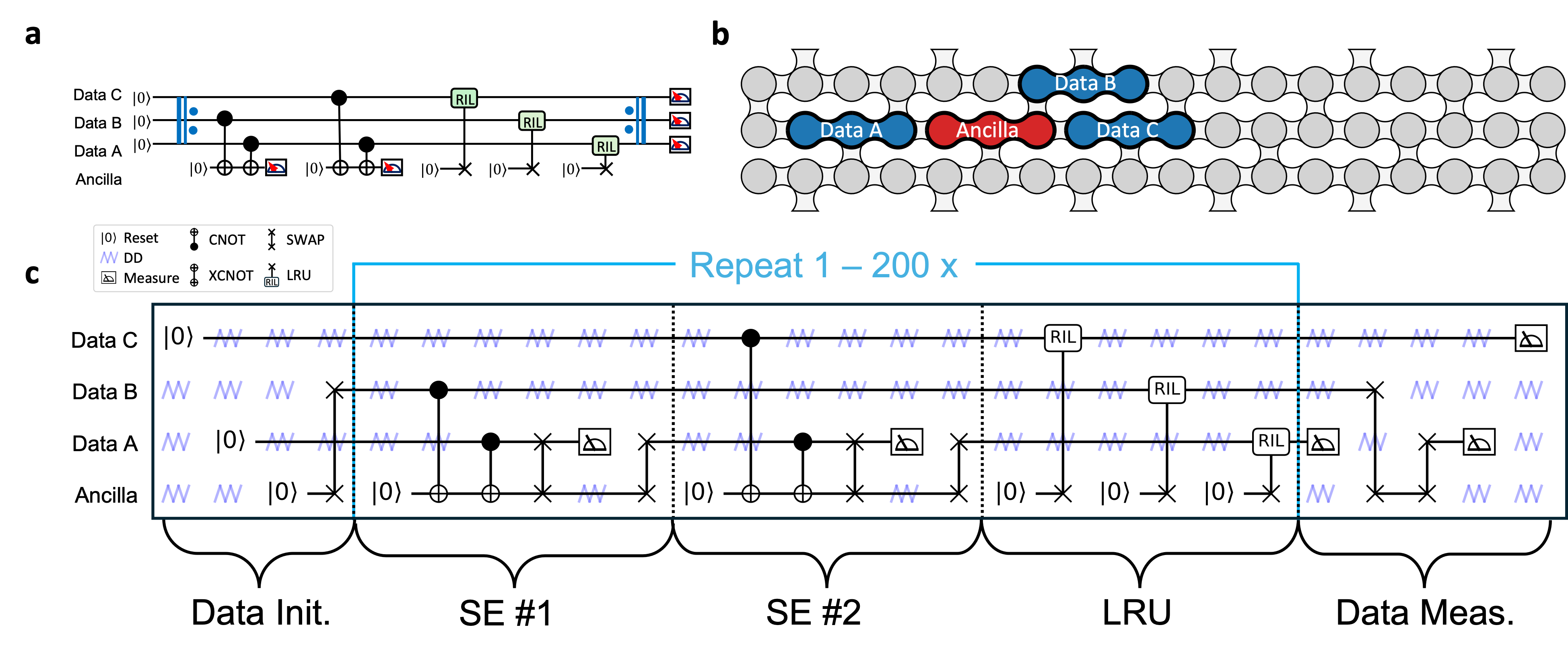}
	\caption{
		\textbf{Full {[}3,1,3{]} circuit and qubit layout}
		\textbf{a}, An idealized {[}3,1,3{]} circuit. 
		\textbf{b}, A cartoon schematic of the distance-3 repetition code device with the qubit layout. 
		\textbf{c}, The full circuit used specifying all swap (drawn as \swapsymbol) and dynamical decoupling (drawn as purple resistor-like symbols) operations.
	}
	\label{figS40}
\end{figure*}

\subsubsection{Alternative readout configuration}\label{alternative-readout-configuration}

For the demonstrations in this work, an alternative readout configuration referred to as ``forward DCS'' is used. 
The readout dot charge sensors (DCS) are formed using the first row of plunger dots (PA3, PB3, etc., labeled in dark green), rather than the nominal measurement dots (MA1, MB1, etc.), because the shorter P-to-P pitch between rows improves charge sensitivity over the P-to-M pitch, and thus readout fidelity. 
To enable this style of measurement, intervening gates between the nominal bath gates (BA, BB, etc.) and the operating DCS measure dot are biased to a highly over-coupled regime, effectively extending the electron bath on both source and drain pathways to reach the DCS. 
Dots ``flooded'' in this manner are highlighted in dark gray.
In the forward DCS configuration, as in the nominal configuration, the DCS operates in a many-electron regime. 
Modifications to the gate geometry and improvements to the measurement chain will render this technique obsolete going forward.

\subsubsection{Qubit orientations}\label{qubit-orientations}

Exchange-only qubits are initialized by preparing a two-electron singlet and a third, uninitialized spin on three consecutive quantum dots. 
The exchange ``axes'' of a single EO qubit, corresponding to a pair of spins undergoing exchange, are referred to by the geometric axes which would result from a qubit Bloch sphere on the EO qubit. 
(We emphasize that this Bloch sphere is not that corresponding to the total spin of the EO qubit; that total-spin Bloch sphere would track the irrelevant gauge-degree of freedom). 
For the EO qubit, the \emph{Z} axis refers to the pair of spins over which the singlet is initially prepared. 
The \emph{N} axis refers to the axis between the middle spin and the initially uninitialized third spin. 
Quantum angular momentum recoupling rules fix this axis is $N = Z\cos(\phi) + X\sin(\phi)$, with $\phi=2π/3$. 
We also refer to $N$ and $Z$ as Bloch-sphere $\pi$ pulses about their corresponding axes, which in turn corresponds to complete swaps of corresponding spin pairs. 
These axis names also correspond, for any given EO qubit, with two exchange gates; for example XA5XA6 refers to a qubit in which the pair of dots surrounded by exchange gate XA5 defines the $Z$ axis, and the pair of dots surrounded by exchange gate XA6 defines the $N$ axis. 
With this notation established, a given set of three dots can equivalently be prepared in the $ZN$ or $NZ$ orientation, and two-qubit gates can be performed between any combination of two adjacent qubit orientations, given appropriate modifications to the pulse sequence. 
It is straightforward to tile our 18-qubit device such that each qubit's $Z$ and $N$ axes only interact with the $Z$ and $N$ axes, respectively, of adjacent qubits. 
For simplicity, this is our preferred configuration, and is used for all demonstrations in this work.

\subsection{{[}3,1,3{]} demonstration}\label{demonstration-313}

For the distance-3 code, four qubits are used. 
Qubit XB6XB5 is used as the ancilla, while XA5XA6, XC7XC8, and XC5XC6 are used as data qubits A, B, and C respectively. 
The full circuit for this experiment is displayed in \figref{figS40}c. 
In this configuration we only use ``Data A'' as our native measurement site during the bulk of the sequence, and perform the necessary swaps to initialize or measure the data qubits.

\subsubsection{{[}3,1,3{]} noise characterization}\label{noise-characterization-313}

\begin{figure*}[ht]
	\centering
	\smartincludegraphics[width=165mm]{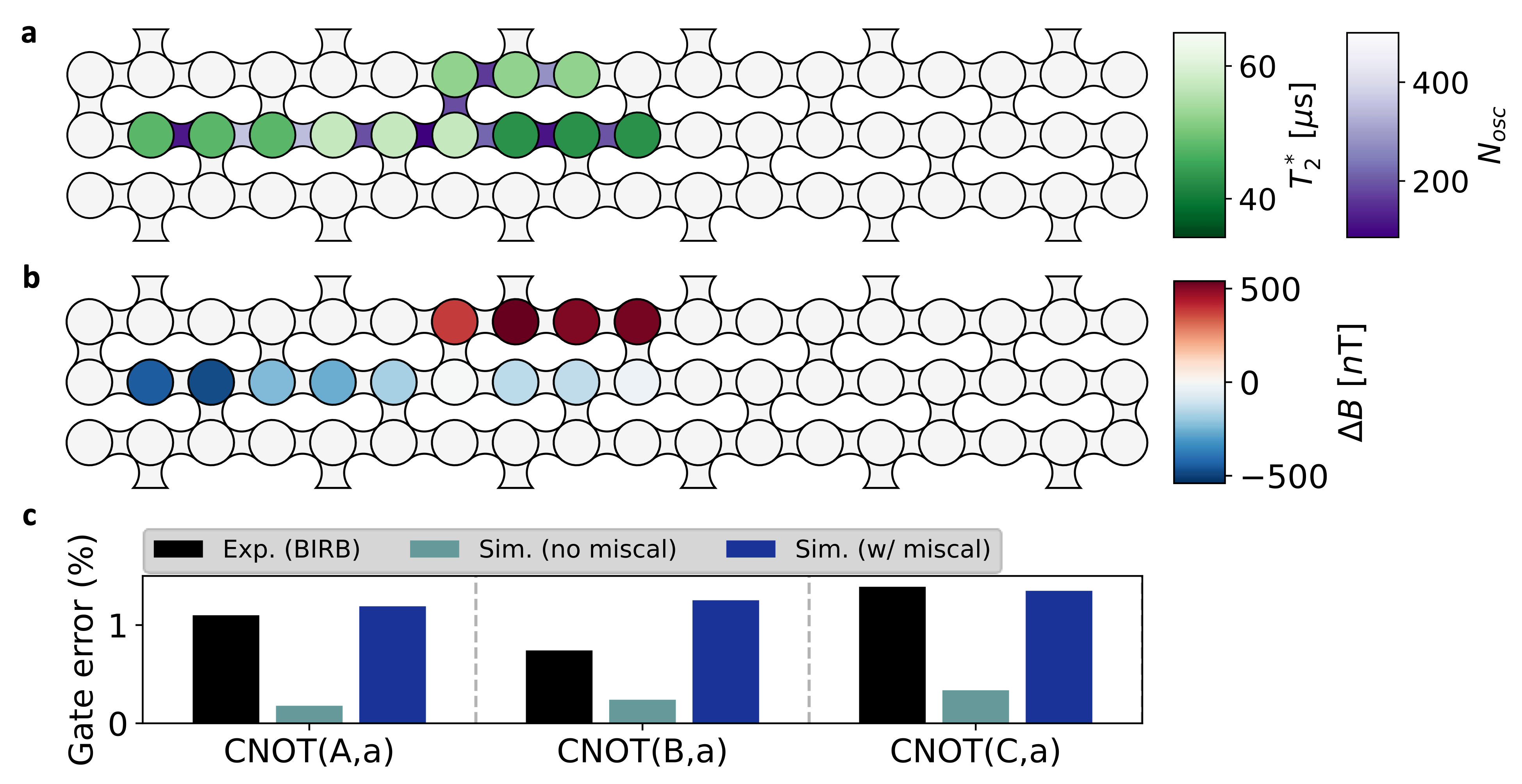} 
	\caption{
		\textbf{Device noise characterization of the QPU used for the distance-3 repetition code.}
		\textbf{a,} Envelope \(T_{2}^{*}\) decay rates averaged by qubit and inverted per dot and charge noise parameter \Nosc\  for each exchange axis. 
		\textbf{b,} Estimated static magnetic gradient across dots with a global magnitude of 18~$\mu$T. 
		Note this is measured with the integrated QPU and includes all associated system noise, as opposed to the data shown in the main text. 
		\textbf{c,} Comparison of CNOT gate errors with ersatz miscalibration between simulation and experiment.
	}
	\label{figS41}
\end{figure*}

\begin{figure*}
	\centering
	\includegraphics[width=165mm]{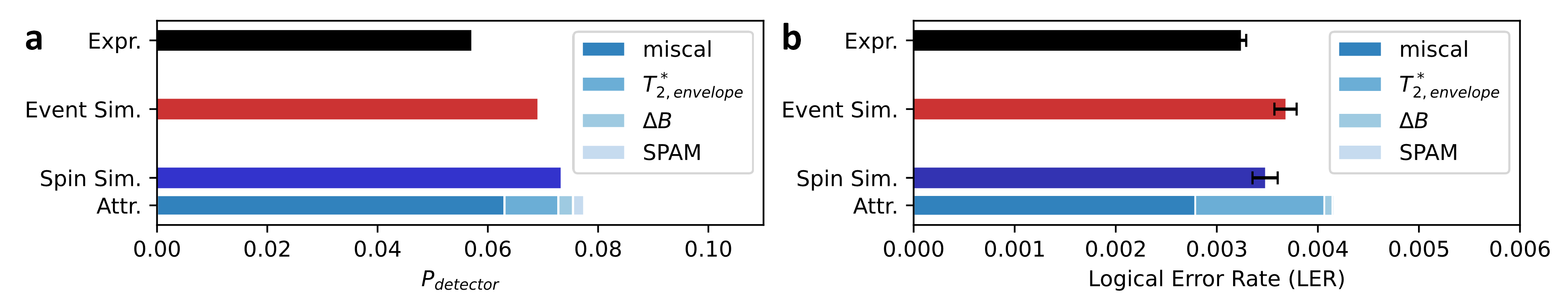} 
	\caption{
		\textbf{Comparison between experiment (black), event simulator (red), and spin simulator (dark blue) for the distance-3 repetition code.}
		Also included is a linear error-attribution for the spin simulations (light blues). 
		Attribution of \Nosc\ error was evaluated but not plotted because it was negligible. 
		\textbf{a,} A comparison of the average detector rate, \(\ts{P}{detector}\), for {[}3,1,3{]} circuits with 10 rounds. 
		\textbf{b,} A comparison of the logical error rate by way of an exponential fit to the round-dependence of the logical error probability. 
		Decoding was performed with the software package pymatching \cite{s_higgott_sparse_2025}.
	}
	\label{figS42}
\end{figure*}

The microscopic device parameters for the distance-3 repetition code were characterized by various experiments, and are summarized in \figref{figS41}. 
Singlet decay experiments were performed across all dot-pairs by performing series of swaps to and from the respective dot-sites and SPAM site, and measuring the oscillation frequency, \(\omega\) and envelope decay \(T_{2}^{*}\). 
We assume the dot-swap error is negligible with respect to magnetic error. 
For spin simulation, we use an average \(T_{2}^{*}\) for qubits of the constituent dot \(T_{2}^{*}\) (see \secref{coherent-spin-simulation}). 
To derive a local magnetic field, we solve the linear system of equations \(|A\cdot \mathbf{B}| = |\mathbf{W}|\) for \(B\), where \(\mathbf{W}\) is a vectorized list of all spin-pair oscillations $(\omega/\gamma_e)$, \(A\) is a matrix that tracks the relevant spins in each vectorized spin-pair, and (in this case) \(\mathbf{B}\) is a vectorized list of the magnitude of the local magnetic field at each dot. 
The derived local magnetic field is treated as a static field throughout the spin simulation. 
Time-domain exchange oscillation experiments estimate the microscopic charge noise parameter \(\Nosc\). 
This value integrates all sources of system control noise and is a distinct quantity from that reported in the main text, which was dominated by intrinsic device noise. 
For budgeting purposes, we use this more inclusive \textit{in situ} parameter. 
With envelope \(T_{2}^{*}\), a static magnetic gradient, and system charge noise, we perform GAQQMap attribution (\secref{gaqqmap}) and a complete randomized benchmarking simulation. 
Both methods estimated similar CNOT gate errors, but were lower than RB error rates reported from experiment. 
We include an ersatz quasi-static miscalibration error of 1.5\% to account for contextual pulse miscalibration. 
This excess noise, referred to as ``extrinsic'' in the main text, is applied uniformly across the repetition code simulation and is sampled per-shot. 
For SPAM errors, we evenly split 1.1\e{-3} between both initialization and measurement error.

\subsubsection{{[}3,1,3{]} error attribution}\label{error-attribution-313}

\begin{figure*}
	\centering
	\includegraphics[width=140mm]{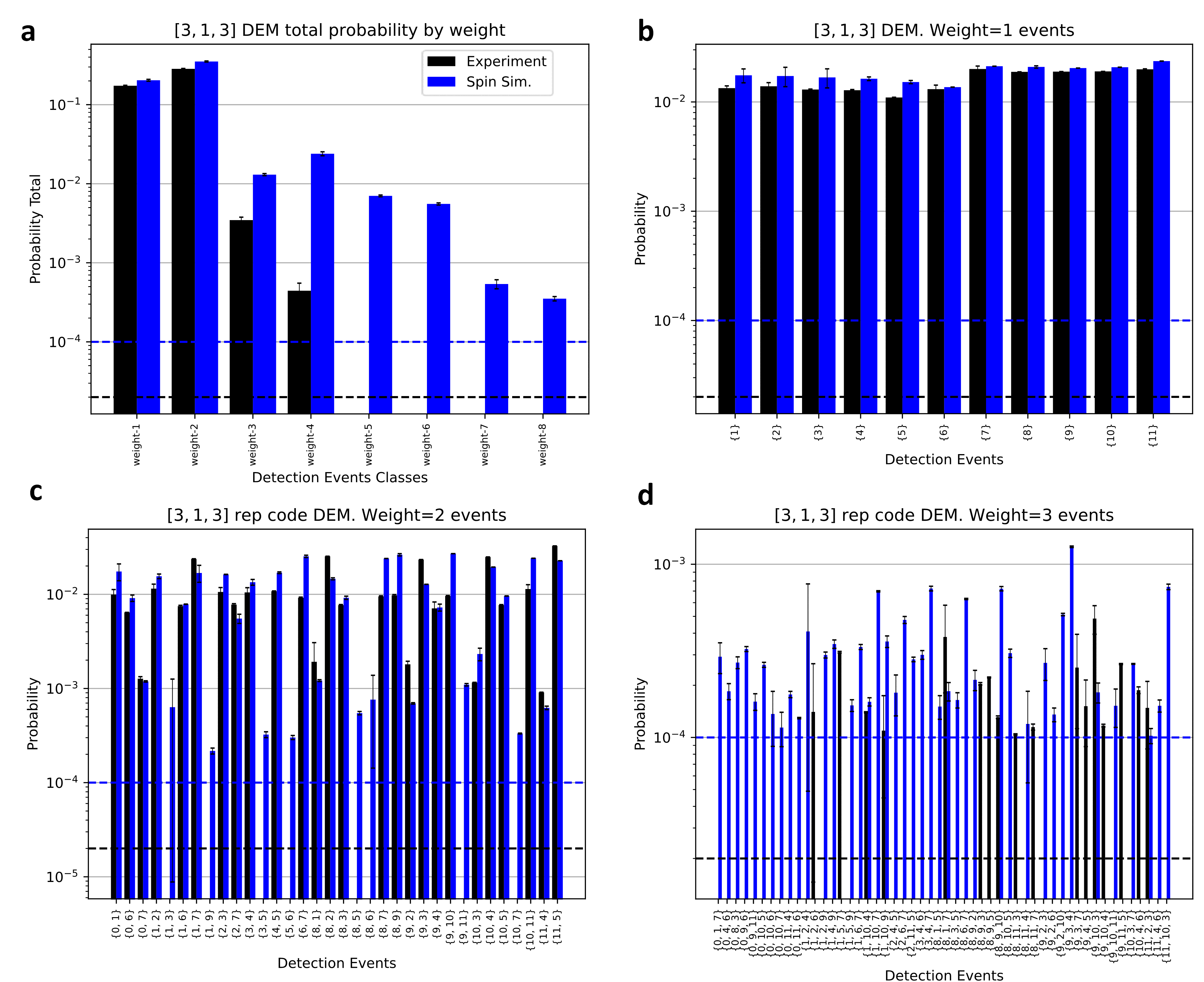} 
	\caption{
		\textbf{DEMs constructed from experiment (black) and pyquest simulation (blue) data for the distance-3 repetition code with 5 rounds of syndrome measurement.} 
		This is a DEM of size $N = 2\times 6 = 12$. 
		\textbf{a,} Comparison of total probabilities for each weight-class of error events. 
		\textbf{b,} Comparison of error probabilities event-by-event for weight-1 errors.
		\textbf{c,} Comparison of error probabilities event-by-event for weight-2 errors. 
		\textbf{d,} Comparison of error probabilities event-by-event for weight-3 errors. 
		Dashed lines correspond to 1/$N$\textsubscript{shot} with $N$\textsubscript{shot} = {[}5\e{4}, 1\e{4}{]} for experiment and pyquest. 
		Overall, experiment and simulation track well when comparing events in aggregate, especially for lower weight events with pyquest overestimating high weight errors.
	}
	\label{figS43}
\end{figure*}

\begin{figure*}
	\centering
	\smartincludegraphics[width=\textwidth]{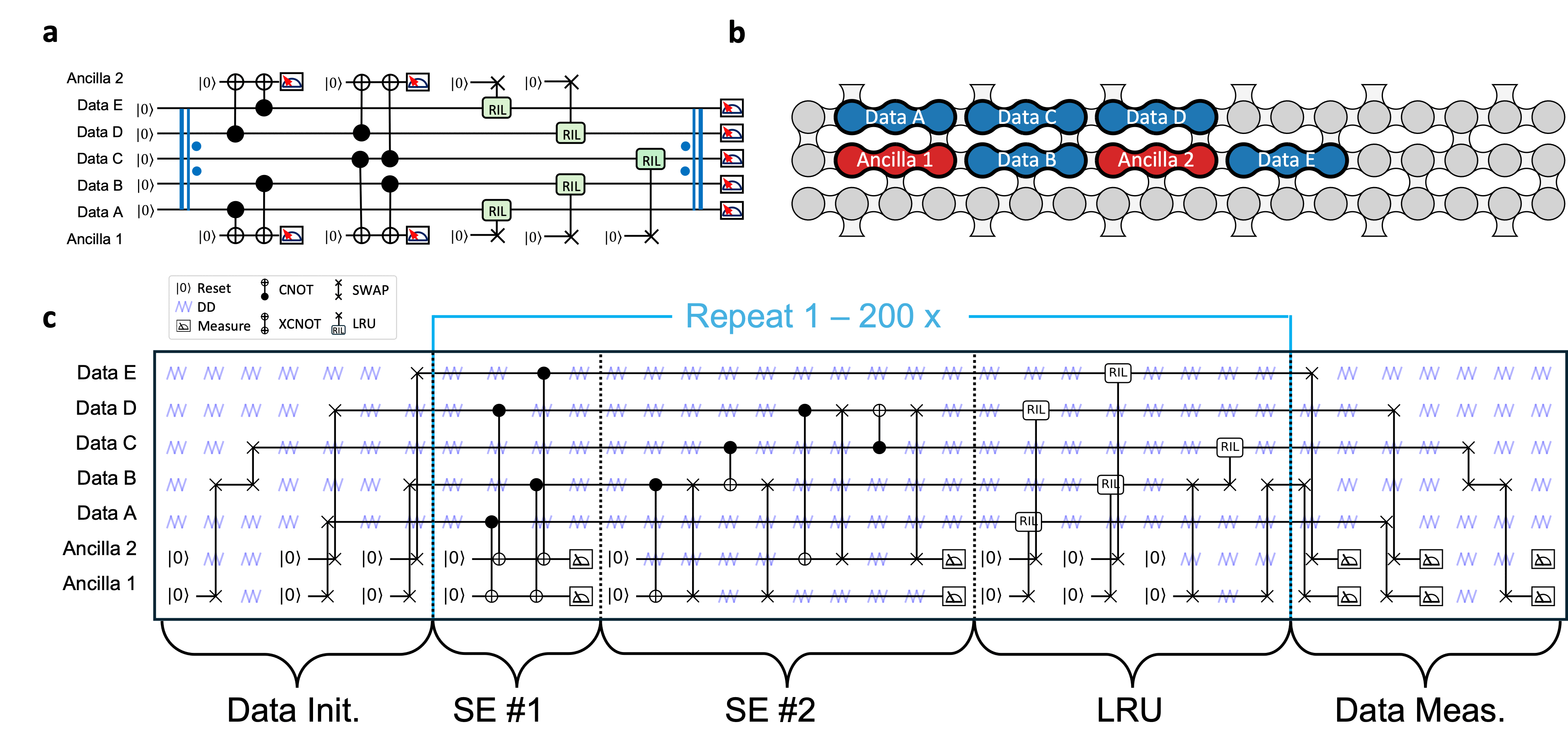}
	\caption{
		\textbf{Full {[}5,1,5{]} circuit and qubit layout.}
		\textbf{a,} The idealized distance-5 repetition code circuit shown in the main text, reproduced here. 
		\textbf{b,} A cartoon schematic of the device used for the \mbox{distance-5} repetition code with the qubit layout overlaid. 
		\textbf{c,} The full circuit used specifying all swap (drawn as \swapsymbol) and dynamical decoupling (drawn as resistor like symbols) operations.
	}
	\label{figS44}
\end{figure*}

In \figref{figS42} we compare simulated and experimental performance of the distance-3 repetition code, using both the event simulator and the spin simulator. 
A first-order (linear) attribution is performed via the spin simulator by simulating the repetition code under the influence of only a single noise source at a time. 
We individually evaluate the impact of \(T_{2}^{*}\),
\(\Delta B\), SPAM, and quasi-static miscal. 
Charge noise, characterized by $N_\text{osc}$, is also included but negligible and therefore not plotted.) 
To get a full picture of how these noise sources contribute to circuit error, one must also compute the non-linear (quadratic, cubic, etc.) relationships, but for simplicity we do not fully explore that here. 
We find that this first-order attribution model slightly overestimates the total error rate produced by a holistic simulation but still provides a reasonable guide.
Figure \ref{figS42} applies this analysis to two metrics of circuit performance: the first metric is the average probability of a syndrome detection of a local error, $\ts{P}{detector}$ (also referred to as the detector event fraction, or DEF); the second metric is the logical error rate \(\epsilon_{L}\) which is fitted to \([1 - \left( 1 - 2\epsilon_{L} \right)^{n}]/2\) when observing the decoded error probability per round \(n\). 
In the distance-3 repetition code, we use pymatching \cite{s_higgott_sparse_2025} to construct a minimum-weight perfect matching decoder, assuming a uniform depolarizing error channel given the error rates of CNOTs, dynamic decoupling exchanges, etc. 
In both metrics, the most dominant first-order noise contribution is the ersatz quasi-static miscalibration, followed by the nuclear magnetic noise term \(T_{2}^{*}\).

\subsubsection{{[}3,1,3{]} DEM analysis}\label{dem-analysis-313}

A DEM analysis of the distance-3 repetition code with five rounds of syndrome measurement is shown in \figref{figS43}.
The excellent agreement between DEMs generated from simulation and experiment for weight-1 and weight-2 events shows that our modeling workflow is doing a good job of predicting simple qubit event errors.
Indeed, in panel c, disagreements only occur when the simulated or experimental probabilities are near their shot-limit (blue and black dashed lines, respectively).

The simulator overestimates high weight errors compared to experiment.
This is not surprising: the ersatz miscalibration used in simulation is an axis-distinct, uniform-in-time error, and as such is capable of creating high-weight events. 
For example, if axis XA5 is very noisy in one simulated shot, then one would expect many repeated errors on data qubit A. 
In experiment, the contextual pulse miscalibration in the real cryo-controller is a non-uniform-in-time in error, and is likely less axis-dependent (e.g. XA5 and XC5 may experience similar errors in similar contexts), leading to mostly low-weight errors.

\subsection{{[}5,1,5{]} demonstration}\label{demonstration-515}

Seven qubits are used in the distance-5 demonstration, as shown in the simplified version of the circuit in \figref{figS44}a. 
The layout of the device is depicted in \figref{figS44}b: Qubits XA6XA5 (corresponds to the gate names of $Z/N$ axes) and XC6XC5 are used as ancilla 1 and 2, while XA9XA8, XB5XB6, XB8XB9, XC9XC8, and XD5XD6 are used as data qubits A-E respectively. 
All qubits are coupled to their neighbors through the X and Y gates shaded in light blue, except Data A and C due to a known issue with XB7 due to a daughterboard flaw. 
We can perform native SPAM at Ancilla 1, 2, and Data B (not used, however), whereas we must add swaps to initialize and measure Data A, C, D, and E.

The detailed circuit used in the repetition code is shown in \figref{figS44}c, where we are now showing all of the qubit-level swaps needed for initialization, certain entangling operations, measurement, and dynamical decoupling that is present throughout the entire circuit.

\subsubsection{{[}5,1,5{]} noise characterization}\label{noise-characterization-515}

\begin{figure*}
	\centering
	\smartincludegraphics[width=165mm]{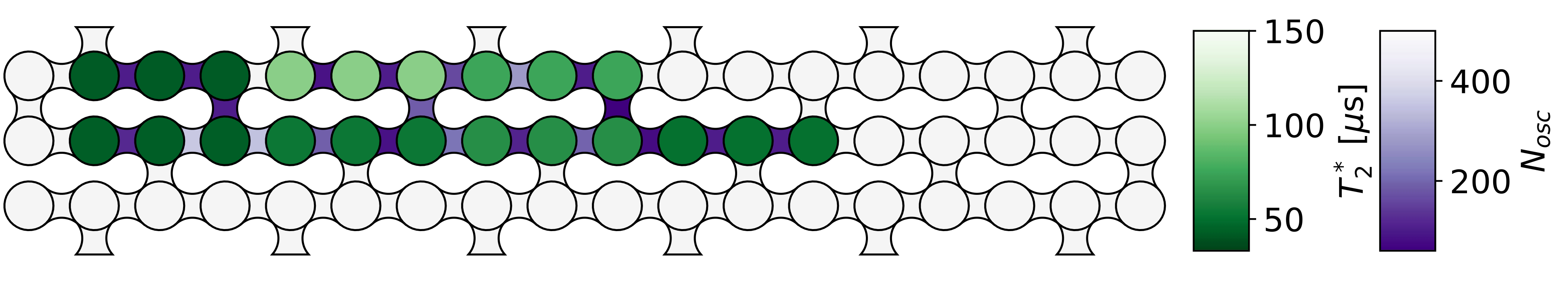}
	\caption{
		\textbf{Device noise characterization for distance-5 repetition code.}
		Effective $T_2^*$ decay rates (accounting for static magnetic gradients) are averaged by qubit and inverted per dot. 
		Charge noise parameters \Nosc\ are measured for each exchange axis. 
		Note this is measured with the integrated QPU and includes all associated system noise, as opposed to the data shown in the main text.
	}
	\label{figS45}
\end{figure*}

\begin{figure*}
	\centering
	\includegraphics[width=165mm]{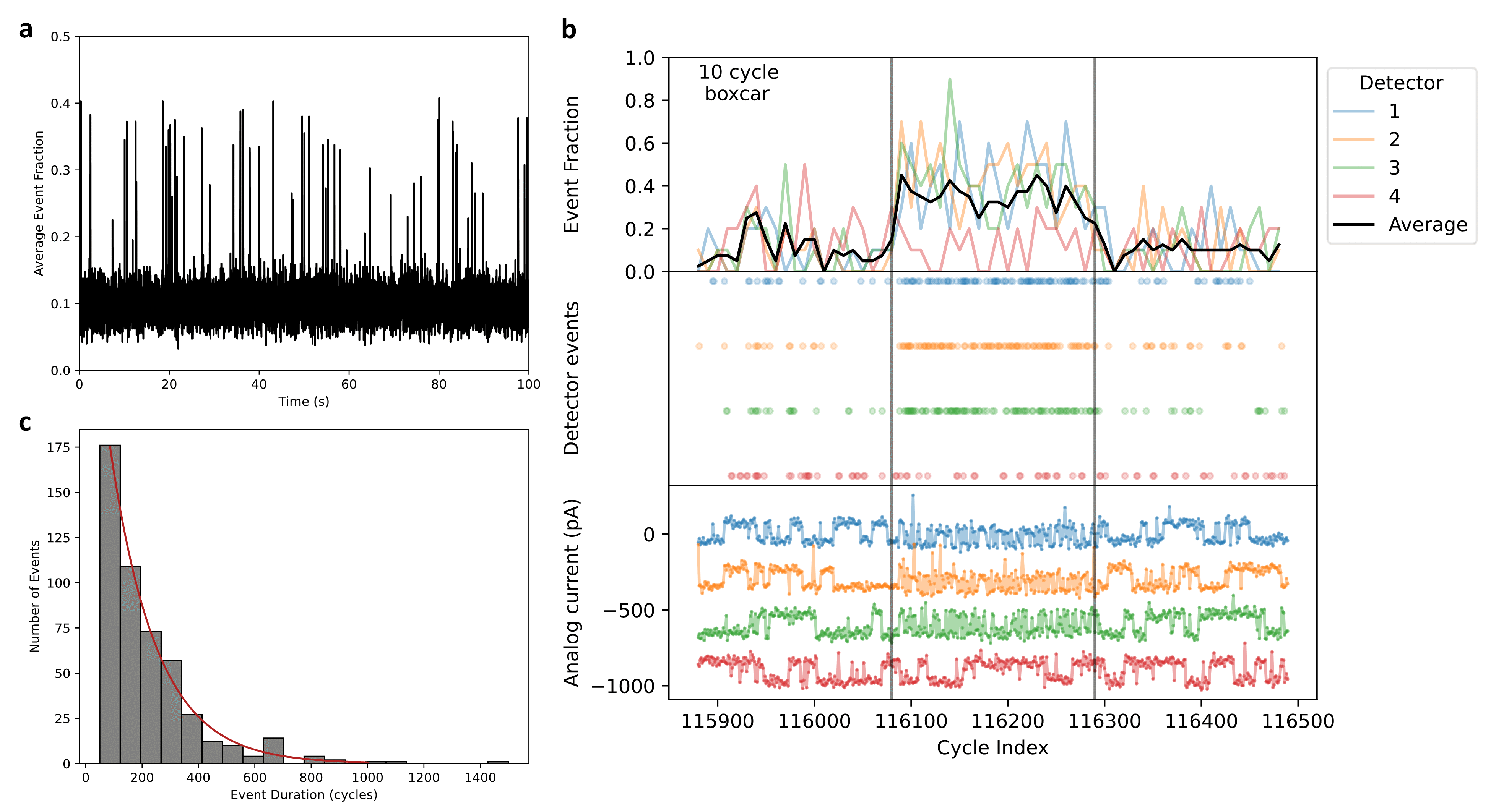} 
	\caption{
		\textbf{{[}5,1,5{]} event fraction spikes.} 
		\textbf{a,} Continuously running distance-5 syndrome extraction. 
		The detector event fraction averaged over all detectors with a 100-cycle boxcar filter. 
		\textbf{b,} A zoomed in view of about 600~cycles. 
		The top subfigure displays a 10-cycle boxcar filter of the individual detector events per cycle (middle subfigure).
		The bottom subfigure shows the raw current values measured for each detector, which are ultimately converted to a 0 or 1 with a predefined threshold and finally a detection event if the measurement outcome flips between rounds. 
		\textbf{c,} Histogram of number of spike events binned by duration, well fit by an exponential with a 164~cycle (14~ms) time constant.
	}
	\label{figS46}
\end{figure*}

The dot-level device parameters for the distance-5 repetition code were characterized in a similar fashion as the distance-3 repetition code measurement, and are summarized in \figref{figS45}.
Instead of measuring singlet return probability of all possible pairs, as was done in the distance-3 setup, we only measure local qubit-defined singlet return probabilities of the three possible dot combinations.
Additionally, due to low frequency oscillations that we attribute to static magnetic field gradients, we combine the envelope and oscillation into one effective first-order parameter given by \({{(T}_{2}^{*})}^{-2}\  = {(T_{2,\ \text{env}}^{*})}^{-2} + \omega^{2}/2\). 
A global static magnetic field of \(160\ \mu T\) was measured via dynamical decoupling spectroscopy \cite{s_kerckhoff_magnetic_2021}.
Time-domain exchange oscillation experiments estimate the microscopic charge noise parameter $N_\text{osc}$. 
We include an ersatz quasi-static miscalibration error of 1.5\% to account for contextual pulse miscalibration, as described in \secref{quasi-static-miscalibration-model}.
As we will discuss in the next section, one of the two SPAM sites provided unreliable and sometimes correlated measurements. 
Given this uncertainty, we assume a SPAM error of 1.2\e{-3} for both sites, which we evenly split between both initialization and measurement error.

\subsubsection{Spikes in detection events}\label{spikes-in-detection-events}

\begin{figure*}[t]
	\centering
	\smartincludegraphics[width=150mm]{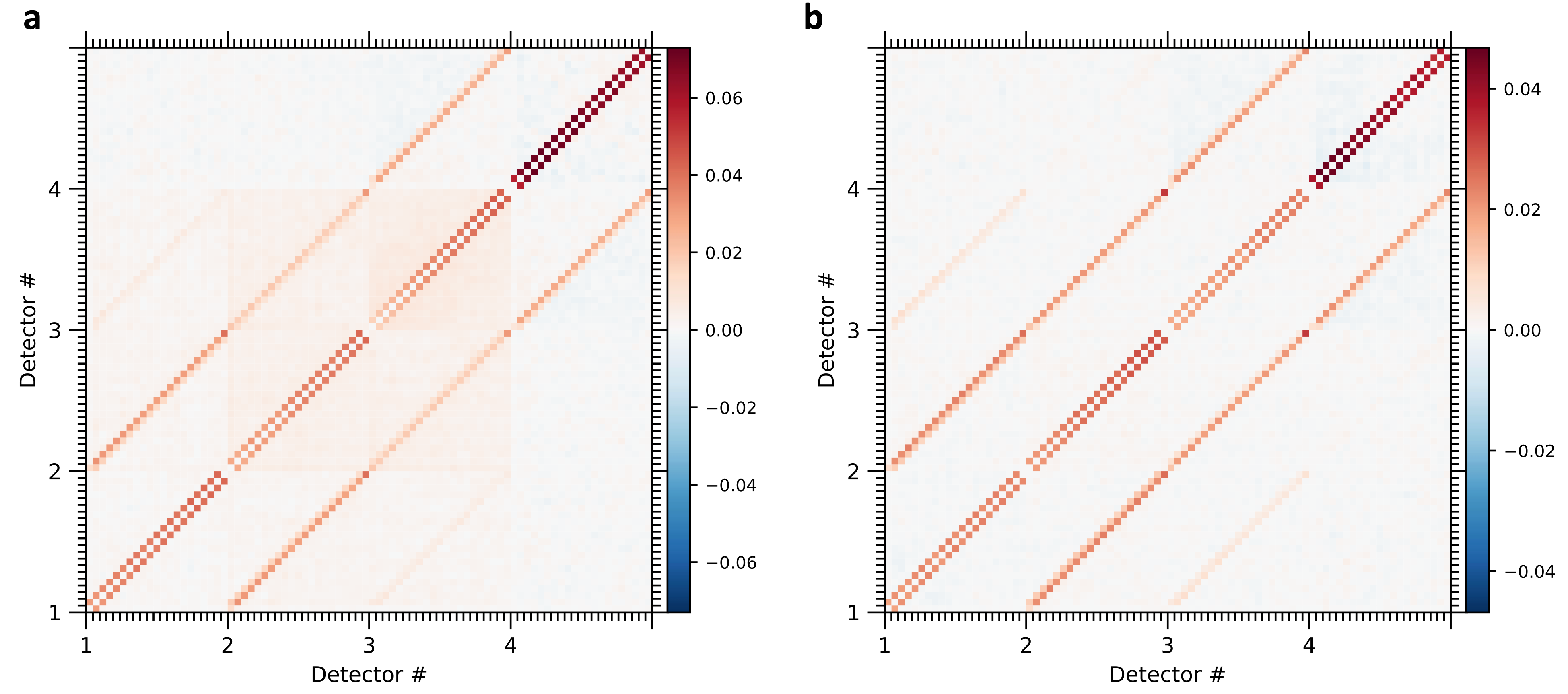} 
	\caption{
		\textbf{{[}5,1,5{]} correlations with and without spike events.} 
		\textbf{a,} $p_{ij}$ of the 20-round distance-5 circuit.
		\textbf{b,} $p_{ij}$ plot after filtering out shots for which a spike event occurred.
	}
	\label{figS47}
\end{figure*}

One particular oddity we observe in the detector error fraction vs time data (\figref{figS46}a, running the experiment continuously rather than for a fixed number of syndrome rounds) is the presence of spikes in detection event fraction that appear to last for several rounds. 
Focusing on a single spike and separating by detector (\figref{figS46}b), we see that detectors 1 to 3 show highly elevated errors, whereas detector 4 appears largely unaffected.
We estimate the length of a spike event by computing when the event fraction deviates by \(1.4826\times\)MAD (median absolute deviation) to mark the start and end of a particular spike event. 
We can then plot the distribution of durations, and we see that it is well described by an exponential with a 164~cycle (14~ms) timescale.

So-called $p_{ij}$ plots can give insight into what type of errors these spike events caused \cite{s_google_exponential_2021}.
In \figref{figS47}, we see that by filtering out shots containing an identified spike event, the large red background covering detectors 1-3 is also removed. 
This is consistent with some event causing all three of those detectors simultaneously having elevated event fractions due to SPAM errors. 
However, we do notice that there exists a space-like correlation between detector 1 and 3 that remains. 
We suspect the same microscopic event that caused the SPAM errors is also the source of these correlations, but further investigation is required to understand the origin.

We did not observe this phenomenon in the {[}3,1,3{]} or {[}{[}4,2,2{]}{]} demonstrations.

\subsubsection{{[}5,1,5{]} DEM analysis}\label{dem-analysis-515}

\begin{figure*}
	\centering
	\includegraphics[width=120mm]{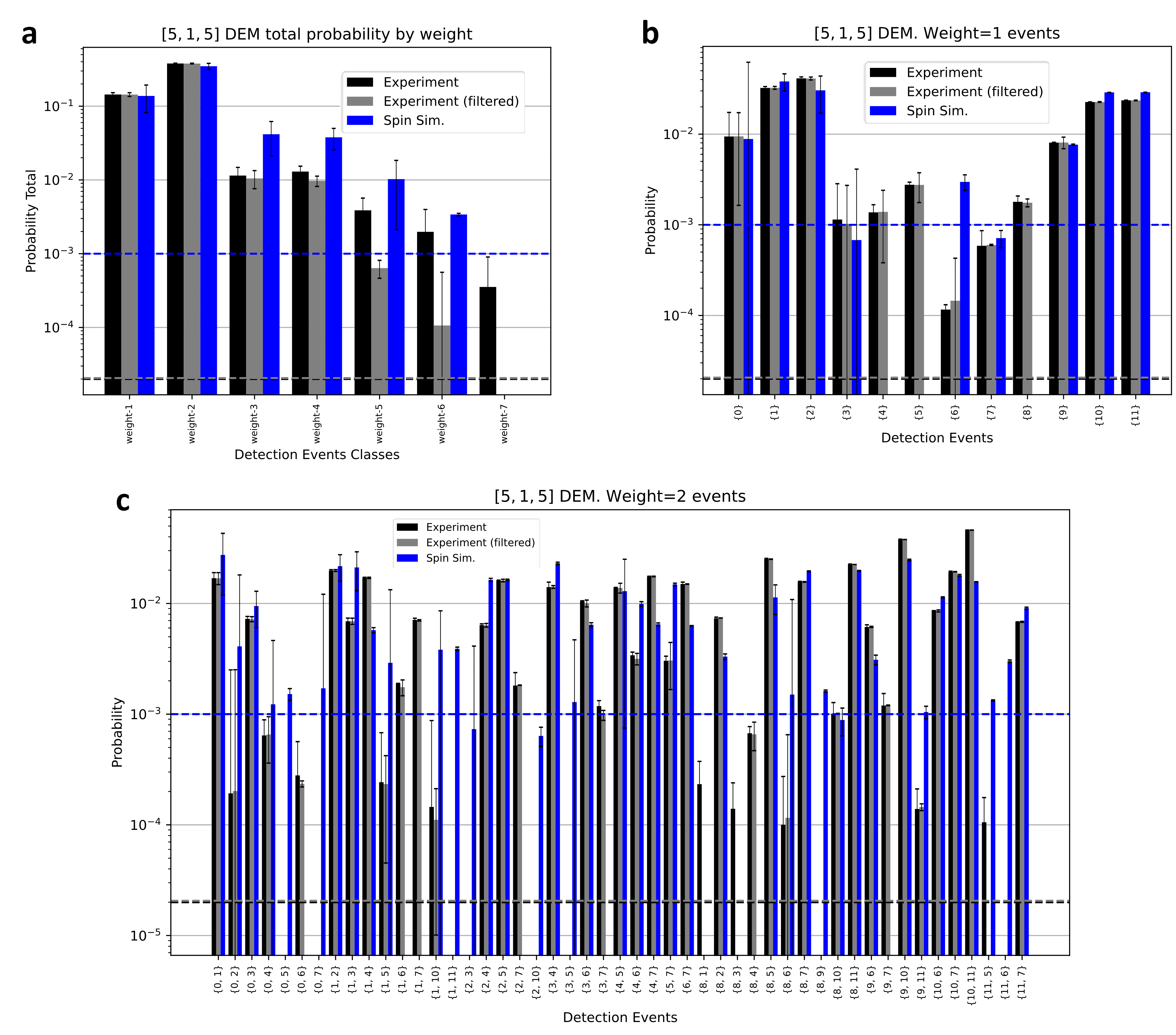} 
	\caption{
		\textbf{DEMs constructed from experiment (black), filtered experiment (grey), and pyquest simulation (blue) data for the distance-5 repetition code with two rounds of syndrome measurement.} 
		This is a DEM of size $N = 4\times 3 = 12$. 
		\textbf{a,} Comparison of total probabilities for each weight-class of error events. 
		\textbf{b,} Comparison of error probabilities event-by-event for weight-1 errors. 
		\textbf{c,} Comparison of error probabilities event-by-event for weight-2 errors.
		Dashed lines correspond to 1/$N$\textsubscript{shot} with $N$\textsubscript{shot} = [5\e{4}, 48510, 1\e{3}] for experiment, filtered experiment, and pyquest.
		Overall, we observe that filtering out the measurement spikes in the experimental data suppresses high-weight errors and that the pyquest simulations on aggregate overestimate high weight errors.
	}
	\label{figS48}
\end{figure*}

We can use DEMs to further analyze the effects of filtering out spurious or anomalous data. 
As described above, the distance-5 repetition code consistently saw measurement spikes that indicated highly correlated errors (as seen in the \(p_{ij}\) matrix). 
Using a DEM analysis, we see that filtering out those measurement spikes suppresses the high-weight error probabilities in aggregate (\figref{figS48}).
Similarly to the {[}3,1,3{]} DEM analysis, the simulator's use of the ersatz miscalibration model leads to an overestimation of events of weight-3 and higher. 
The overall similarities between the {[}3,1,3{]} and {[}5,1,5{]} DEM analyses is also a good indication that our intrinsic and extrinsic error sources are well characterized.

\subsection{{[}{[}4,2,2{]}{]} demonstration}\label{demonstration-422}

\begin{figure*}
	\centering
	\smartincludegraphics[width=\textwidth]{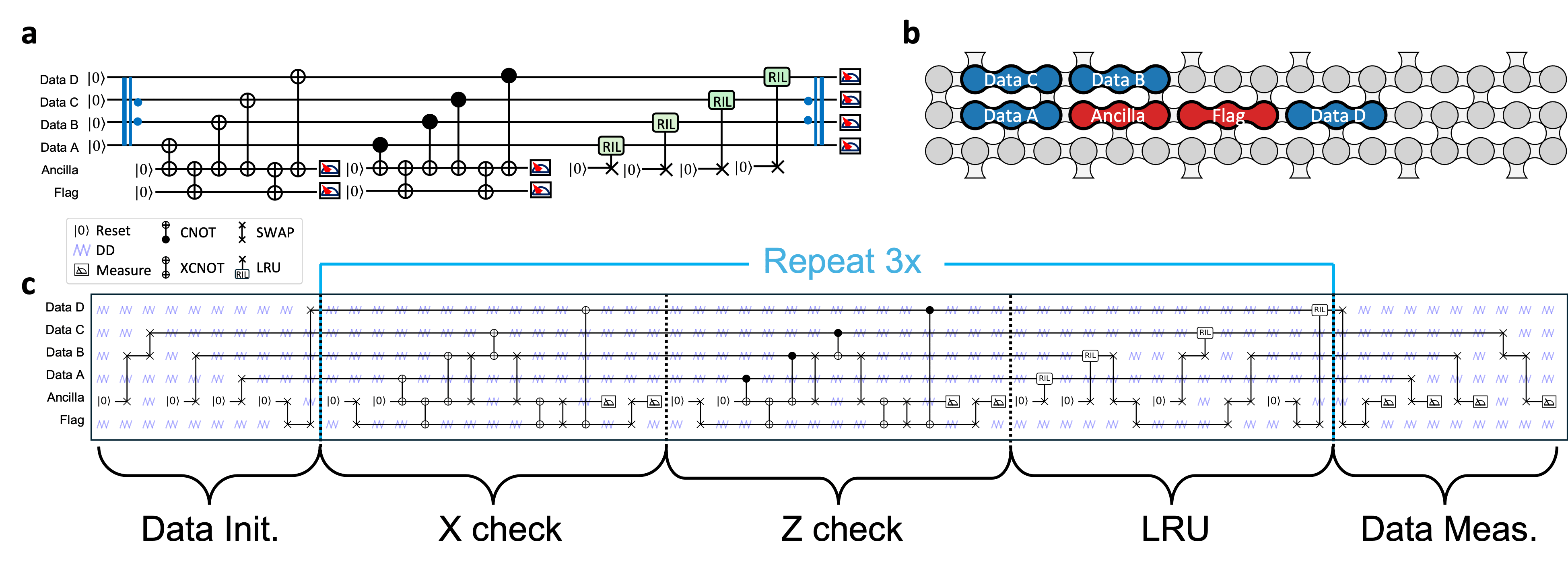}
	\caption{
		\textbf{Full {[}{[}4,2,2{]}{]} circuit and qubit layout.} 
		\textbf{a,} The idealized {[}{[}4,2,2{]}{]} circuit. 
		\textbf{b,} A cartoon schematic of the {[}{[}4,2,2{]}{]} device with the qubit layout. 
		\textbf{c,} The full {[}{[}4,2,2{]}{]} circuit used to specify all swap (drawn as \swapsymbol) and dynamical decoupling (drawn as resistor-like symbols) operations.
	}
	\label{figS49}
\end{figure*}

For the {[}{[}4,2,2{]}{]} demonstration, six qubits were used, from the top and middle rows of an 18 qubit device (\figref{figS49}b) Qubits XB6XB5 and XC5XC6 are used as the ancilla and flag qubits, while XA5XA6, XB9XB8, XA8XA9, and XD6XD5 are used as data qubits A-D respectively. 
Data qubits A, C, and the flag qubit are oriented with axes $ZN$ from left to right, while data qubits B, D and the ancilla qubit are oriented with axes $NZ$ from left to right. 
For better performance on this particular device, the circuit is compiled to use a single measurement site without consequence to error propagation.

Compared to the repetition code demonstrations, the {[}{[}4,2,2{]}{]} demonstration was performed in a system using a later version of the cryo-controller but with comparable electronic performance, a later revision of the daughterboard with improved signal integrity, and custom transimpedance amplifiers (\secref{transimpedance-amplifier}). 
With respect to quantum performance, this system exhibited better one-qubit and two-qubit error rates and similar SPAM errors.

The {[}{[}4,2,2{]}{]} quantum circuit is shown in \figref{figS49}a and c. 
After all four data qubits are initialized, successive rounds of syndrome extraction and leakage removal are applied. 
Each round consists of an \emph{XXXX} parity check, a \emph{ZZZZ} parity check, and leakage reduction units (LRUs). 
The parity check measurements of the ancilla are each immediately preceded by a measurement of the flag qubit, which validates the ancilla outcome. 
When the data qubits are initialized to logical states in the \(\pm\) basis, the order of the $Z$ and $X$ parity checks is reversed to ensure that the first parity check efficiently projects the prepared state into the code space.
The quantum operations are performed sequentially, with dynamical decoupling pulses \cite{s_sun_full-permutation_2024} on all idle qubits interleaved with operational pulses.

\begin{figure*}
	\centering
	\smartincludegraphics[width=165mm]{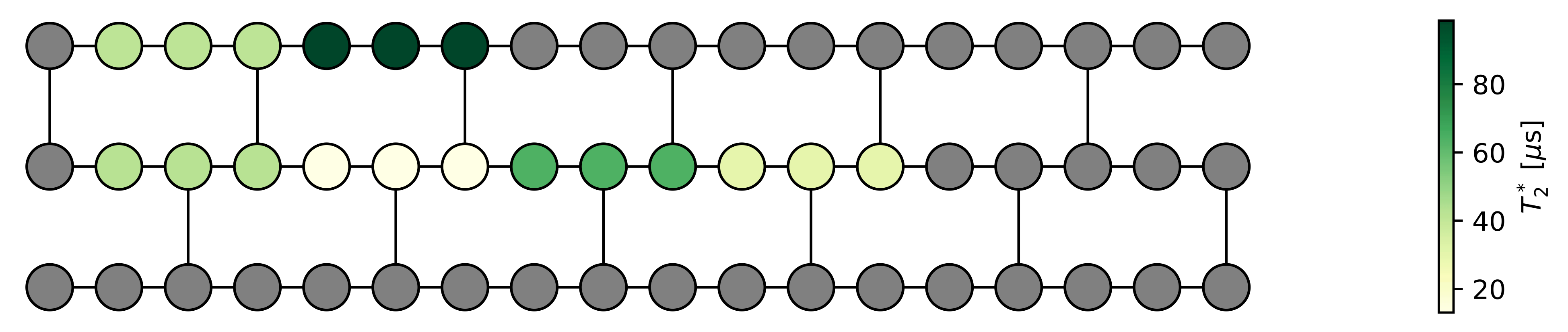}
	\caption{
		\textbf{Map of the effective $T_2^*$ values used in simulation of the {[}{[}4,2,2{]}{]} circuit.} 
		For each software-defined qubit, the three unique dot-pair \(T_{2}^{*}\) values are averaged into a single qubit-wide value.
	}
	\label{figS50}
\end{figure*}

In the {[}{[}4,2,2{]}{]} code, the logical codewords for the $Z$ basis are:
\begin{align*}
	\ket{00}_{L} &= \frac{1}{\sqrt{2}}\left( \ket{0000} + \ket{1111} \right),
	\\
	\ket{01}_{L} &= \frac{1}{\sqrt{2}}\left( \ket{0011} + \ket{1100} \right),
	\\
	\ket{10}_{L} &= \frac{1}{\sqrt{2}}\left( \ket{0101} + \ket{1010} \right),
	\\
	\ket{11}_{L} &= \frac{1}{\sqrt{2}}\left( \ket{0110} + \ket{1001} \right).
\end{align*}
The data qubits are initially prepared in a computational state, such as \(|0000\rangle\). 
In this case, the \emph{XXXX} stabilizer is applied first, projecting it into an equal mixture of the ideal \(\left| 00 \right\rangle_{L}\) state and \(\left| 00 \right\rangle_{L}\) with an initial \emph{Z} error.
Analogous projections occur for other starting states.

$X$-basis codewords are found by application of transversal Hadamard to each qubit. 
Similar to the 0/1 basis, the initial \emph{ZZZZ} stabilizer application projects the initial state into the code space.

\subsubsection{{[}{[}4,2,2{]}{]} noise characterization}\label{noise-characterization-422}

For the {[}{[}4,2,2{]}{]} experiment, an unexpected temperature cycle of the device occurred before full qubit characterization data could be obtained. 
The microscopic magnetic noise parameters \(T_{2}^{*}\) were characterized by experiment, and are summarized in \figref{figS50}. 
Dynamical decoupling spectroscopy, used as an \textit{in situ} measure of the local magnetic field for each qubit, showed an average global magnetic field of 20 $\mu$T and minimal magnetic gradients. 
To capture any small magnetic gradients not visible to this technique, we used effective \(T_{2}^{*}\) values instead of the envelope of the fit in our simulations.

For the microscopic charge noise parameter $N_\text{osc}$, we used a placeholder value of 200 for every axis. 
This is a pessimistic number (see main text Fig.~3), which likely overestimates the amount of charge noise in the device. 
But it is a necessary precaution, as no $N_\text{osc}$ measurements were taken in the final tuned-up state of the device before the fridge was temperature cycled.

For SPAM errors, based on singlet/triplet readout histograms we assigned a low initialization error of 1.2\e{-5} in the coherent spin simulations.
Assignment fidelity experiments were run multiple times on the {[}{[}4,2,2{]}{]} readout site and the best performance was about 1.2\% infidelity, with triplet readout often better than singlet readout.
However, there was a slow drift in the optimal SPAM threshold current, and some assignment fidelity experiments showed closer to 1\% error for triplet readout and \textgreater2\% error for singlet readout. 
Due to the long duration of the {[}{[}4,2,2{]}{]} experiment and the slow drift, we used this low-water mark as the measurement fidelity in the spin and event simulations.

\begin{figure*}
	\centering
	\includegraphics[width=165mm]{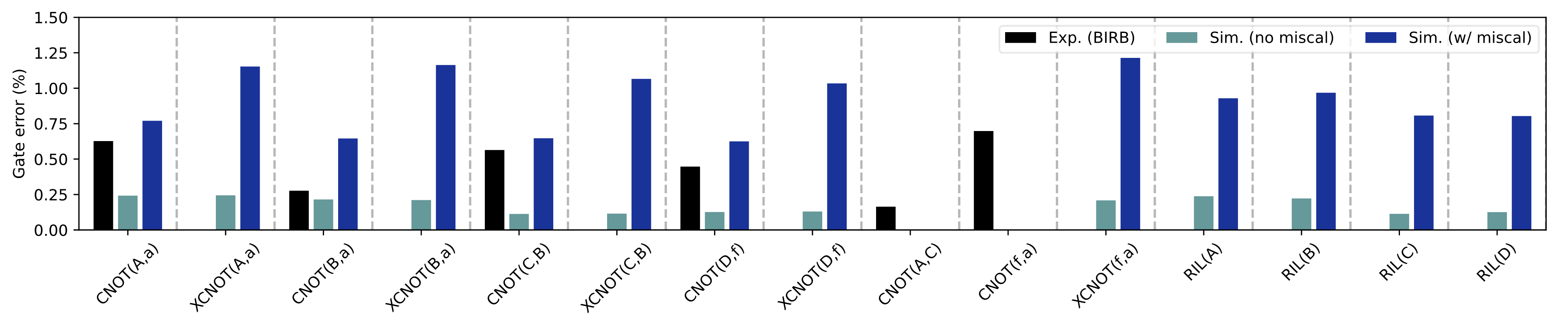} 
	\caption{
		\textbf{Comparison of experimental two-qubit gate errors (obtained via BIRB) and simulated two-qubit gate errors (obtained via GAQQmap, \secref{gaqqmap}) in the {[}{[}4,2,2{]}{]} demonstration}. 
		The data qubits are listed as A, B, C, D, the ancilla qubit as ``a'' and the flag qubit as ``f'' (as defined in \figref{figS49}). 
		Not all gates used in the circuit were experimentally characterized (i.e. missing black bars), and two CNOTs were experimentally characterized even though they did not occur in the circuit [CNOT(A,C) and CNOT(f,a)].
	}
	\label{figS51}
\end{figure*}

Unlike the {[}3,1,3{]} and {[}5,1,5{]} experiments, we did not calibrate the ersatz miscal parameter to the one-qubit and two-qubit BIRB values. 
Instead, we ran full circuit simulations with the above magnetic, charge, and SPAM noise parameters, and added a quasi-static miscalibration value (0.9\%) that reproduced experimental detector probability rates. 
We then performed GAQQmap attribution (\secref{gaqqmap}) of the two-qubit gates in our spin simulator, and used those error rates in the event simulator. 
This led to excellent agreement in circuit throughput and fidelity predictions, and improved the agreement in two-qubit gate error rates (in three out of the four gates in which that comparison could be performed, see \figref{figS51}). 
Note that the simulated error rates without any miscalibration added are all below the experimentally observed two-qubit BIRB errors. 
This highlights that, just like in the system used for the repetition code experiments, the intrinsic magnetic and charge noise of the device are not dominant error sources for two-qubit gates.

\subsubsection{{[}{[}4,2,2{]}{]} post-selection}\label{post-selection-422}

\begin{table*}
	\centering
	\caption{
		\textbf{Codeblocks and basis of the {[}{[}4,2,2{]}{]} code.} 
		The sixteen basis states of the four-data-qubits are listed in parenthesis from (1) to (16), and are grouped into  four  codespaces {[}A{]} through {[}D{]} based on the stabilizer eigenvalues (\(Z_{1234}\) and \(X_{1234}\), i.e. $Z$-check and $X$-check ancilla outputs in the absence of errors).
	}
	\noindent\makebox[\textwidth]
	{ 
		\small
		\begin{tabular}{|c|c|c|c|c|}
			\hline
			& \textbf{Codespace A} & \textbf{Codespace B} & \textbf{Codespace C} & \textbf{Codespace D} \\ 
			$(Z_{1234}, X_{1234})$ & $(1,1)$ & $(1,-1)$ & $(-1,1)$ & $(-1,-1)$ \\
			\hline
			$|00\rangle_L$ & (1) $|0000\rangle + |1111\rangle$ & (5) $|0000\rangle - |1111\rangle$ & (9) $|0001\rangle + |1110\rangle$ & (13) $|0001\rangle - |1110\rangle$ \\
			$|01\rangle_L$ & (2) $|0011\rangle + |1100\rangle$ & (6) $|0011\rangle - |1100\rangle$ & (10) $|0010\rangle + |1101\rangle$ & (14) $|0010\rangle - |1101\rangle$ \\
			$|10\rangle_L$ & (3) $|0101\rangle + |1010\rangle$ & (7) $|0101\rangle - |1010\rangle$ & (11) $|0100\rangle + |1011\rangle$ & (15) $|0100\rangle - |1011\rangle$ \\
			$|11\rangle_L$ & (4) $|0110\rangle + |1001\rangle$ & (8) $|0110\rangle - |1001\rangle$ & (12) $|1000\rangle + |0111\rangle$ & (16) $|1000\rangle - |0111\rangle$ \\
			\hline
		\end{tabular}
	}
\end{table*}

We prepared four different initial states on the data qubits: $\ket{0000},\ket{0110},\ket{{+}{+}{+}{+}},$ and $\ket{{-}{-}{+}{+}}$.  
The first two states use a circuit which first performs an $X$-stabilizer, then measures the final data qubits in the Z-basis (native to our architecture). 
The last two states use a circuit which first performs a $Z$-stabilizer, then measures the final data qubits in the $X$-basis (by using a single-qubit $H$ gate before the measurement).
These states do not begin in a specific codespace, but rather a superposition of A and B (C) for the $Z$-basis ($X$-basis) initial states.
Therefore, the first stabilizer round will project the initial state into one of \emph{two} possible states, one in each of the possible codespaces. 
For example, the initial state \(|0110\rangle\) will project either into state 4 or into state 8, depending on the ``sign'' assigned by the $X$-check (with an equal chance of either outcome).

Without a flag qubit, the {[}{[}4,2,2{]}{]} code can detect any weight-1 event (i.e. single-qubit error) on the data qubits. 
However, when a single $Z$ fault on the ancilla qubit occurs between the second and third entangler of either stabilizer, it propagates an error ($X$ or $Z$, depending on the stabilizer) onto two data qubits. 
This changes the logical state, but keeps the total $Z$ or $X$ parity the same and is therefore an undetectable weight-1 error in our circuit. 
To make our circuit robust to \emph{all} single qubit errors, we therefore add an additional ancilla, called the ``flag'', that entangles our original ancilla qubit around this dangerous location. 
When a $Z$ error occurs on the ancilla halfway through a stabilizer round, the flag qubit will detect it, improving the logical fidelity of the circuit.

To quantify the error-detecting capability of the {[}{[}4,2,2{]}{]} code, we post-select only the experiment shots where no error was detected. 
Each such shot must meet four requirements:

\begin{enumerate}
	\def\labelenumi{\arabic{enumi}.}
	\item
	\emph{All flag measurements are} \(|0\rangle\)\emph{.}
	\item
	\emph{All $X$ syndrome measurements are identical and consistent with the targeted codespace.}
	\item
	\emph{All $Z$ syndrome measurements are identical and consistent with the targeted codespace.}
	\item
	\emph{The parity of the measured data qubits is consistent with the targeted codespace.}
\end{enumerate}

For requirement 4, we use parities computed from data-qubit measurements to determine if an error occurred. 
This is analogous to using the measurement of the data qubits in a repetition code to infer errors that occur during or after the last round of syndrome extraction.

If any of the above checks fail, an error is considered to have occurred. 
For example, if we started with the state $\ket{0000}$, the four requirements might be violated in the following ways:
\begin{enumerate}
	\def\labelenumi{\arabic{enumi}.}
	\item
	\emph{The flag measurement during the second $X$ stabilizer is triggered.}
	\item
	\emph{The first $X$ stabilizer is $-1$ and the second $X$ stabilizer is 1.}
	\item
	\emph{The first $Z$ stabilizer is $-1$  ($\ket{0000}$ must live in a $Z_{1234} = 1$ codespace).}
	\item
	\emph{The final measured state is $\ket{0001}$.}
\end{enumerate}
The ``circuit throughput'' is defined as the probability that a shot passes all of the post-selection rules.

\begin{figure*}
	\centering
	\smartincludegraphics[width=140mm]{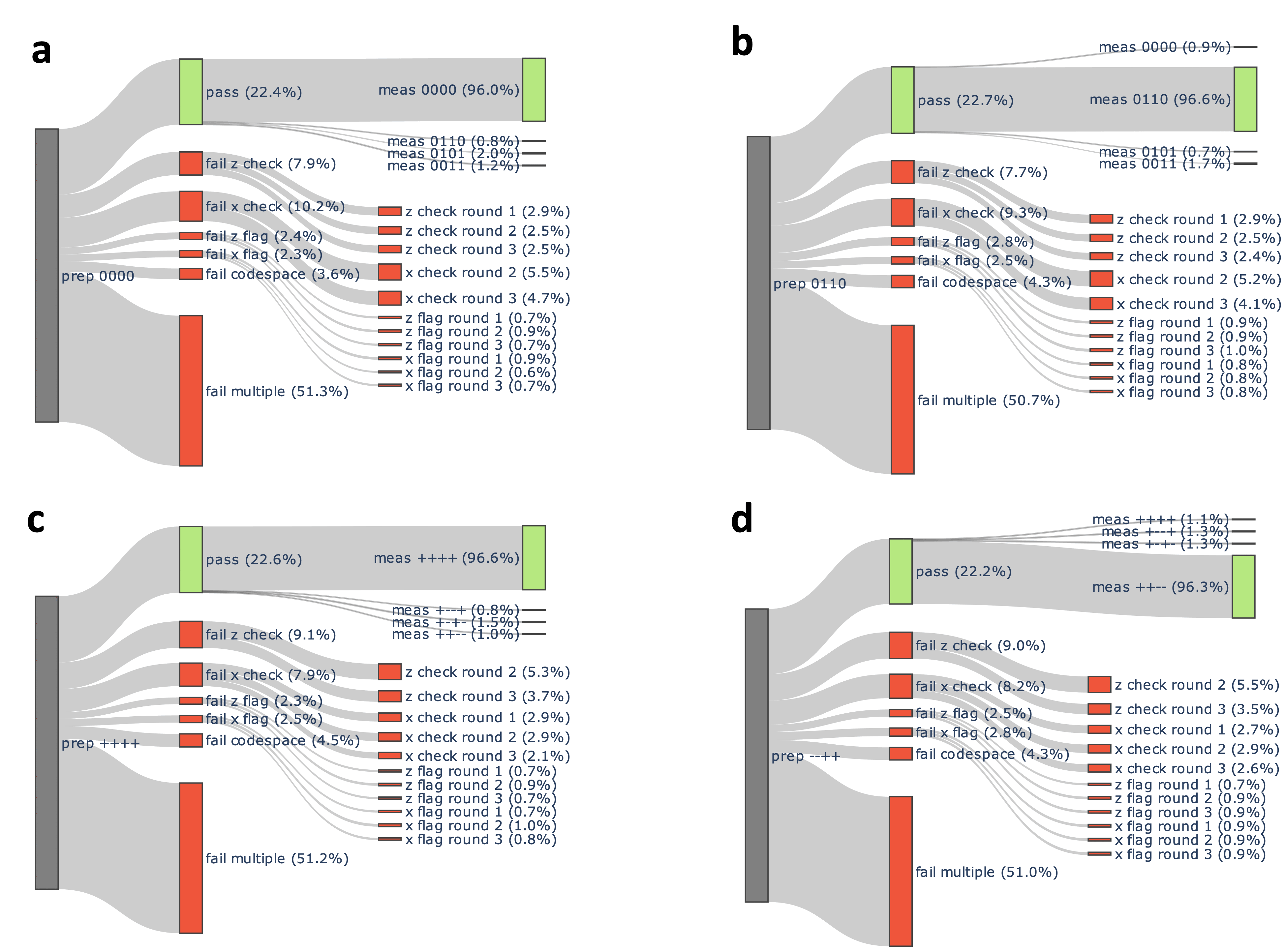}
	\caption{
		\textbf{Classification of all 10k experimental shots of each prepared state}. 
		The percentages associated with the ``pass'' state of the circuit throughput check (green) are normalized to the number of passing shots, while all other rates (red) are normalized to the 10k shots.
		\textbf{a,}~Prepare $\ket{0000}$.
		\textbf{b,}~Prepare $\ket{0110}$.
		\textbf{c,}~Prepare $\ket{{+}{+}{+}{+}}$.
		\textbf{d,}~Prepare $\ket{{-}{-}{+}{+}}$.
	}
	\label{figS52}
\end{figure*}

From these post-selected shots, we then check if the final measured state is in agreement with the state that was prepared, represented by two correct states. 
For example, if we prepared the state $\ket{{-}{-}{+}{+}}$ both the state $\ket{{-}{-}{+}{+}}$ and $\ket{{+}{+}{-}{-}}$ are considered a success. 
The ``circuit fidelity'' is then defined as the probability that a post-selected shot agrees with the prepared state. 
Figure~\ref{figS52} summarizes the throughput and fidelity outcomes of all 10k shots of each prepared state. 
Shots in which multiple post-selection rules were violated were only coarsely categorized, for figure simplicity.

\subsubsection{{[}{[}4,2,2{]}{]} DEM analysis}\label{dem-analysis-422}

\begin{figure*}
	\centering
	\includegraphics[width=140mm]{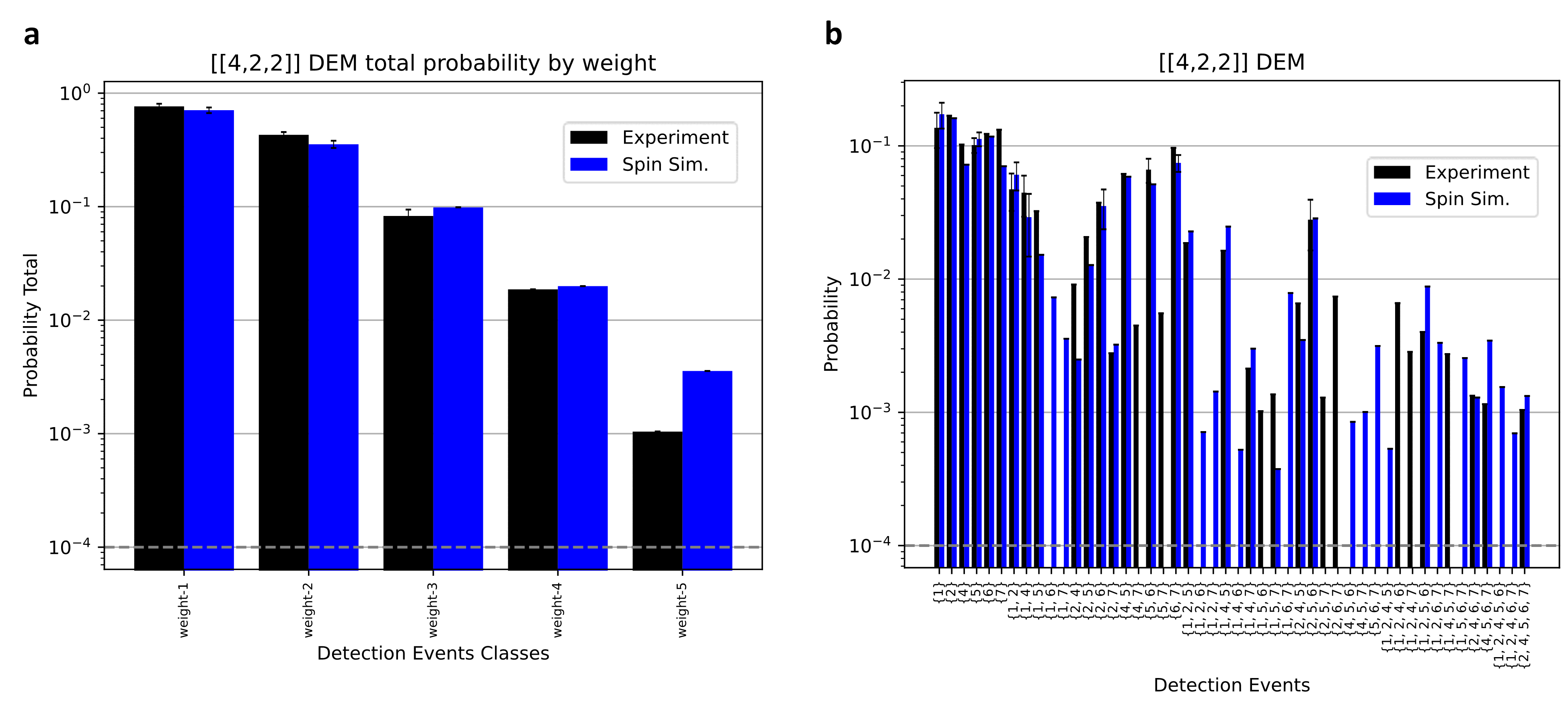} 
	\caption{
		\textbf{DEMs constructed from experimental (black) and simulated (blue) data for the {[}{[}4,2,2{]}{]} code with 3-rounds of syndrome detection.} 
		This is a DEM of size $N = 2\times 4 = 8$. 
		\textbf{a,}~Comparison of total error probabilities for each weight-class of error events.
		\textbf{b,}~Comparison of error probabilities event-by-event. 
		Grey dashed line corresponds to 1/$N$\textsubscript{shot}, $N$\textsubscript{shot}=10,000. 
		Overall, experiment and simulation track well when comparing events in aggregate, especially for lower weight events with pyquest overestimating high weight errors.
	}
	\label{figS53}
\end{figure*}

Directly comparing DEMs constructed from experimental data and simulation data event-by-event we find discrepancies, especially as the weight increases. 
For example, in \figref{figS53}a the \mbox{weight-1} {[}{[}4,2,2{]}{]} events between experiment and simulation match up decently well but diverge as the weight increases above weight-3 because of the ersatz miscalibration model used by the simulator (as discussed in the {[}3,1,3{]} and {[}5,1,5{]} DEM analysis sections as well). 
This uniform DEM behavior between the classical codes ({[}3,1,3{]} and {[}5,1,5{]}) and a quantum-error detecting code ({[}{[}4,2,2{]}{]}) further reinforces our confidence in the quality of our qubit error characterization approach.

One topic not directly addressed in our DEM analysis is a description of what types of events leakage creates. 
This is a rich theoretical topic and is critical to understanding the scalability of the exchange-only architecture. 
We have found no evidence that leakage (fully present in both simulation and experiment) is adversely affecting the DEM analysis.
This is likely thanks to our generous use of RIL gates. 
Future work studying leakage in a quantum code would be a valuable effort. 
For example, specific RIL gates can be removed from a circuit and any clear DEM changes could be analyzed from a leakage-spreading perspective. 
Initial experiments doing just this were performed, but further analysis is needed.

\end{document}